\documentclass[12pt, letterpaper, twoside]{article}

\usepackage[square,sort,comma,numbers]{natbib}
\usepackage{lineno}
\usepackage{lmodern}
\usepackage{amsmath,amssymb}
\usepackage{amsthm}
\usepackage{hyperref}
\usepackage{amsfonts}
\usepackage{verbatim}
\usepackage{url}
\usepackage{graphicx}
\usepackage{float}
\usepackage{booktabs}
\usepackage{mathrsfs}
\usepackage{epstopdf}
\usepackage{cancel}
\usepackage{multirow}

\usepackage{MnSymbol}
\usepackage[dvipsnames]{xcolor}
\usepackage[margin=1in]{geometry}
\usepackage[font=small]{caption}
\usepackage{algorithm}
\usepackage{algpseudocode}
\algrenewcommand\algorithmicrequire{\textbf{Input:}}
\algrenewcommand\algorithmicensure{\textbf{Output:}}
\modulolinenumbers[5]
\allowdisplaybreaks

\title{Advances in Scaling Community Discovery Methods for Large Signed Graph Networks}

\author{Maria Tomasso \and Lucas Rusnak \and  Jelena Te\v{s}i\'{c}}

\date{Feb 15 2022}

\begin{document}


\maketitle

\begin{abstract}
{Community detection is a common task in social network analysis (SNA) with applications in a variety of fields including medicine, criminology, and business.  Despite the popularity of community detection, there is no clear consensus on the most effective methodology for signed networks.  In this paper, we summarize the development of community detection in signed networks and evaluate current state-of-the-art techniques on several real-world data sets.  First, we give a comprehensive background of community detection in signed graphs. Next, we compare various adaptations of the Laplacian matrix in recovering ground-truth community labels via spectral clustering in small signed graph data sets. Then, we evaluate the scalability of leading algorithms on small, large, dense, and sparse real-world signed graph networks. We conclude with a discussion of our novel findings and recommendations for extensions and improvements in state-of-the-art techniques for signed graph community discovery in real-world signed graphs.}
{Sign Graph Clustering, Community Discovery, Sparse Networks}\\
Classification Here
\end{abstract}

\section{Introduction}
\label{sec:intro}

The rise of social media interactions has illuminated an increasing necessity for a robust understanding of social network analysis, and social network theory has provided explanations for a variety of social phenomena ranging from individual creativity to corporate profitability. Community discovery has proven valuable in many areas of application, including detection of bot activity and fraud in criminology, identification of customer segmentation in marketing, characterization of \emph{ astroturfing} in political science, detection of cancers via diagnostic imaging, and quantification of environmental hazards in public health \cite{karatas_application_2018}. In an era dominated by social media communication, the community detection tools developed specifically from social network theory can help researchers understand trends and propagation patterns within online communities \cite{2020Lia}. 

Users are generally represented as vertices (nodes) in a graph, while edges are defined based on users' friendships and interactions with posts or re-posts; they are generally based on any social interaction on a media platform between users. A plethora of methods has been proposed to transform social media interactions in a simple graph network where edges exist along user interactions, but do not exist if the connection is unknown. With the increased richness of social media interactions and additional information in the types of interactions that can exist in the social networks today (e.g., reviews, comments, shares, friends, blocked users), researchers have turned to richer interpretations of the edges in graph networks (signs, weights), recently turning their attention to the use of signed graph networks \cite{esmailian_community_2015}. A community within a network is defined as a partitioning of nodes such that nodes within the same cluster are similar and usually strongly connected, while nodes in different clusters are dissimilar and usually weakly connected; in essence, similar nodes should be grouped together. In real data sets, community structure is almost always present to some degree \cite{2002girvan}. Community detection in unsigned networks traditionally relies on the absence of connections between vertices (e.g., users) to determine if they belong in different communities. The presence of negative links provides affirmative evidence of their dissimilarity, allowing the use of richer signed network analysis for community detection. When negative edges are included in a network, we can study social dynamics and stability in respect to friendship and enmity in more depth \cite{antal_social_2006,Leskovec2010b}, or expand to new application domains such as the behavior of the brain \cite{saberi_topological_2021}.

In this paper, we present an overview of the work that has been done on community detection in signed networks to date. The methods are divided into two top level categories: methods adapted from unsigned methods and methods that work only for signed graphs, with additional subcategories. We then compare state-of-the-art methodologies on small signed networks with known ground-truth communities and compare their ability to recover the ground-truth labels based on edge signs alone. Finally, we evaluate the scalability in terms of effectiveness and efficiency of leading clustering methodologies on real signed networks \cite{snapnets} as they increase in complexity.  Signed graph definitions are outlined in Section~\ref{sec:graph}, while Section~\ref{sec:adapted_unsigned} describes unsigned clustering methods adapted for signed graphs, and Section~\ref{sec:signedNew} reviews novel methods that utilize signed graph characteristics such as balance or the random walk gap.  

\paragraph{Prior Surveys:} A comprehensive survey on mining techniques for signed graphs \cite{tang_survey_2016} includes several community detection methods. The survey had a much broader scope on signed graph analysis, from node ranking, classification, and embedding, over link and sign prediction, to information diffusion and recommendations in signed graphs \cite{tang_survey_2016}. In the same year, a survey of spectral clustering methods for unsigned and signed graphs was published \cite{2016gallier}.  This survey provides a thorough overview of spectral methods and Laplacian variants for both unsigned and signed graphs. In this paper, we focus on community detection in signed graphs and provide a comprehensive examination of spectral methods and non-spectral methods for community detection with respect to incremental development and suitability based on data characteristics. \\
\paragraph{Reference Searching:} In the first stage of the literature review, both forward and backward reference searching were used to find significant contributions to the field. After forward and backward reference searching, a systemic literature review was conducted to find any publications on signed graph clustering that were previously missed. The following search terms were used: \emph{“signed" AND “graph" AND “clustering"} and \emph{“signed" AND “graph" AND “community" AND “detection"}. All results were saved and manually reviewed for relevance. 


\subsection{Signed Graph Definitions and Methods}
\label{sec:graph}

\begin{figure}[!th]
\centering
\includegraphics[width=3.5in]{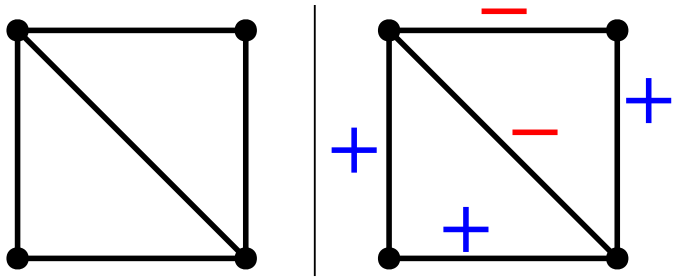}
\caption{Left: Unsigned graph with 4 vertices and 5 edges; \\Right: Signed graph with the same underlying unsigned graph: edges labeled as $+$ or $-$.}
\label{fig:Signed}
\end{figure} 

A \textbf{graph} $\mathcal{G}$ consists of two disjoint sets: a set of \emph{ vertices v} $, v \in \mathcal{V}$ and a set of \emph{ edges e}, $e \in \mathcal{E}$.  In this paper, graph and network can be assumed to be synonymous. Graphs can be \textbf{directed} or \textbf{undirected}. In a directed graph, an edge may connect node i to node j without node j necessarily being connected to node i.  In an undirected graph, if node i is connected to node j, then node j must be connected to node i.  A graph can also be \textbf{weighted}, which means that each edge has a 'weight' attribute that can represent the strength of the connection.  In a \textbf{signed} graph, edges are assigned  +1 or -1 weights, as illustrated in Figure~\ref{fig:Signed}. In graph theory, a signed graph is \textbf{balanced} if the product of edge signs around every cycle is positive. In a \textbf{complete} graph, every pair of vertices is connected by an edge.  A complete graph is \textbf{weakly balanced} if and only if it can be divided into multiple sets of mutual friends, with complete mutual antagonism between each pair of sets.

The \emph{ graph density} $d$ of undirected graphs is the ratio of the number of edges $e$ with respect to the maximum possible edges in a fully connected graph with $v$ vertices; see Eq.~\ref{eq:d}. 
 \begin{equation}
d = \frac{2 \cdot e}{v \cdot (v-1)}
\label{eq:d}
\end{equation}
 
A \textbf{dense} graph is a graph with a number of edges close to the maximum number of edges. With density scores of $0.483$, $0.782$, and $0.225$ respectively, Highland Tribes \cite{1954Read}, Sampson \cite{sampson}, and Correlates of War \cite{COW} are three examples of real-world signed and dense graphs. A \textbf{sparse} graph has very few edges relative to the number of nodes. Most social media networks have a high number of vertices (users) $v$ and relatively small number of edges $e$ as they are only connected to a small fraction of the overall community with density score $0.1$ \cite{snapnets}.

\subsection{Background}

A \emph{community} in graph theory is a set (cluster) of nodes that are similar and generally densely connected to other nodes within the cluster and dissimilar and sparsely connected to those outside of the cluster relative to given graph or data metrics. However, in signed graph theory a \emph{community} has the additional stipulation that the cluster is densely-positive and sparsely-negative within its connected component, and densely-negative and sparely-positive to the nodes outside the cluster. As a signed network graph innately represents expressed opinions between entities (vertices) through edge signs between them, the signed graph balancing model proved successful in social science in the 20th century. Social balance theory, described in Section~\ref{ssec:balance}, is a branch of signed graph theory proposed by \cite{Heider} and developed by \cite{Har0} in the 1940s and 1950s. It allows certain well-behaved graphs to be 'perfectly' partitioned so that negative edges exist only between clusters, and positive edges exist only within clusters.  In the real world, however, such well-behaved data sets are rare. Modularity-based metrics seek densely connected clusters using optimization techniques. Yang et al. introduced FEC, an algorithm for signed graphs that was designed for densely connected networks \cite{yang2007}.  FEC treats the sign and density of edges as clustering attributes. Gomez et al. refined modularity metrics and extended existing methods to signed graphs that were directed, weighted, or contained loops \cite{2009Gomez}. Both FEC and the Gomez method were designed for smaller signed graphs with dense connections: approaches were prohibitively slow even on our smallest networks, and fail to produce any results. 

Prior to the explosion of Social Network analysis methods, all signed networks were assumed to be small and relatively densely connected; be it inter-personal relationships or warring counties such as the modeling of diplomatic relations in the Middle East \cite{moore_1978}, South Asia \cite{moore_1979}, and Allied and Axis powers during World War II \cite{axelrod_bennett_1993}. Real world datasets and particularity social networks tend to follow power law degree distribution, as most nodes are only directly connected to a small percentage of the overall network and only few nodes are highly connected.  While spectral clustering is a powerful technique for the detection of graph communities, the eigenvalues of signed graphs present a substantial obstruction in the development of a parallel spectral theory that is meaningful for the data. In \cite{2021dallamico} is it observed that these sparse graphs with a power-law degree distribution present two primary issues with unsigned spectral clustering: (1) the eigenvalues of a sparse network tend to spread, which can obscure the largest and smallest eigenvalues and makes the informative eigenvalues difficult to isolate; and (2) high heterogeneity in the degree distribution modifies the $i^{th}$ entry of the informational eigenvectors in proportion with the degree of node $i$, known as "eigenvector pollution" by the authors \cite{2021dallamico}. 

In this paper, we survey the signed graph community discovery methods of the 21st century. These techniques generally fall into two categories, and we explore them in the following sections. In Section~\ref{sec:adapted_unsigned}, we review the signed graph adaptations from algorithms developed for unsigned graphs using discrete optimization techniques. We review novel methods that utilize signed graph characteristics such as balance or the random walk gap in Section~\ref{sec:signedNew}. In Section~\ref{sec:exp1} we provide a comprehensive study on twelve methods on real-world datasets of varying complexity and summarize the effectiveness each. Section~\ref{sec:exp2} concludes with a discussion on timing and scalability to inform the creation of synthetic data to further test the algorithms.

\section{Adaptations of Unsigned Spectral Clustering Methods to Signed Graph Clustering}
\label{sec:adapted_unsigned}
\subsection{Clustering for Unsigned Graphs}
\label{ssec:unsignedobjectives}

Before examining methods for signed graphs, the algorithms developed for unsigned graphs must be understood.  The simplest non-trivial case in undirected graph clustering is 2-way partitioning.  This involves separating the nodes of the graph into two groups such that nodes in the same group are strongly connected and nodes in opposite groups are weakly connected.  To accomplish a 2-way partitioning, two items are needed: (1) a criterion that defines a 'good' partition, and (2) a method to efficiently optimize the criterion.

\begin{figure}[!ht]
\centering
\includegraphics[width=3in]{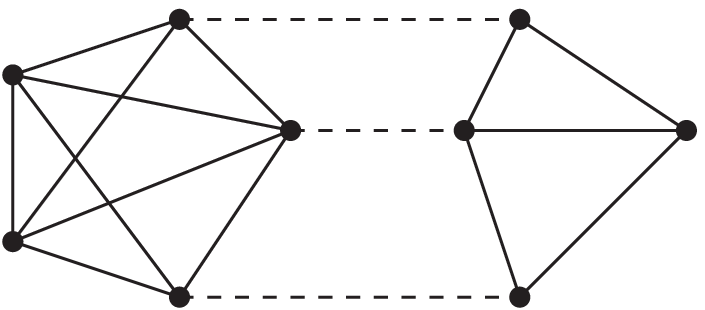}
\includegraphics[width=3in]{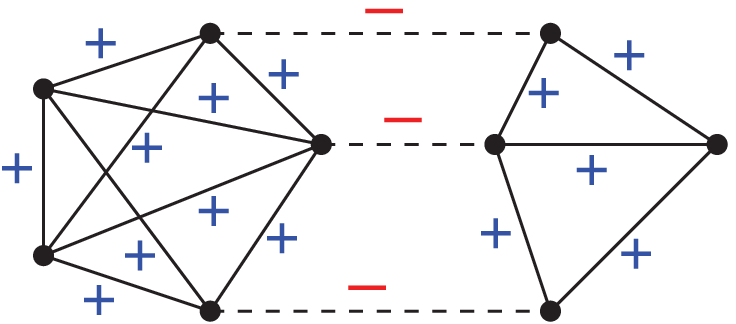}
\caption{(left) Unsigned graph cut set, where cuts are dashed edges, and the criteria scores described in Section~\ref{ssec:unsignedobjectives} are: $cut(X,Y) = 3$, $rcut(X,Y)=1.35$, and $ncut(X,Y)=.933$; (right) An equivalent balanced signed graph cutset with edges  labeled as $+$ or $-$. Dashed lines illustrate the ideal Harary cut.}
\label{fig:SGCutEx}
\end{figure}

\paragraph{Criteria for 2-way Partitioning} Many criteria have been proposed for 2-way graph partitioning.  The first measure we will introduce for assessing 2-way clustering of a graph is the graph cut. For an unsigned graph G with disjoint clusters X and Y, the \emph{2-way graph cut} is defined as $cut(X, Y) = \sum_{i \in X j \in Y} A_{ij}$.  The cut is essentially the number of edges or, for a weighted graph, the sum of weights between clusters.  Since the typical goal of clustering is to group densely connected nodes together, choosing X and Y to minimize the cut is a good first step.  Unfortunately, the cut does not account for the size of clusters, and the optimal solution can separate few or single vertices if applied as is.  To remedy this, the \emph{ratio cut} is introduced. For an unsigned graph G with disjoint clusters X and Y, the \emph{2-way ratio cut} is defined as $rcut(X, Y) = cut(X, Y)(\frac{1}{|X|}+\frac{1}{|Y|})$, as illustrated in Figure~\ref{fig:SGCutEx}(left).  The ratio cut takes the size of the clusters into consideration by minimizing the graph cut relative to the sizes of each cluster.  Shi and Malik refine the ratio cut to consider the strength of the connection of the nodes in X and Y to the rest of the graph with the \emph{normalized cut} \cite{shi_normalized_2000}. For an unsigned graph G with disjoint clusters X and Y, the \emph{2-way normalized cut} is defined as $ncut(X, Y) = cut(X, Y)(\frac{1}{vol(X)}+\frac{1}{vol(Y)})$, where $vol(P)$ represents all of the weights of all edges adjacent to nodes in the cluster $P$. By including volume in the normalized cut objective, the cut is minimized relative to both the size of the clusters and the connectivity of the graph.

\paragraph{Extending the Criteria to k-way Partitions} Real networks have more than two communities and a need for efficient k-way partitioning algorithms.  The 2-way partitioning algorithms provide a simple recursive technique to perform k-way partitioning \cite{shi_normalized_2000}. First, partition the graph into two clusters. Then recursively run the 2-way partitioning algorithm separately on the subgraph for each cluster. While this technique can be efficiently computed, it ignores the higher-order spectral information of the graph.  As an alternative, k-way generalizations of the ratio cut and normalized cut have been introduced and are defined as follows: for an unsigned graph G with disjoint clusters $X_1$,...,$X_k$, the \emph{k-way ratio cut} is defined as  $rcut(X_1,...,X_k) = \sum_{i=1}^k \frac{cut(X_j, \bar{X_j})}{|X_j|}$ and  the \emph{k-way normalized cut} is defined as $ncut(X_1,...,X_k) = \sum_{i=1}^k \frac{cut(X_j, \bar{X_j})}{vol(X_j)} $. From \cite{shi_normalized_2000}, we know that 2-way partitions can be solved efficiently for unsigned graphs.  Unfortunately, the same is not true for k-way partitions, and finding a global optimum is NP-complete for most graphs.  Thus, approximation methods are used to estimate solutions to the k-way criteria.  Researchers initially tried to use greedy algorithms and gradient descent to find solutions to k-way clustering problems, but these approaches often failed to find a global optimum due to the high dimensionality of graph data and nonlinearity of the criteria.  As an alternative, Shi and Malik developed a technique for approximating k-way normalized cut solutions by formulating them as generalized eigenvalue problems \cite{shi_normalized_2000}.  This approach became known as spectral clustering; twenty years later, it is still considered to be a foundational algorithm in graph clustering.

\subsubsection{Spectral Clustering}
\label{sssec:spectral} 

Spectral clustering begins by finding the Laplacian of the matrix representation of the network. Since several variants of the Laplacian exist, there are multiple versions of the spectral clustering technique.  After finding the Laplacian, the eigenvalues are computed. Note that the algorithm assumes that all eigenvalues of the Laplacian are non-negative (i.e., the Laplacian is positive semidefinite), and that the eigenvalues of the Laplacian can be efficiently computed. After the eigenvalues are found, they are plotted in increasing order, and the eigengap, the largest 'early' increase in sequential eigenvalues, is identified.  The location of the eigengap provides options for the value of $k$, the number of communities in the graph.  After identifying the number of clusters, k-means can be applied to cluster the communities \cite{shi_normalized_2000}. The Laplacian matrix of an unsigned graph $G$ is defined as $L = D - A$. $D$, the degree matrix, is a diagonal matrix such that the $(i,i)^{th}$ entry represents the degree of vertex $v_i$. $A$, the adjacency matrix, contains edge weight information such that entry $(i, j)$ represents the weight of the edge between vertices $v_i$ and $v_j$.  If no such edge exists, the entry is 0. The spectral clustering algorithm is described in Alg.~\ref{alg:USC} and consists of four steps: (1) calculate the Laplacian $L$ (or the normalized Laplacian); (2) calculate the first $k$ eigenvectors (the eigenvectors corresponding to the $k$ smallest eigenvalues of $L$); (3) consider the matrix $U$ formed by the first $k$ eigenvectors; the $i^{th}$ row defines the features of graph node $i$; (4) cluster the graph nodes based on $u_i$ features using k-means clustering as outlined in Section~\ref{sssec:KMeans} and in Alg.~\ref{alg:KMeans}. Since minimizing the normalized cut is NP-complete, the goal of the original and all subsequent spectral clustering algorithms is to find an approximate discrete solution efficiently. The two central problems of spectral clustering are the criterion that determines if a partition is 'good' and how partitions fitting the above criterion can be efficiently computed.  

\emph{ Laplacian Variants} The standard graph Laplacian matrix as defined in Section \ref{sssec:spectral} is symmetric and positive semidefinite, meaning the eigenvalues are real and non-negative \cite{von_luxburg_tutorial_2007}.  Additionally, two normalized variants of the Laplacian are commonly used in clustering, and they are defined as follows:  the {\emph symmetric normalized Laplacian} is $L_{sym} = D^{-1/2}LD^{-1/2}$, and the {\emph random walk Laplacian} is $L_{rw} = D^{-1}L$.  For undirected graphs, both $L_{sym}$ and $L_{rw}$ are positive semidefinite and have real, non-negative eigenvalues \cite{von_luxburg_tutorial_2007}.

\emph{ Eigenvalue computation} is often expensive and prone to error for very large matrices, so if reasonable bounds on the problem are known (i.e., the maximum number of clusters), the problem can be reduced to finding the k smallest eigenvalues. The eigengap heuristic in spectral clustering  indicates the number of clusters to use, but for noisy data sets the eigengap may be relatively small and difficult to detect.  If the eigengap is large, however, the first k eigenvalues can be computed relatively efficiently through the use of Krylov subspaces or the power method \cite{von_luxburg_tutorial_2007}. 

\begin{algorithm}
\caption{Normalized Spectral Clustering \cite{shi_normalized_2000}}\label{alg:USC}
\begin{algorithmic}
\State \textbf{Input:} Adjacency matrix $A \in \mathbb{R}^{n \times n}$ of signed graph $\Sigma$; $k$ number of clusters to construct:
\State \textbf{Step 1:}  Compute the normalized Laplacian matrix $L_{n \times n}$ and diagonal matrix D of $A$.
\State \textbf{Step 2:} Compute the first $k$ eigenvectors $l_1,...,l_k$ corresponding to the $k$ smallest eigenvalues of $L_{n \times n}$ .
\State \textbf{Step 3:} Construct $U_{n \times k} \in \mathbb{R}^{n \times k}$ as the matrix containing the vectors $l_1,..., l_k$ as columns.
\State \hskip 1em  Let $u_i$ be the vector corresponding to the $i$-th row of $U_{n \times k}$, $i=1,...,n$, $u_i \in \mathbb{R}^{k}$. 
\State \hskip 1em  $u_i$ defines the k-dimensional features of graph node $i$ in $U_{n \times k}$,  $i=1,...,n$. 
\State \textbf{Step 4:} Cluster the graph nodes  $i=1,...,n$ based on $u_i$ features using k-means clustering described in Alg.~\ref{alg:KMeans}. 
\State \textbf{Output:} Cluster labels $l=1, ..k$ for all n nodes based on $u_i$ vector closeness to final clusters centroids $C_1, ... C_k$.
\end{algorithmic}
\end{algorithm}

\subsubsection{k-means Clustering and Weighted Kernel k-means Clustering}
\label{sssec:KMeans}

\emph{ The k-means algorithm} is used to partition a given set of observations into a predefined number ($k$) clusters. The algorithm starts with a set of $k$ center-points and goes through multiple iterations to find optimal cluster centroids $C_1,...,C_k$.  Here, we present the k-means++ algorithms used in experimental comparisons in Section~\ref{sec:exp1} and Section~\ref{sec:exp2}. The k-means++ algorithm distributes the initial centroids over the given data to minimize the probability of bad outcomes\cite{2007Arthur} by a very simple randomized seeding technique, as illustrated in Algorithm~\ref{alg:KMeans}. 

\begin{algorithm}[!ht]
\caption{k-means++ \cite{2007Arthur}}\label{alg:KMeans}
\begin{algorithmic}
\State \textbf{Step 1:} Select centroids $C_1, C_2, ... C_k$ by taking uniformly a random data point from the data $X$ and mark it as centroid $C_1$
\For           select centroid $C_i, i \in [2, k]$
\State Choose $C_i$ $C_i = x, max_{x \in X}( \frac{D(x)^2}{\sum_{x \in X} D(x)^2}), D(x)= min_{l \in [1, i-1]}(d(x,C_l))$
\EndFor
\State \textbf{Step 2:} Iteratively compute new centroids for all data $X$ 
\State Iteration 0: $t = 0, t = limit$
\While $t \ge limit$ 
\State $t = 0$ 
\For $i, i \in [1,k]$
\State $S'_i = \big \{ x_p : \big \| x_p - C_i \big \|^2 \le \big \| x_p - C_j \big \|^2 \ \forall j, 1 \le j \le k \big\}$
\State Previous centroid value: $C'_i = C_i$
\State New centroid value: $C_i = \frac{1}{|S'_i|} \sum_{x_j \in S'_i} x_j$
\State $t =  max(t, d(C_i,C'_i))$
\EndFor
\EndWhile
\State \textbf{Step 3:} Assign $x, x \in X$ cluster label $j$ so that $min_{i \in [1,...k]}(\big \| x - C_i \big \|^2)= \big \| x - C_j \big \|^2$
\end{algorithmic}
\end{algorithm}

During each update step $t$ in Algorithm~\ref{alg:KMeans}, all observations $x$ are assigned first to their nearest center-point $S^t_i$. Next, the center-points $C_i^{t+1}$ are repositioned by calculating the mean of the assigned observations to the respective center-points. As shown in  Algorithm~\ref{alg:KMeans}, this update process reoccurs until the center-point update distance $d(C^{(t+1)}_i,C^{(t)}_i)$ is smaller than the specified $limit$. There is only a finite number of possible assignments for the amount of centroids and observations available. As each iteration has to result in a better solution, the algorithm always ends in a local minimum. k-means++ approximately can be computed in $O(\log n)$ time \cite{2007Arthur}.

\begin{algorithm}[!ht]
\caption{Batch Weighted Kernel K Means Clustering \cite{dhillon_kernel_2004}}\label{alg:wkernelkmeans}
\begin{algorithmic}
\State \textbf{Input:} The kernel matrix $K$, the number of clusters $k$, and the weights for each input $w$, the inital clusters $\pi_1^{(0)}, \ldots, \pi_k^{(0)}$ (optional), and maximum number of iterations $t_{max}$  (optional)
\State \textbf{Step 1:} Initialize the $k$ clusters $\pi_1^{(0)}, \ldots, \pi_k^{(0)}$ arbitrarily if initial clusters were not provided.
\State \textbf{Step 2:} Set $t=0$
\State \textbf{Step 3:} For each point $a_i$ and every cluster $c$, compute the distance between $a_i$ and the centroid of $c$ as:  $d(a_i, m_{c}) = K_{ii} - \frac{2 \sum_{a_j \in \pi_c^{(t)}} w_jK_{ij}}{\sum_{a_j \in \pi_c^{(t)}} w_j} + \frac{\sum_{a_j,a_l \in \pi_c^{(t)}} w_jw_lK_{jl}}{(\sum_{a_j \in \pi_c^{(t)}}w_j)^2}$
\State \textbf{Step 4:} Find the updates index for each point $a_i$ as $c^*(a_i) = argmin_c d(a_i, m_c)$ and update the clusters with $\pi_c^{(t+1)} = \{a : c^*(a_i) = c\}$
\State \textbf{Step 5:} If not converged and $t < t_{max}$, increment $t$ by 1 and go to Step 3. 
\State \textbf{Output:} Cluster labels $\pi_1^{(t+1)}, \ldots,\pi_k^{(t+1)}$
\end{algorithmic}
\end{algorithm}

\pagebreak{}

The \emph{ weighted kernel k-means clustering} family of algorithms improves k-means clustering by introducing the weighted kernel approach, which maps the data to a higher-dimensional space and allows the separation of nonlinear components \cite{dhillon_kernel_2004}. Kernel function can be polynomial, Gaussian, or Sigmoid, and the correct choice depends on target data characteristics.  For weighted kernel k-means clustering, we need to choose the kernel matrix $K$ first. If an input matrix is given, it is the weighted kernel matrix. If a standard graph partitioning objective is being used, we obtain the initial clusters using one of the following initialization methods: random, spectral, negative $\sigma$ shift, or METIS \cite{METIS}, a fast, multi-level graph partitioning algorithm that produces equally sized clusters. After we obtain initial clusters, we make kernel matrix $K$ positive definite by adding to the diagonal. Finally, we oscillate between running batch weighted kernel k-means and incremental weighted kernel k-means (local search) until convergence. The sensitivity of the approach lies in the selection of the kernel matrix. The described spectral approach of finding eigenvectors, then performing clustering on features derived from eigenvectors, and its extensions and improvements proved to be highly effective on unsigned graphs. In the next section, we present the adaptations of unsigned graph clustering by applying spectral methods to signed graphs.

\subsection{Clustering for Signed Graphs}
\label{ssec:spectral_signed}

Unsigned graph clustering methods described in Section~\ref{ssec:unsignedobjectives} can be applied to signed graphs by either dropping all negative edges and hoping that the positive interactions produce the clusters, or all edges may be treated as positive, which returns the equivalent clustering of the underlying graph ignoring all sentiments; either way a lot of information is lost.  In this section we present state-of-art modifications to signed graph clustering methods, and new methods that build upon unsigned graph clustering baseline.   

\subsubsection{Modification of Laplacian Matrix for Signed Spectral Clustering}

Spectral clustering assumes that all the eigenvalues of the Laplacian are \emph{ nonnegative} and \emph{ real}. The standard Laplacian matrix of a signed graph is indefinite \cite{kunegis_spectral_2010} and will not yield real, non-negative eigenvalues.  Moreover, accurate and efficient eigenvalue computation for large and sparse matrices is an open problem without signed graph extension. Thus, any method seeking to adapt spectral clustering to signed graphs must ensure that (1) eigenvalues are real and non-negative, and (2) the new procedure is scalable to large networks.  How do we compute a Laplacian for a signed graph and ensure that it is positive semidefinite? The Laplacian matrix introduced in \cite{kunegis_spectral_2010} is indefinite, and modifications were made using the signed degree matrix, with the {\em signed Laplacian matrix} of a graph $G$ as $\bar{L} = \bar{D} -A$, where $\bar{D}$ is the signed degree matrix given by $\bar{D}_{ii} = \sum_{j \sim i}|A_{ij}|$. Kunegis et al. \cite{kunegis_spectral_2010} demonstrated that this signed Laplacian is positive semidefinite and, in some cases, positive definite, thus guaranteeing this Laplacian is a suitable basis for spectral clustering. Moreover, spectral clustering using the signed Laplacian is shown to be equivalent to the k-way signed ratio cut problem, which counts positive edges between clusters and negative edges within clusters \cite{kunegis_spectral_2010}. A more natural signed graph Laplacian that possesses the expected relationship to its underlying incidence matrix as well as the signed-path property on the adjacencies was first presented in \cite{2013zaslavsky}.

\emph{Symmetric normalized Laplacians} tend to yield better results than unnormalized Laplacians for graphs with skewed degree distributions \cite{kunegis_spectral_2010}. Kunegis et al. propose two ways of normalizing signed Laplacians. First, they define the random walk normalized Laplacian for signed graphs as $\bar{L}_{rw} = I - \bar{D}^{-1}A$ and show that the $\bar{L}_{rw}$ matrix is positive semidefinite \cite{kunegis_spectral_2010}. Second, they define the symmetric normalized Laplacian for signed graphs as ${\bar{L}}_{sym}$ = $ {\bar{D}}^{-1/2}\bar{L}\bar{D}^{-1/2}$ = ${I} - {\bar{D}}^{-1/2}A\bar{D}^{-1/2}$, where ${I}$ is the identity matrix. The signed Laplacian matrix of a graph is positive-definite if and only if the graph is unbalanced \cite{kunegis_spectral_2010}. For a signed graph G, the signed graph cut is given by $scut(G) = 2 \cdot cut^{+}(X, Y) + cut^{-}(X, X) + cut^{-}(Y, Y)$. The signed ratio cut is given by $SignedRatioCut(X, Y) = scut(X, Y)(\frac{1}{|X|}+\frac{1}{|Y|})$. The signed normalized cut is given by $SignedNormalizedCut(X, Y) = scut(X, Y)(\frac{1}{vol(X)}+\frac{1}{vol(Y)})$, where $vol(X)$ and $vol(Y)$ represent the sum of the degrees of the nodes in X and Y, respectively. 

The \emph{balanced normalized signed Laplacian} is proposed by Zheng et al. \cite{zheng_spectral_2015} as an extension of the normalized signed Laplacian defined in \cite{kunegis_spectral_2010}, with an embedding map rather than an index of partitions. This yields additional information on the similarity between nodes rather than simply assigning cluster labels.  Additionally, the authors argue that an embedding map is more likely to yield an approximate global solution rather than local optima.  Zheng et al. take a two-step approach: (1) the Rayleigh quotient of the random walk normalized Laplacian is used as an objective function to achieve the embedding, and (2) an objective function derived from the normalized signed cuts in \cite{kunegis_spectral_2010} is used to complete clustering. 

\emph{Geometric Laplacian means} are proposed as a way to address  shortcomings in \cite{zheng_spectral_2015} and its inability to recover ground-truth labels in real data sets. The arithmetic mean of the positive-edge and negative-edge Laplacians introduces noise to the embedding of the data points, and with the arithmetic mean the smallest eigenvectors of the Laplacian do not necessarily correspond to the smallest eigenvalues.  The authors propose the use of the geometric mean of the positive-edge and negative-edge Laplacians to remedy these issues, although they concede that the geometric mean is more computationally expensive than the arithmetic mean and not well-suited to large, sparse networks \cite{mercado2016}. The latest work modifies the Laplacian by combining the positive and negative Laplacians using the \emph{matrix power means} \cite{mercado2019}.  This approach further improved the results in Section~\ref{sec:exp1}.

\subsubsection{Modification of k-way Signed Ratio Cut Criteria for Signed Clustering} \label{ssec:spectral}

Chiang et al. introduce the \emph{ balance normalized cut}, a criterion for k-way clustering problems that is analogous to the normalized cut \cite{chiang_scalable_2012}. The balance normalized cut is motivated by the failure of the signed k-way ratio cut (Eq.~\ref{eq:kcut}) to be minimized by any representation of partitions ${\{{x_1},\ldots, {x_k}\}}$. Additionally, the k-way ratio cut (Eq.~\ref{eq:kcut}) inherently has less available information about each node than the 2-way signed ratio cut when $k > 2$.  If there are only two clusters, $c_1$ and $c_2$, and we know that node i and node j both do not belong to $c_1$, then they both belong to $c_2$ and are therefore in the same cluster. However, if $k > 2$, we cannot infer that if two nodes are both excluded from one cluster, they must share another cluster. Without modification, minimizing the k-way signed ratio cut will not yield an optimal solution as proved by the authors \cite{chiang_scalable_2012}. 

\begin{equation}
\sum_{c=1}^k \frac{{x_c^T}\bar{L}{x_c}}{{_c^T}{x_c}}
\label{eq:kcut}
\end{equation}

Chiang et al. propose a series of new objectives that extended well to k-way partitioning \cite{chiang_scalable_2012}.  In the following definitions, ${x_c}$ represents the set of points assigned to cluster $c$; $A$, $A^+$, and $A^-$ represent the full adjacency matrix, the positive edge-only adjacency matrix, and the negative edge-only adjacency matrix respectively; $D$, $D^+$, and $D^-$ represent the diagonal matrix, the positive edge-only diagonal matrix, and the negative edge-only diagonal matrix; and $L = D-A$, $L^+=D^+-A^+$, and $L^-=D^--A^-$.
The \emph{positive ratio association} maximizes the number of positive edges within each cluster relative to the cluster's size, equal to the following:
\begin{equation}
\underset{\{{x_1},...,{x_k}\} \in I}\max \sum_{c=1}^{k}\frac{{x_c^T}A^{+}{x_c}}{{x_c^T}{x_c}}. 
\end{equation}
The \emph{negative ratio association} minimizes the number of negative edges within each cluster relative to the cluster's size:
\begin{equation}
\underset{\{{x_1},...,{x_k}\} \in I}\min \sum_{c=1}^{k}\frac{{x_c^T}A^{-}{x_c}}{{x_c^T}{x_c}}. 
\end{equation}
The \emph{positive ratio cut} minimizes the number of positive edges between clusters: 
\begin{equation}
\underset{\{{x_1},...,{x_k}\} \in I}\min \sum_{c=1}^{k}\frac{{x_c^T}L^{+}{x_c}}{{x_c^T}{x_c}}. 
\end{equation}
The \emph{negative ratio cut} maximizes the number of negative edges between clusters:
\begin{equation}
\underset{\{{x_1},...,{x_k}\} \in I}\max \sum_{c=1}^{k}\frac{{x_c^T}L^{-}{x_c}}{{x_c^T}{x_c}}. 
\end{equation}
The \emph{balance ratio cut} combines the positive ratio cut with the negative ratio association and simultaneously minimizes the number of positive edges between clusters while minimizing the number of negative edges within clusters: \begin{equation}
\underset{\{{x_1},...,{x_k}\} \in I}\min \sum_{c=1}^{k}\frac{{x_c^T}(D^{+}-A){x_c}}{{x_c^T}{x_c}}.
\end{equation}
The \emph{balance ratio association} combines the negative ratio cut with the positive ratio association and simultaneously maximizes the number of positive edges within clusters while maximizing the number of negative edges between clusters: 
\begin{equation}
\underset{\{{x_1},...,{x_k}\} \in I}\max \sum_{c=1}^{k}\frac{{x_c^T}(D^{-}+A){x_c}}{{x_c^T}{x_c}}
\end{equation} 
The \emph{balance normalized cut} is very similar to the balance ratio cut, except it normalizes the clusters by volume instead of number of nodes:
\begin{equation}
\underset{\{{x_1},...,{x_k}\} \in I}\min \sum_{c=1}^{k}\frac{{x_c^T}(D^{+}-A){x_c}}{{x_c^T}\bar{D}{x_c}}.
\end{equation}
Similarly, the \emph{balance normalized association} can be derived from the balance ratio association:
\begin{equation}
\underset{\{{x_1},...,{x_k}\} \in I}\max \sum_{c=1}^{k}\frac{{x_c^T}(D^{-}+A){x_c}}{{x_c^T}\bar{D}{x_c}}.
\end{equation}
Minimizing the balance normalized cut is equivalent to maximizing the balance normalized association. Thus, the choice between balance normalized cut and association is inconsequential \cite{chiang_prediction_2014}.  Chiang et al. proposed a multilevel framework that refines results by first dividing vertices into levels, and then applying their modified version of spectral clustering to each level. Here, the hierarchical approach to graph clustering increases algorithm scalability, and an 100 million edge graph is partitioned in under 4000 seconds \cite{chiang_scalable_2012}.

\subsubsection{Modified Generalized Eigenvalue Method for Signed Spectral Clustering}
\label{ssec:sponge}

Spectral methods in ~\ref{ssec:spectral} use eigenvalues from one matrix. In this section, we describe the generalized eigenproblem methods that uses eigenvalues from two matrices, SPONGE \cite{cucuringu_sponge_2019}. The SPONGE algorithm is a method for k-way clustering in signed graphs that scales well to large graphs \cite{cucuringu_sponge_2019}. The objective is to decompose the network into disjoint groups, such that individuals within the same group are connected by as many positive edges as possible, while individuals from different groups are connected by as many negative edges as possible. The algorithm relies on a generalized eigenproblem formulation to find the k smallest eigenvectors before k-means clustering.  The approach was inspired by constrained clustering \cite{2001wagstaff}, and the authors provide theoretical guarantees in the setting of a signed stochastic blockmodel \cite{cucuringu_sponge_2019}. For a given signed graph G, the objective function for SPONGE is derived from the normalized cut of the positive-edges subgraph $ncut(G^+)$ and the inverse normalized cut of the negative-edges subgraph $(ncut(G^-))^{-1}$.  The trade-off and regularization parameters $\tau^+$ and $\tau^-$ are introduced, and the previous metrics are merged into the new objective function in Eq~\ref{eq:obj0}. 

\begin{equation} 
\min_{C_1, \ldots, C_k} \sum _{i=1}^k\frac{cut_{G^+}(C_i, \bar{C_i})+\tau^-vol_{G^-}(C_i)}{cut_{G^-}(C_i, \bar{C_i})+\tau^+vol_{G^+}(C_i)}
\label{eq:obj0}
\end{equation}

$C_1, \ldots, C_k$ represents a partitioning of $G$. The authors demonstrate that the prior objective function is equivalent to the discrete optimization problem in Eq~\ref{eq:obj1}.

\begin{equation} 
\min_{C_1, \ldots, C_k} \sum_{i=1}^k \frac{x_{C_i}^T(L^+ + \tau^-D^-)x_{C_i}}{x_{C_i}^T(L^- + \tau^+D^+)x_{C_i}}
\label{eq:obj1}
\end{equation}

The discrete optimization problem in Eq~\ref{eq:obj1} is NP-hard, so the authors relax the discreteness constraint on the $x_{c_i}$'s and allow solutions that are in $\mathbb{R}^n$.  New vectors $z_1, \ldots, z_k \in \mathbb{R}$ are introduced such that $z_i^T(L^-+\tau^+D^+)z_i = 1)$ and $z_i^T(L^-+\tau^+D^+)z_i = 0 for i \neq j$, i.e., they are orthonormal with respect to $L^-+\tau+D^+$.  Finally, the objective function can be rewritten as in Eq.~\ref{eq:obj2}. 

\begin{equation}
\min_{z_i^T(L^-+D^+)z_j = \delta_{ij}} \sum_{i=1}^k \frac{z_{i}^T(L^+ + \tau^-D^-)z_{i}}{z_{i}^T(L^- + \tau^+D^+)z_{i}}
\label{eq:obj2}
\end{equation}

Objective function in Eq.~\ref{eq:obj2} can be formulated as the generalized eigenproblem whose solution is given by the eigenvectors of $(L^-+\tau^+D^+)^{-1/2}(L^++\tau^-D^-)(L^-+\tau^+D^+)^{-1/2}$ \cite{cucuringu_sponge_2019}.  The authors use LOBPCC \cite{Knyazev2001TowardTO} to solve for the eigenvectors corresponding to the k smallest eigenvectors and cluster the resulting node embedding using k means++ \ref{alg:KMeans}.  The output of the k-means++ step is the final cluster labeling for the graph from the $SPONGE$ procedure.  Alternately, $SPONGE_sym$ uses the symmetric signed Laplacian, defined as $L^{(+/-)}_{sym}=(D^{(+/-)})^{-1/2}L^{(+/-)}(D^{(+/-)})^{-1/2}$ in the prior equations. $SPONGE$ and $SPONGE_{sym}$ compared favorably against other leading signed spectral clustering algorithms in experiments performed by the authors, and we evaluate it further in Section \ref{sec:exp1}.

\section{Novel Clustering Methods for Signed Graphs} 
\label{sec:signedNew}

In this section, we review all clustering algorithms developed specifically for signed graphs. Early work in this field was heavily constrained by computational technology and focused on smaller, denser networks.  While effective at the time, we note that some of these methodologies do not necessarily translate well to the large, sparse networks that are the focus of most modern research.

\subsection{Blockmodels}
\label{ssec:block}

\begin{figure}[!ht]
\centering
\includegraphics[width=3in]{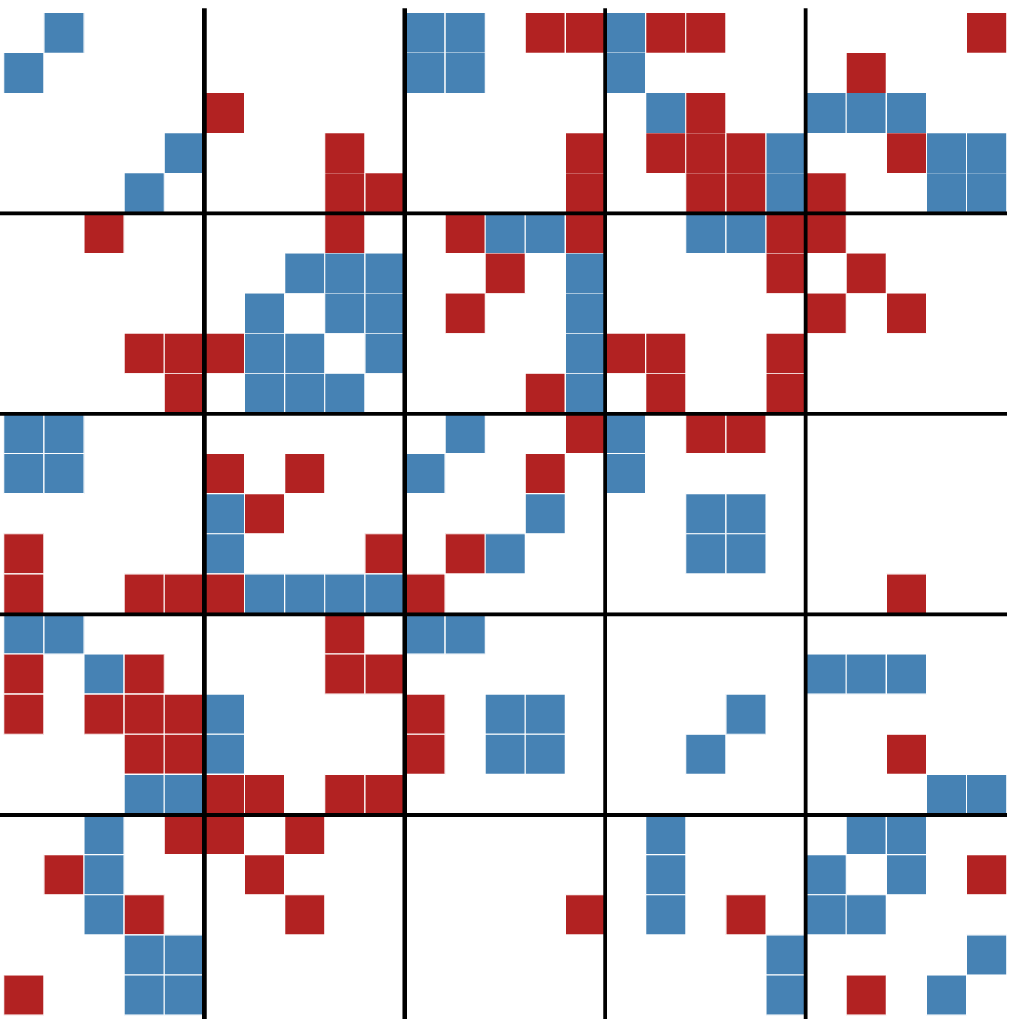}
\hfill
\includegraphics[width=3in]{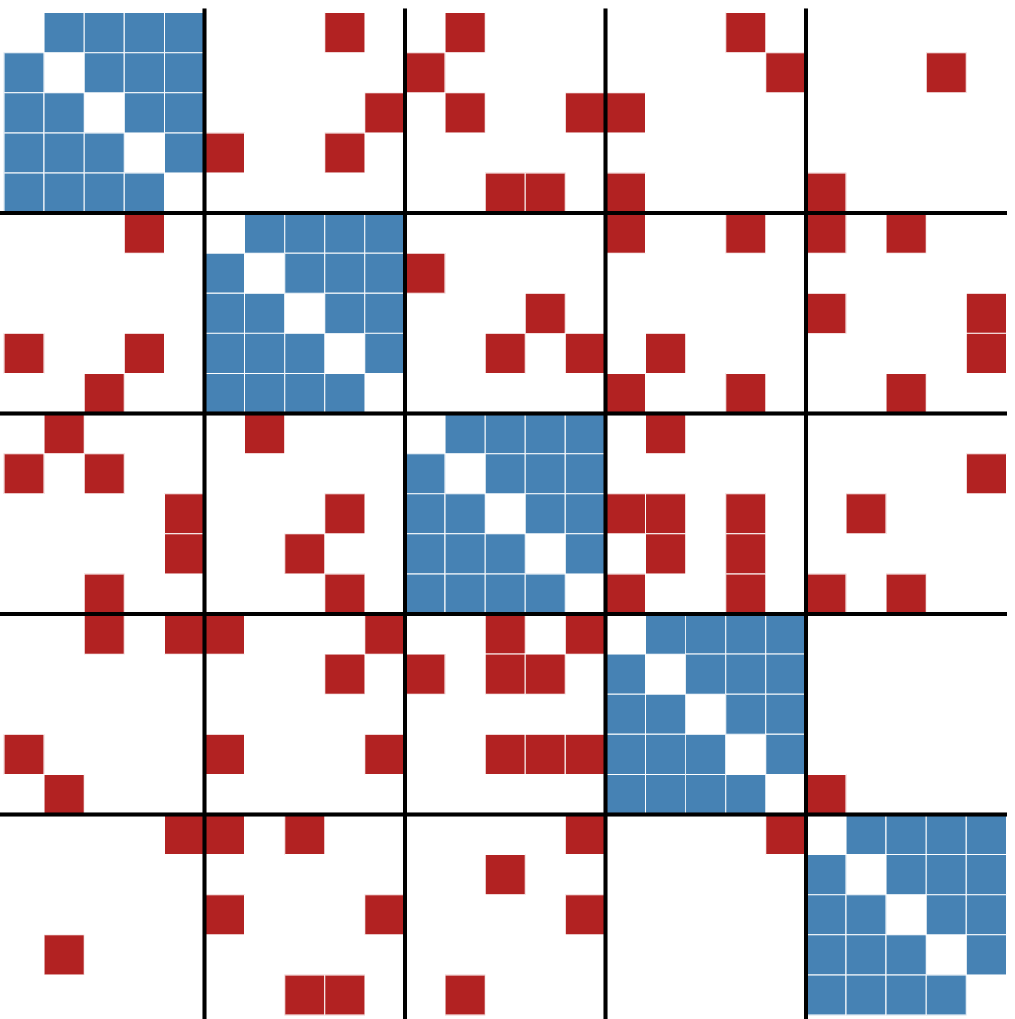}
\caption{Blockmodel illustration for two graphs with 5 community labels. Positive edges are blue and negative edges are red. Both blockmodel illustrations are sorted by assigned community labels: (a) random community assignment visualized with blockmodel illustration shows no clean community separation; (b) if the distinct communities are present in community assignment, it is visible in blockmodel as large positive connected component (blue blocks).
}\label{fig:blockmodel}
\end{figure}

Let ${S}$ be a set and let $\{R_i\}_{i=1}^{m}$ be a set of binary relations on S.  Individuals $a, b \in S$ are said to be \textbf{structurally equivalent} if and only if for any $c \in {S}$ and any $R_i \in \{R_i\}_{i=1}^{m}$, $aR_ic \iff bR_ic$ and $cR_ia \iff cR_ib$. In graph theory terms, the structural equivalence of two nodes means that both nodes are adjacent to exactly the same set of nodes with the same edge weights. Since this occurrence is very rare in real data sets, the authors used the concept of a blockmodel to relax the definition of structural equivalence. In a \emph{ blockmodel}, if the same permutation is applied to both the rows and the columns of the adjacency matrix, the underlying network structure is not changed. Blockmodels seek to \emph{ permute the data} in such a way that submatrices of all 0s exist within the adjacency matrix, as illustrated in Figure~\ref{fig:blockmodel}. The adjacency matrix is then divided into blocks, with a block being assigned a value of 0 if all entries within it are 0 and 1 otherwise.

  Nodes in the same block are assumed to be structurally equivalent if the value of the block is equivalent, thus relaxing the prior definition of structural equivalence to fit real-world data sets. 

\paragraph{Lean Fit} Breiger et al. build on the concept of blockmodels to develop their system for clustering signed data \cite{BREIGER1975328}. A blockmodel is said to be a \emph{ lean fit} to a matrix $M$ if and only if there exists a permutation of M, yielding a permuted matrix $M^*$ that can be blocked in such a way that (1) zeroblocks in $M^*$ correspond to 0s in the blockmodel, and (2) blocks in $M^*$ containing at least one nonzero value have a corresponding value of 1 in the blockmodel. While a \emph{ lean fit} falls short of the algebraic definition of structural equivalence, the authors rationalize this decision by arguing that maintaining a social tie requires effort, while no work is required in the absence of a tie \cite{BREIGER1975328}. Thus, it is appropriate to assign any block with nonzero values an overall value of 1 in the blockmodel.  The authors further emphasize that nonzero blocks do not need to be true cliques or fully connected subgraphs. 

\paragraph{Limitations} For a graph with $n$ vertices, there are $n!$ possible permutations (and thus blockmodels) of the vertices. The first limitation is that exhaustively checking all possible blockmodels quickly becomes impractical, even on relatively small graphs. The second limitation is that once a blockmodel is chosen, the blockings must still be enumerated and checked for a lean fit.  The third limitation is the upper limit on number of blockings and the resulting clustering interpretation. The CONCOR ({CON}vergence of iterated {COR}relations) algorithm repeatedly applies bipartitions to the raw data until a hierarchical clustering at the appropriate level of granularity is established \cite{BREIGER1975328}.  Ordinary product moment correlation coefficients between the columns of the input matrix are computed at each iteration and stored in a correlation matrix. The process is repeated on the correlation matrix until convergence is reached, i.e., there is no change in the matrix between iterations. Once convergence is reached, a clear bipartition emerges.  The process is then repeated on each partition to identify sub-clusters. CONCOR does not optimize a specific metric; it exhaustively checks all possible blockmodels and checks blockings for a lean fit.  This makes the algorithm prohibitively slow when applied to large and sparse signed graphs \cite{BREIGER1975328}.

\subsection{Random Walk Models}
\label{ssec:randomwalk}

A \emph{ random walk} on a graph is a process that begins at a node and moves to one of the nodes to which it is connected.  When the graph is unweighted, the node to which the walk moves is chosen uniformly at random among the neighbors of the present node. Harel et al. introduced the random walk clustering algorithm for positively weighted edges in the graph \cite{2001harel}.  The method requires only $O(n log n)$ time, and one of its variants needs only constant space \cite{2001harel}. The Fast Clustering for Signed Graphs (FCSG) algorithm employs a random walk gap approach to extract cluster structure information within the graph for positive edges only and for the entire graph \cite{hua_fast_2020}. A random walk gap is defined as the difference in cumulative transition probabilities between nodes in the positive-only subgraph versus the unsigned graph.  The FCSG Algorithm \ref{alg:FCSG} calls the RWG Algorithm \ref{alg:RWG} as a subroutine and uses the RWG matrix to reweight the edges.  Then, an iterative procedure is used to merge nodes connected by a positive edge until no positive edges remain.  Nodes that have been merged together are assigned to the same cluster. The FCSG algorithm gives better results than existing algorithms based on the performance criteria of imbalance and modularity \cite{hua_fast_2020}.

For Algorithm \ref{alg:RWG} the transition probabilities for the all-positive subgraph are normalized using the unsigned version of the input graph. If nodes $i$ and $j$ are not connected by a path with a length less than or equal to $k$, the k-step transition probability is zero.  The matrix is $D$ is defined as $D = ((H^{G^{\prime \prime}}_{ij}-H^{G^{\prime}}_{ij})/H^{G^{\prime}}_{ij})_{n\cdot n}$ where $H^{G^{\prime \prime}}_{ij}$ and $H^{G^{\prime}}_{ij}$ represent the k-step transition probabilities between $i$ and $j$ on the all-positive subgraph and unsigned version of the input graph respectively.  If $H^{G^{\prime}}_{ij} = 0$, the normalized transition probability is undefined.

\begin{algorithm}
\caption{RWG algorithm \cite{hua_fast_2020}}\label{alg:RWG}
\begin{algorithmic}
\State \textbf{Input:} The adjacency matrix W of a signed graph
\State \textbf{Step 1:} Compute one step transition probabilities $\theta^\prime$ and $\theta^{\prime \prime}$
\State \textbf{Step 2:} Compute k-step transition probabilities $H^{\prime(k)}$ and $H^{\prime \prime(k)}$
\State \textbf{Step 3:} Compute the sum of transition probabilities $H^{G^\prime}$ and $H^{G^{\prime \prime}}$
\State \textbf{Step 4:} Compute normalized transition probabilities $G$
\State \textbf{Step 5:} Adjust the normalized transition probabilities  $D$ to generate $D^*$ 
\State \textbf{Step 6:} Generate the RWG matrix $H$
\State \textbf{Output:} RWG matrix $H$
\end{algorithmic}
\end{algorithm}

Random Walk Gap (RWG) algorithm is outlined in Alg.~\ref{alg:RWG}, and its underlying assumption is that \emph{the positive-only subgraph of the network must be a single connected component}. This places a large constraint on running the algorithm on a real data set.  If it is not a single connected component, clustering will only be performed on the greatest connected component, and all nodes outside of the greatest connected component will not be placed into a cluster. This condition is usually not met, and it results in many vertices being left out in experiments described in Section~\ref{sec:exp1} and Section~\ref{sec:exp2}. 

\begin{algorithm}
\caption{FCSG algorithm \cite{hua_fast_2020}}\label{alg:FCSG}
\begin{algorithmic}
\State \textbf{Input:} A RWG matrix H and graph G
\State Create a new weighted signed graph, $G^*$ with weights $W^*$, by updating the weights of $G$ using the following formula: $W^* = (w_{ij}^*)$ where $w_{ij}^* = w_{ij} x h_{ij}$ if $(i,j) \in E^+$
\While {there are positive edges in $G^*$}
\State Select the edge $(i,j)$ with the greatest weight 
\State Let $i^\prime = min(i,j)$
\State Fuse $i$ and $j$ into a single node $i^\prime$
\State Merge the edges that linked to both $i$ and $j$ and shared a common node.  The weight of the new edge is the sum of the weights.
\EndWhile
\State All points that have been merged are labeled as a cluster
\State \textbf{Output:} Cluster labels $C_1, ... C_k$

\end{algorithmic}
\end{algorithm}

The first significant limitation of the proposed Algorithm~\ref{alg:FCSG}  for signed graphs is the underlying assumption that the spanning tree can be constructed using \emph{ only positive edges} in the signed graph. The algorithm starts from the premise that, for some values $\alpha \in [0,1]$ and $d > 0$, if node $i$ and node $j$ belong to the same cluster and $dist(i,j) \leq d$, then the probability that a random walk originating at $i$ will reach $j$ before leaving the cluster is at least $\alpha$. While this principle is sound for unsigned graphs, it cannot be easily generalized to signed graphs, as its underlying assumption is that the spanning tree can be constructed using \emph{ only positive edges} only in the signed graph. To satisfy this requirement, we have to take the greatest connected component of the all-positive subgraph of the input signed graph. Figure \ref{fig:Highland}(center) illustrates two all-positive subgraphs for Highland Tribes, and it is clear there is no all-positive spanning tree. As a result, four vertices in a community (in blue) are left out of the analysis. In summary, the first limitation leads to some vertices being unlabeled in the final analysis, as shown in Section~\ref{sec:exp2}. 

The second significant limitation is the assumption of the \emph{small world hypothesis}, a theory that most users are linked by no more than 5 degrees of separation in a social network.  This becomes critical in step 4 of Algorithm \ref{alg:RWG}. The authors assume that $H^{G^{\prime}}_{ij} > 0$ for $k \geq 5$ due to the small-world hypothesis, but, for graphs with a diameter exceeding 5, this does not hold. For this reason, the parameter used in the random walk gap matrix calculation ${L}$ must be greater than or equal to the diameter of the all-positive subgraph of the input graph.  The authors recommend that ${L}$ be set to 5 and warn that the algorithm begins to degrade in quality if ${L}$ is greater than or equal to 10. 

\subsection{Heider Balance Theory Based Methods}
\label{ssec:balance} 

\begin{figure}[!th]
\centering
\includegraphics[width=6in]{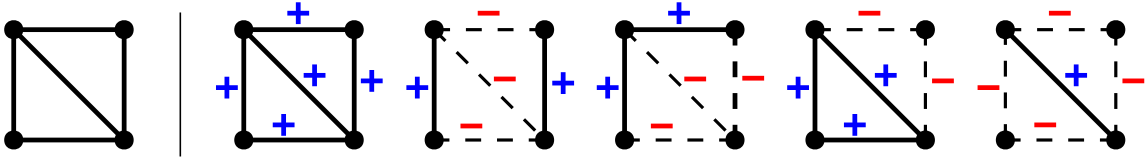}
\caption{For an underlying unsigned graph with 4 vertices and 5 edges in (a), there are 5 different balanced signed graphs in (b). Alg.~\ref{alg:fullybalanced} can be applied to these 5 signed graphs. We remove the negative (dashed) edges from the fully balanced graph, and the result is a bi-cut set of clusters with positive (solid) edges only.}
\label{fig:HararyCut}
\end{figure} 

\subsubsection{Clustering for Balanced Signed Graphs} 

Social balance theory was introduced by Fritz Heider in 1946 \cite{Heider} and mathematically formalized by Cartwright and Harary in 1956 \cite{Har0}. A signed graph is \textbf{balanced} if and only if (1) all of its edges are positive, or (2) the nodes can be partitioned into two distinct clusters so that all edges within a cluster are positive and all edges between clusters are negative. Such a partition is known as a \textbf{Harary cut}, as illustrated in Figure~\ref{fig:HararyCut} for a balanced signed graph with 4 nodes and 5 edges. The clustering approach for the \emph{ balanced signed graphs} reduces to a 2-step process as illustrated in Algorithm~\ref{alg:fullybalanced}.  

\begin{algorithm}[!ht]
\caption{Clustering a balanced signed graph}\label{alg:fullybalanced}
\begin{algorithmic}
\State \textbf{Input:} A balanced signed graph $\Sigma$
\State \hskip 1em  Step 1: Apply a Harary-cut and partition the signed graphs into 2 sets.
\State \hskip 1em  Step 2: Apply unsigned spectral clustering to each of the subsets. 
\State \textbf{Output:} distinct signed graph clusters
\end{algorithmic}
\end{algorithm}

While clustering adaptation for balanced signed graphs is simple, balanced graphs are rare in real-world data. A Harary cut provides a natural cut to examine clusters with similar sentiments. A new robust signed graphic generalization of normalized cuts using nearest Harary cuts recently appeared in \cite{2020Cloud}.

\subsubsection{Clustering for Weakly Balanced Signed Complete Graphs} 

A \emph{ fully connected} network (complete graph) is \textbf{weakly balanced} if and only if (1) all of its edges are positive, or (2) the nodes can be partitioned into k distinct clusters so that all edges within a cluster are positive, and all edges between clusters are negative \cite{2013Altafini}. If a complete graph is \emph{ balanced} or \emph{ weakly balanced}, it is said that an underlying community structure exists in the graph. The signed clustering extension for \emph{ weakly balanced graphs} is outlined in Algorithm~\ref{alg:weaklybalanced}. 

\begin{algorithm}
\caption{Clustering a weakly balanced signed graph \cite{2013Altafini}}\label{alg:weaklybalanced}
\begin{algorithmic}
\State \textbf{Input:} Weakly balanced signed graph $\Sigma$
\State \hskip 1em  Step 1: Group vertices so that positive edges are within clusters and negative edges are between k clusters.
\State \hskip 1em  Step 2: Apply unsigned spectral clustering to each of the subsets. 
\State \textbf{Output:} distinct signed graph clusters
\end{algorithmic}
\end{algorithm}

Note that weakly balanced partitioning reduces to Algorithm~\ref{alg:fullybalanced} for $k=2$, as illustrated in Figure~\ref{fig:SGCutEx}(right). Weakly balanced graphs are rare within social networks, and identifying the network beyond a special case is computationally prohibitive. We require clustering techniques that can be applied to signed graphs in any state of balance. 

\subsubsection{Augmentation-induced Cluster Balance for Clustering} Hseih et al. add new edges between unconnected nodes to achieve balance.  After creating a fully connected, maximally balanced graph based on the initial data, the eigenvectors corresponding to the k greatest eigenvalues of the completed adjacency matrix are computed.  Finally, k-means clustering is run on the eigenvectors to assign nodes to clusters \cite{hseih2012}. 

\begin{algorithm}
\caption{Clustering via matrix completion \cite{hseih2012}}\label{alg:hseih}
\begin{algorithmic}
\State \hskip 1em \textbf{Input:} A signed matrix $G$ and number of clusters $k$ $\Sigma$
\State \hskip 1em Step 1: Impute the sign of missing edges in a way that minimizes frustration of the complete graph $\hat{G}$.
\State \hskip 1em Step 2: Find the eigenvectors $l_1,..., l_k$ corresponding to the k greatest eigenvalues of the adjacency matrix of $\hat{G}$.
\State \hskip 1em Step 3:  Construct $U_{n \times k} \in \mathbb{R}^{n \times k}$ as the matrix containing the vectors $l_1,..., l_k$ as columns.
\State \hskip 1em Step 4: Cluster the graph nodes  $i=1,...,n$ based on $u_i$ features using k-means clustering described in Alg.~\ref{alg:KMeans}. 
\State \textbf{Output:} Cluster labels for all n nodes.
\end{algorithmic}
\end{algorithm}

\subsubsection{Semi-Supervised Signed Network Clustering} 

Semi-Supervised Signed Network Clustering approach SSSnet \cite{2022sssnet} uses modified social network analysis and triangle balancing heuristics ("friend of my friend is my friend") \cite{Leskovec2010b} to address the issue for cluster discovery based on a modified version of Heider balance theory \cite{Heider}.  The authors of SSSnet transform the paradigm that "the enemy of my enemy is my friend" and assert that the relationship should be neutral.  First, a signed mixed-path aggregation (SIMPA) scheme is used to create the node embedding.  Next, the node embedding is used to generate cluster assignment probabilities, and clustering is achieved by training with a weighted sum of a supervised and unsupervised loss function and the unsupervised loss function is a probabilistic balanced normalized cut. SSSnet produces more robust results when the labeled data and node input features are available in the training step. If node input features are not available, SSSnet constructs the node features from the graph structure \cite{2022sssnet}. In this paper, we focus on comparing community discovery methods that are retrieve community information \emph{solely} on the signed graph structure. In that light, we do not consider external vertex features for SSSnet implementation in Section~\ref{sec:exp1} and Section~\ref{sec:exp2}. We also do not use SSSnet data-driven training step to optimize the supervised loss function to a specific dataset in Section~\ref{sec:exp1} and Section~\ref{sec:exp2}. We compare the performance of SSSnet to the other unsupervised methods based solely on the information SSSnet can retrieve from signed graph structure. Our analysis of SSSnet extends to the community and ground truth recovery SSSnet can retrieve in an unsupervised manner from the previously unseen signed graphs that have no external vertex features. 

\subsubsection{Graph clustering in balance feature space} 

Sharma et al. \cite{2021sharma} study the network design problem of maximizing balance of a target community given a fixed number of edge-deletions. They demonstrate the NP-hardness of this problem while also exhibiting that it is also non-monotone and non-submodular. These computational issues were overcome using the spectral relation of balance with the Laplacian spectrum of the network. Since the spectral approach lacks approximation guarantees, a greedy approach was also implemented with bounds on the approximation quality. The bounds are achieved using pseudo-submodularity, and the effectiveness was established on sample dataset. Optimized nearest balanced states were characterized by Rusnak and Te\v{s}i\'{c} in \cite{2020Cloud} with the introduction of the \emph{frustration cloud} which relaxes the NP-hardness of determining the frustration index to provide additional context on the likelihood a consensus balanced state could be reached from a given signed graph. Exact values for edge deletions are computed, but instead of being deleted, they are trained to change sentiment to report back a balanced state. These balanced states are then aggregated over statistically significant samples to quantify the change a vertex or an edge would contribute to a consensus decision. Graph Balancing was recently explored as an alternative to spectral clustering \cite{2020Cloud,2022Cluster}. The concept of the frustration cloud builds on the graph balance theory, relaxes the notion of a single frustration index to a family of minimally balanced graphs of a given signed graph, and derives the numeric features of the vertex, status, and influence in a signed graph based on the balance theory \cite{2020Cloud}. 

The \emph{status} of a vertex is the likelihood a vertex will appear in the majority over all sampled balancings, while the \emph{influence} of a vertex is the likelihood the edges incident to the vertex will appear in the majority. Thus, influence is always less than or equal to status and, when plotted against each other they appear in between the status axis and the line $y=x$. It was shown in \cite{2020Cloud} that these two metrics are very different where status can detect ``promotability'' while influence can detect those making the decisions regarding promotion. The authors have also demonstrated that status and influence attributes capture the spectrum of signed spectral clustering (Fig.~\ref{fig:Highland}(right)) and indicate a possible direction for moving away from eigenvector computation. This is accomplished by sampling spanning trees to detect a minimal set of signs that obstruct balance, thus leveraging the difference between underlying bases of balance and unbalanced signed graphs. The study on more degenerate, adversarial networks is necessary as the next step to determine if consensus-based attributes \cite{2020Cloud,2022Cluster} can provide insight into networks where spectral clustering fails.

\section{Signed Graph Clustering for Known Communities: A Comparison}
\label{sec:exp1}
\begin{table}[!ht]
\setlength\tabcolsep{2pt}
\begin{tabular}{ll||l||l}
\textbf{Table label} & & \textbf{Figure label} & \textbf{Method Description} \\ \hline \hline
ground & truth & & grond truth community labeling \\ \hline
\multirow{3}{*}{Laplacian} & \_none & lap\_none & spectral clustering using the signed graph Laplacian \cite{Kunegis2009}\\ 
& \_sym & lap\_sym & spectral clustering using the symmetric Laplacian \cite{Kunegis2009}\\
& \_sep & lap\_sep &  spectral clustering using the symmetric separated Laplacian \cite{Kunegis2009}\\ \hline
\multirow{2}{*}{Balanced Cuts} & \_none & BNC\_none & balanced normalized cuts \cite{chiang_scalable_2012}\\
 & \_sym & BNC\_sym & symmetric balanced normalized cuts \cite{chiang_scalable_2012} \\ \hline
\multirow{2}{*}{SPONGE} & \_none  & SPONGE\_none & baseline SPONGE implementation \cite{cucuringu_sponge_2019} \\ 
 & \_sym & SPONGE\_sym &  symmetric SPONGE \cite{cucuringu_sponge_2019}\\ \hline
\multirow{2}{*}{Power Means} & GM & GM & geometric means \cite{mercado2016} \\
& SPM & SPM & matrix power means \cite{mercado2019}\\ \hline
FCSG & & FCSG & Fast Clustering for Signed Graphs \cite{hua_fast_2020}\\ \hline
SSSnet & & SSSnet & SSSnet: Semi-Supervised Signed Network Clustering \cite{2022sssnet}\\ \hline
{graphB}\_km & & graphB\_km & k-means clustering in graph Balancing space \cite{2020Cloud}\\
\end{tabular}
\caption{Legend: indexing twelve methods in Tables \ref{tab:ARIHighland} -- ~\ref{tab:exp1ARI} and Figure~\ref{fig:labeledGraph}.}\label{tab:legend}
\end{table}

In this section, we compare the effectiveness, strengths and weaknesses of \emph{twelve} state-of-the-art signed graph clustering approaches when ground-truth is available for five different datasets. Each of the twelve approaches listed in Table~\ref{tab:legend} and the high-level attributes of each of the datasets are listed in Table~\ref{tab:exp1}. Detailed findings on the signed clustering performance and the description of each of the dataset used for comparison are presented in Section~\ref{ssec:Highland}, Section~\ref{ssec:Sampson}, Section ~\ref{ssec:CoW}, and Section~\ref{ssec:Sports}.  We conclude the section with a discussion on method performances, taking labeling and signed graph attributes in consideration over all methods and all datasets,in Section~\ref{ssec:all}. 

\paragraph{Approach} We have applied \emph{twelve} different signed graph clustering methods on four datasets. Tables \ref{tab:ARIHighland} -- ~\ref{tab:exp1ARI} and Figure~\ref{fig:labeledGraph} use the indices for the methods outlined in Table~\ref{tab:legend}. The Python package \emph{signet} \cite{signet_repo} was used to run Laplacian, Balanced Cuts, and SPONGE methods; see Table~\ref{tab:legend} for the list of methods. Power Means\cite{mercado2019}, SSSnet \cite{2022sssnet}, and graphB\cite{graphB} implementations were provided by the authors as an open source. 

Fast Clustering for Signed Network implementation and its limitations are described in Section~\ref{ssec:randomwalk}. The authors did not provide an implementation, and the paper did not discuss efficiency strategies \cite{hua_fast_2020}. Our in-house implementation follows the paper guidelines when possible, as outlined in  Alg.~\ref{alg:RWG} and Alg.~\ref{alg:FCSG}. Random Walk Gap (RWG) algorithm has underlying assumption that \emph{the positive-only subgraph of the network must be a single connected component}. This places a large constraint on running the algorithm on a real data set, and reduces FCSG scores. We implemented FCSG in a Python + NetworkX package, and have released a python wrapper we developed to efficiently compare methods and produce ARI matrices for all methods \cite{graphC}. Note that the reproducibility of the reported results differs from the original paper. 

\paragraph{Datasets} We evaluate the performance of the twelve signed clustering methods on the following five datasets:  (1) {\bf highland} \cite{1954Read} models agreeable and antagonistic relationships between tribes in the Eastern Central Highlands of New Guinea in Section~\ref{ssec:Highland}(2) {\bf sampson} \cite{sampson}, which models sentiment over time between novice monks in a New England monastery captured by Sampson\cite{sampson}; (3) CoW \cite{COW} captures Second World War Allies, among 50 nations from Correlates of War data in Section~\ref{ssec:CoW}; (4) {\bf football} \cite{2013greene} tracks Twitter interactions between players belonging to the English Premier League clubs in  Section~\ref{ssec:Sports}; and (5) {\bf olympics} models Twitter interactions between athletes competing in 2012 London Olympics \cite{2013greene}. Five datasets are selected as the ground labels are known, and they vary in number of community labels, number of vertices, edges, and percentage of negative edges. Table~\ref{tab:exp1} summarizes all five graph characteristics: number of vertices, positive, negative edges, and vertex degree statistics. Table~\ref{tab:exp1} also summarizes graph attributes: density score $d$ (Eq.\ref{eq:d}) and the number of balanced triangles over the total number of triangles in the graph $bal_3$.  Finally, we have included information about the ground truth labeling in Table~\ref{tab:exp1}: the number of communities; pos\_in - the percentage of positive edges in the ground truth communities; and neg\_out - the percentage of negative edges between ground truth communities.

\subsection{Signed Graph Clustering for Clearly Defined Communities in a Dense Graph: Highland's Tribe}
\label{ssec:Highland}

\begin{figure}[!ht]
  \centering
  \includegraphics[width=2.2in]{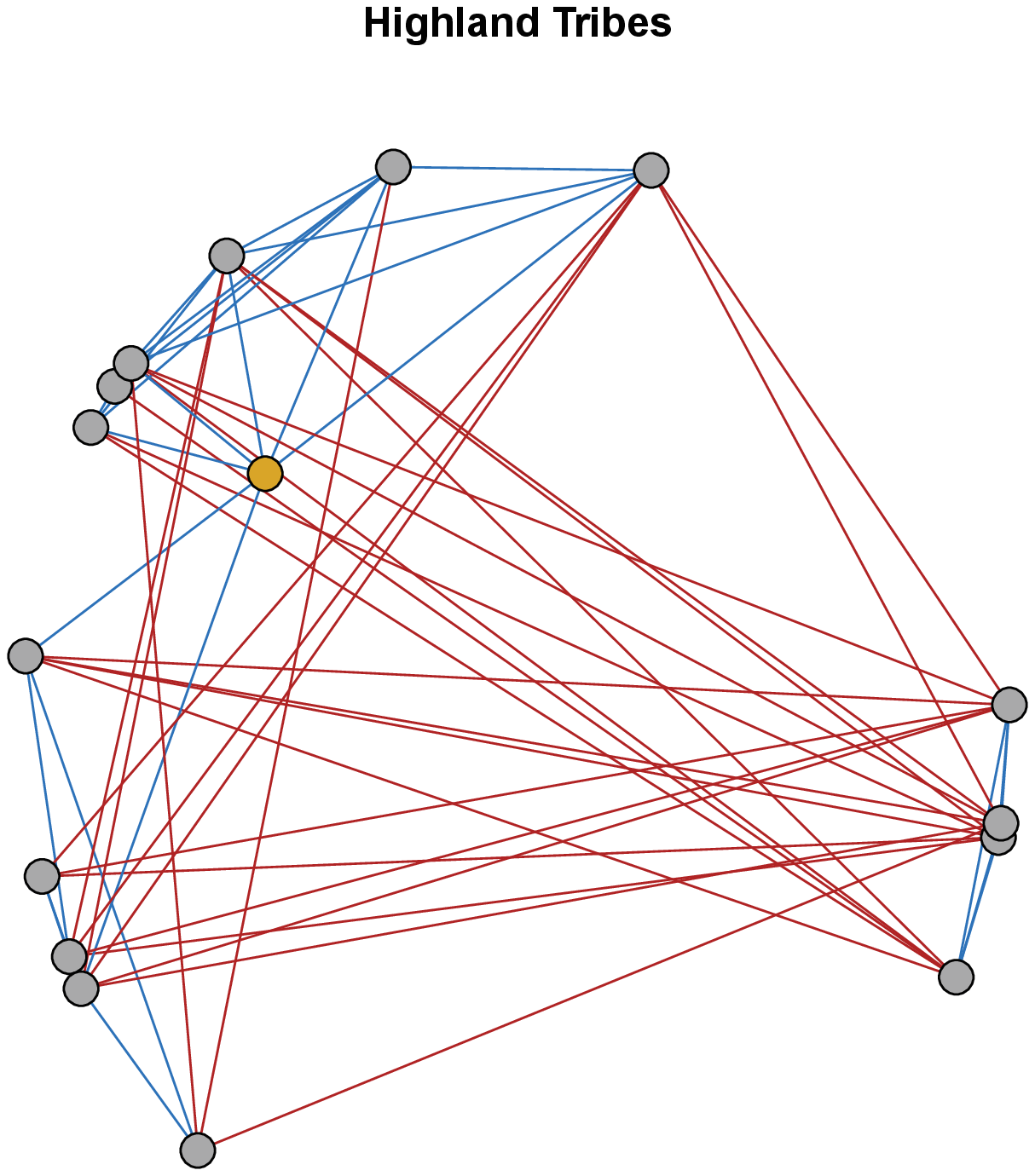}
  \includegraphics[width=1.8in]{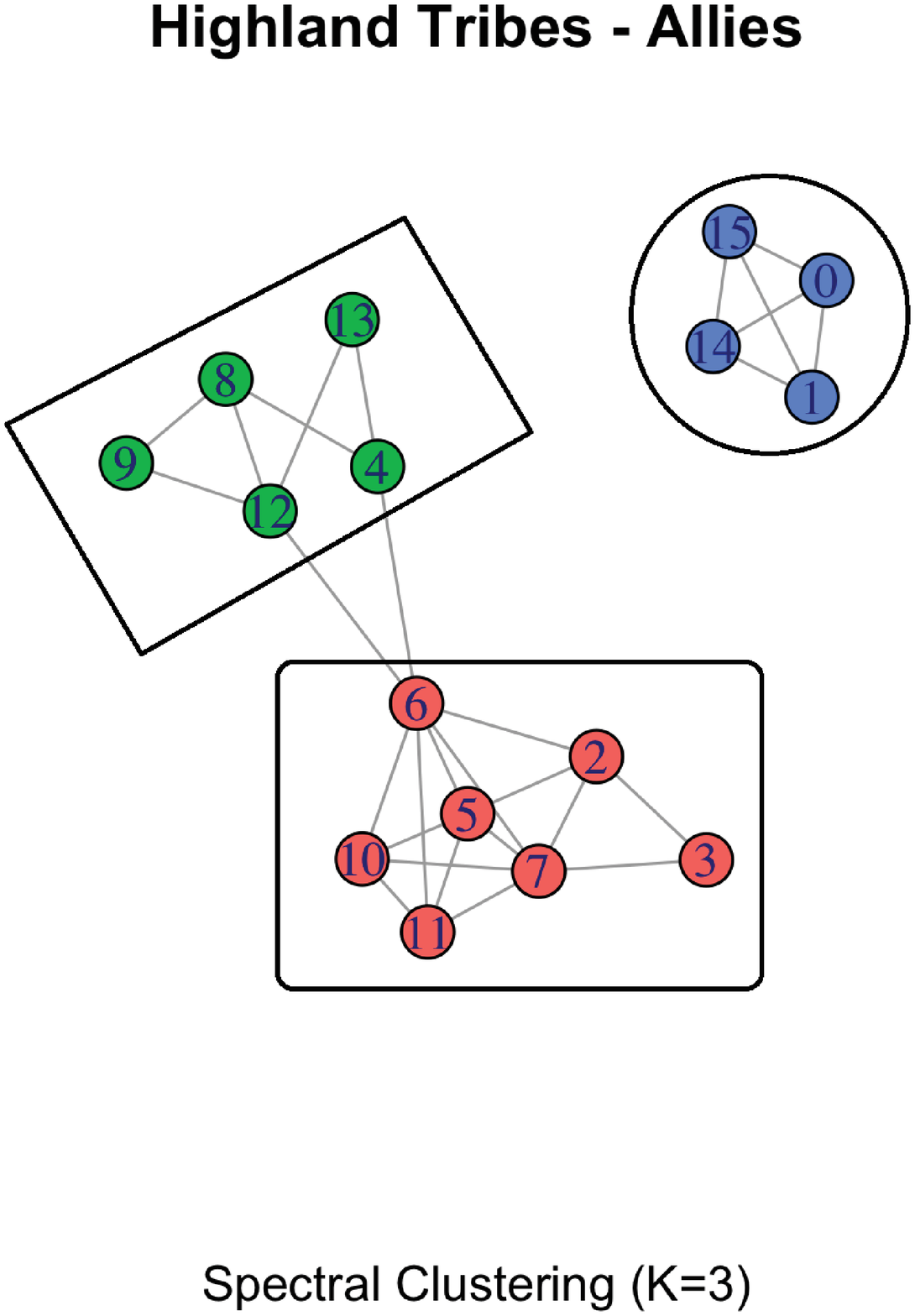}
  \includegraphics[width=2in]{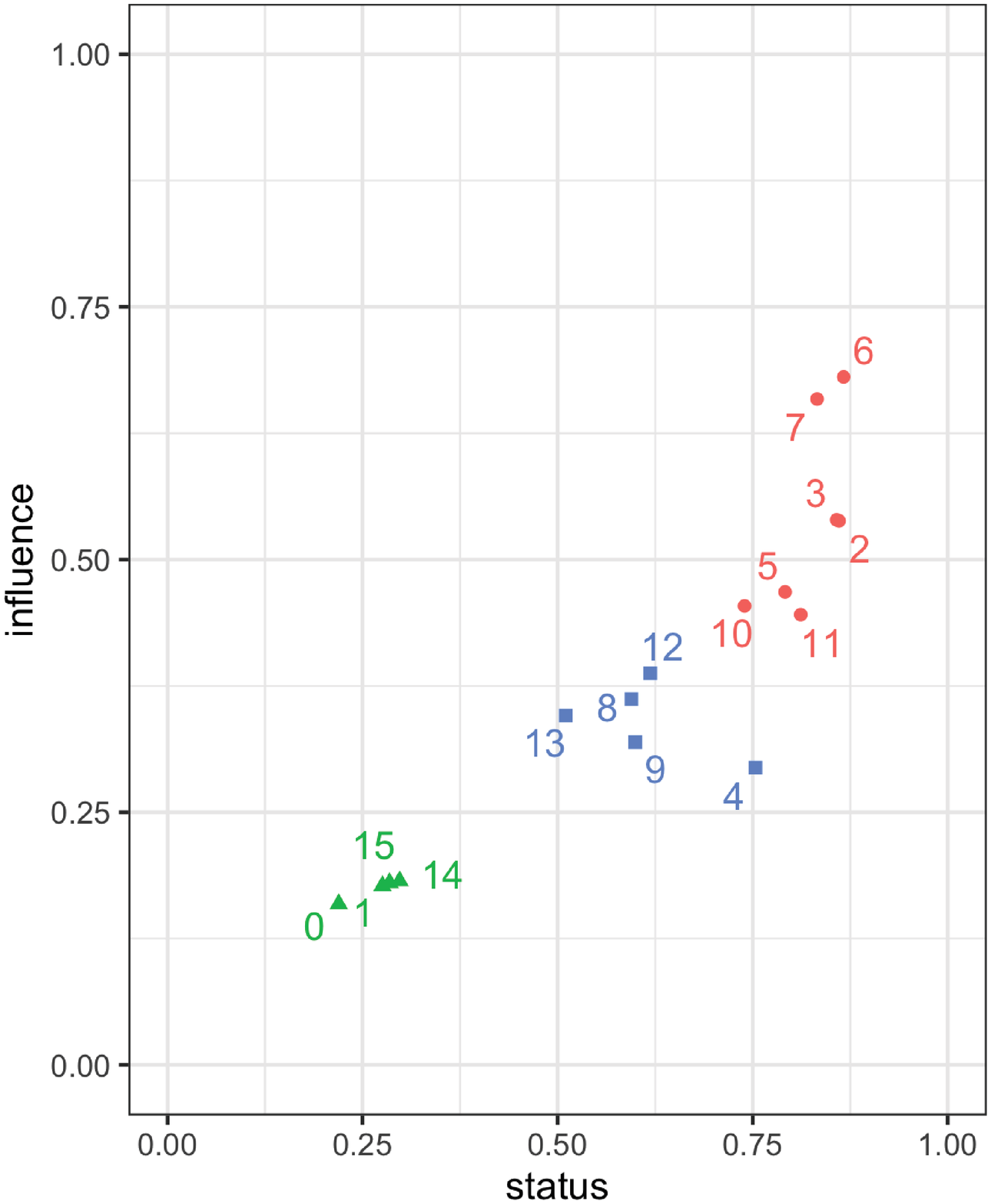}
  \caption{Signed graph representing Highland Tribes ({\bf highland}) has  16 vertices, 29 positive edges in blue, and 29 negative edges in red (left); the dataset has 3 ground truth communities (center); graph vertex correspondence of Laplacian symmetric (lap\_sym) method \citep{kunegis_spectral_2010} to graphB clustering in status/influence space \cite{2020Cloud}. Methods retrieve identical results.}  \label{fig:Highland}
\end{figure}

 The Highland Tribes data set describes agreeable and antagonistic relations between 16 tribes of the Eastern Central Highlands of New Guinea.  Each tribe is a node, with agreeable tribes connected by a positive edge and antagonistic tribes connected by a negative edge. The Highland Tribes data set \cite{1954Read} captures the alliances between sixteen tribes of the Eastern Central Highlands of New Guinea as depicted in Fig.~\ref{fig:Highland} (left). There are three communities in the Highland Tribes data, shown in  Fig.~\ref{fig:Highland} (left). The golden node in Fig.~\ref{fig:Highland} (left) has no adjacent negative edges and belongs to two communities. 

\begin{table}[!ht]
\setlength\tabcolsep{2pt}
\begin{tabular}{|ll||l|l|l|l|l|l|l|l|l|l|l|l|l|}
\hline
\textbf{highland} &  & ground & \multicolumn{3}{c|}{Laplacians} & \multicolumn{2}{c|}{Balanced Cuts} & \multicolumn{2}{c|}{SPONGE}& \multicolumn{2}{c|}{Power Means} & FCSG & SSS & graphB\\
 &  & truth & \_none & \_sym & \_sep & \_none & \_sym & \_none & \_sym & GM & SPM &  & net & \_km \\ \hline \hline
ground & truth & 1 & 1 & 1 & 0.26 & 1 & 1 & 1 & 1 & 0.4 & 0.78 & 1 & 1 & 1\\ \hline \hline
\multirow{3}{*}{Laplacian} & \_none & 1 & 1 & 1 & 0.26 & 1 & 1 & 1 & 1 & 0.4 & 0.78 & 1 & 1 &1 \\ 
& \_sym & 1 & 1 & 1 & 0.26 & 1 & 1 & 1 & 1 & 0.4 & 0.78 & 1 & 1 & 1\\ 
& \_sep & 0.26 & 0.26 & 0.26 & 1 & 0.26 & 0.26 & 0.26 & 0.26 & -0.06& 0.13 & 0.26 & 0.26 &0.26\\ \hline
Balanced  & \_none & 1&1 &1 & 0.26 &1 &1 &1   &1  & 0.4 & 0.78 & 1 & 1 & 1\\ 
Cuts & \_sym &1 &1 &1 &0.26 &1 &1 &1 &1  &0.4 & 0.78 & 1 & 1 & 1\\ \hline
\multirow{2}{*}{SPONGE} & \_none  &1 &1 &1 & 0.26 &1 &1 &1   &1  & 0.4 & 0.78 & 1 & 1 & 1\\ 
 & \_sym   &1    &1 &1 & 0.26 &1 &1 &1   &1  & 0.4 & 0.78 & 1 & 1 & 1\\ \hline
Power & GM & 0.4 & 0.4 & 0.4 & -0.06 & 0.4 & 0.4 & 0.4 & 0.4 & 1 & 0.42 & 0.4 & 0.4 & 0.4\\
Means & SPM & 0.78 & 0.78 & 0.78 & 0.13 & 0.78 & 0.78 & 0.78 & 0.78 & 0.42 & 1 & 0.78 & 0.78 & 0.78 \\ \hline
FCSG & & 1 & 1 & 1 & 0.26 & 1 & 1 & 1 & 1  & 0.4  & 0.78 & 1      & 1          & 1  \\ \hline
SSSnet & &	1 & 1 & 1 & 0.26 & 1 & 1 & 1 & 1 & 0.4 & 0.78 & 1 & 1 & 1\\ \hline
graphB\_km & &	1 & 1 & 1 & 0.26 & 1 & 1 & 1 & 1 & 0.4 & 0.78 & 1 & 1 &1\\ \hline
\end{tabular}
\caption{Detailed analysis of Adjusted Rand Index (ARI) for all twelve methods for {\bf highland} dataset. 75\% of the methods entirely recover the ground truth.}\label{tab:ARIHighland}
\end{table}

The graph constructed from Highland Tribes data, {\bf highland}, has a high number of balanced triangles (86.8\% from Tab.~\ref{tab:exp1}), is of small size (3 communities, 16 vertices), and exhibits high clusterability. Ground truth labels line up with high \% of the positive edges within labeled communities (93\%) and negative edges between them (100\%), as illustrated in Fig.~\ref{fig:Highland} (left). The signed graph has a density score 0.48, and 87\% of triangles in the graph are balanced. Graph's narrow spread of vertex degree (mean is 7; median is 7.5) and high pos\_in and neg\_out scores for ground communities indicate that the graph coherently represents the communities. Results in Table~\ref{tab:ARIHighland} show that 75\% of the methods evaluated achieve a perfect score 1.0 on this dataset. Detailed analysis of Adjusted Rand Index (ARI) for all twelve methods for Highland dataset is presented in Table~\ref{tab:ARIHighland}. In Figure~\ref{fig:Highland}(right), we illustrate the vertex correspondence of Laplacian symmetric (lap\_sym) method \citep{kunegis_spectral_2010} to graphB k-means clustering in status/influence space \cite{2020Cloud}. These methods retrieve identical results, and community separation is clear in status-influence space. Spectral clustering with symmetric separated Laplacian and both power means methods failed to recover the ground truth. Notably, the ARI values comparing the symmetric separated Laplacian labels with the geometric means and matrix power means labels are -0.06 and 0.13 respectively see Table \ref{tab:ARIHighland}, indicating very little similarity in the outcomes between these methods.  This experiment shows that when small signed graphs reflect ground community labeling, most of the methods perform well. 

\subsection{Signed Graph Clustering for Communities Not Reflected in a Signed Dense Graph: Sampson's Monk Survey} 
\label{ssec:Sampson}

\begin{figure}[!ht]
\centering
\includegraphics[width=2.5in]{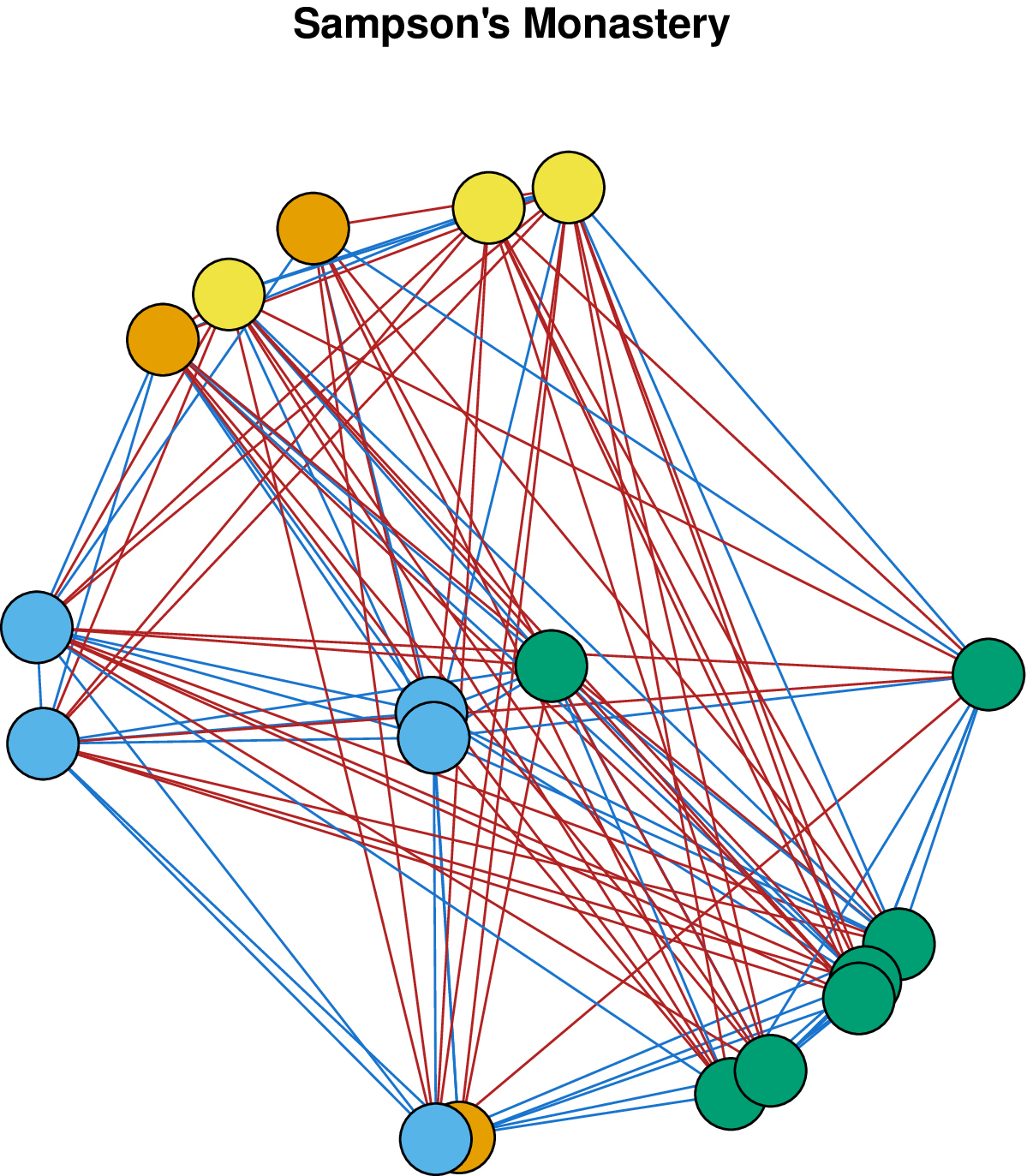}
\includegraphics[width=2.5in]{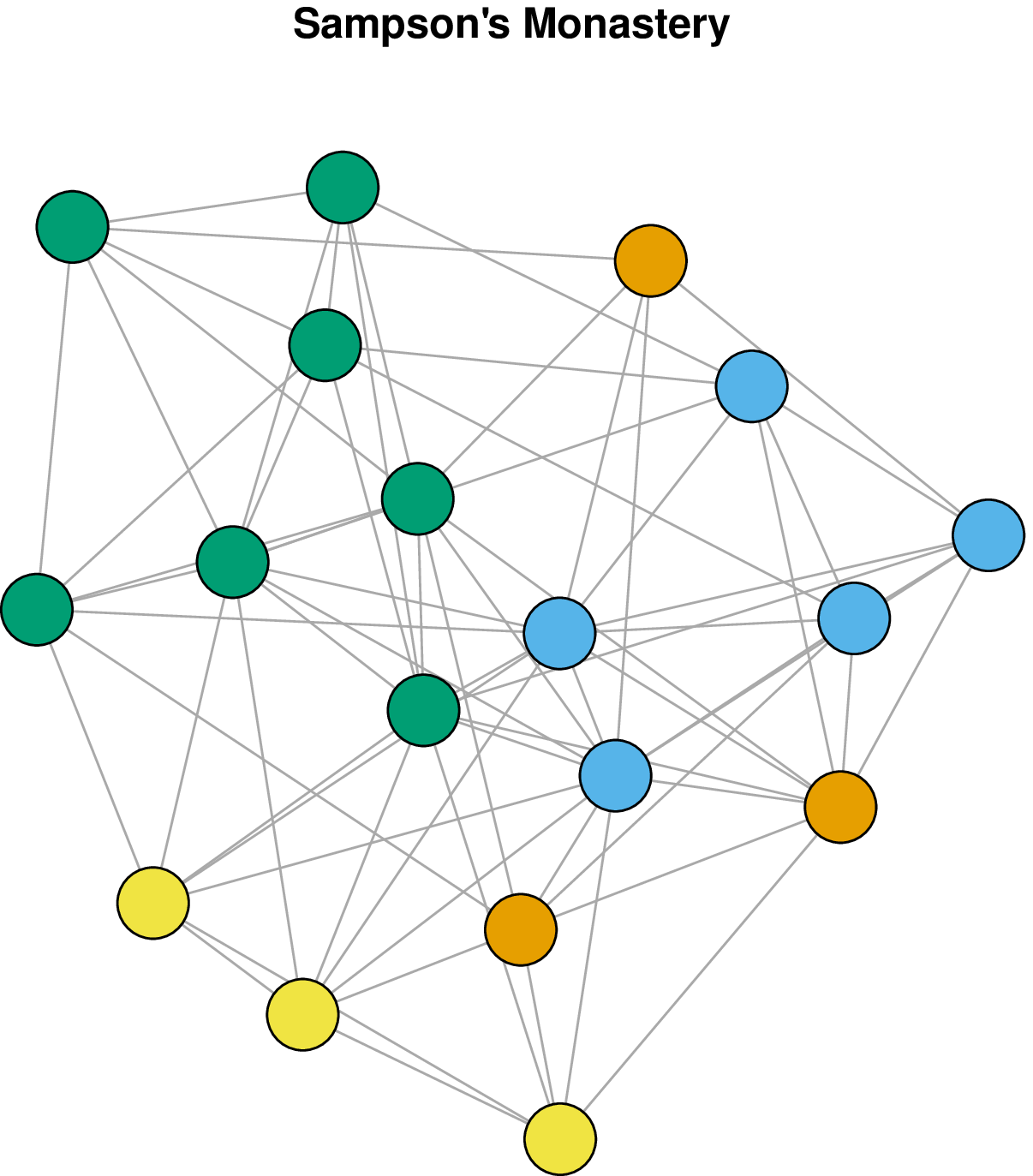}

\caption{Widely used network structure for {\bf sampson} dataset with 18 vertices, 63 positive edges in blue and 49 negative edges in red \cite{sampson} (left); The ground truth grouping is not distinguishable  even when we look at positive edges between nodes (right)} \label{fig:Sampson}
\end{figure}

The Sampson's Monastery data set has eighteen nodes that represent eighteen monks as illustrated in Fig.~\ref{fig:Sampson}(left). Four non-overlapping communities are labeled as the Young Turks, the Loyal Opposition, the Waverers, and the Outcasts \cite{sampson}. Data was collected during the implementation of the Vatican II, an influential change in the Catholic Church that was controversial among the monks.  The Sampson data set captures this dissent among the monks.  The Young Turks began training before the Vatican II and were resistant to change, while the Loyal Opposition were more open to new ideas and began after the Vatican II .  The Waverers did not take a strong stance for or against the Vatican II, while the Outcasts were rejected by both the Young Turks and the Loyal Opposition. Note that the Outcasts and the Waverers are not defined by their positive ties to a group or ideology, but by their ambivalence or their rejection from the mainstream opinions. 

\begin{table}[!ht]
\setlength\tabcolsep{2pt}
\begin{tabular}{|ll||l|l|l|l|l|l|l|l|l|l|l|l|l|}
\hline
\textbf{sampson} &  & ground & \multicolumn{3}{c|}{Laplacians} & \multicolumn{2}{c|}{Balanced Cuts} & \multicolumn{2}{c|}{SPONGE}& \multicolumn{2}{c|}{Power Means} & FCSG & SSS & graphB\\
 &  & truth & \_none & \_sym & \_sep & \_none & \_sym & \_none & \_sym & GM & SPM & & net & \_km \\ \hline \hline
ground & truth & 1 & \underline{\bf 0.61} & 0.33 & 0.25 & 0.51 & 0.51 & 0.52 & \underline{0.6} & 0.32 & 0.37 & 0.31 & 0.13 & 0.41\\ \hline \hline
\multirow{3}{*}{Laplacian} & \_none & \underline{\bf 0.61} & 1 & 0.31 & 0.4  & {\bf 0.69} & 0.39 & {\bf 0.74} & {\bf 0.7} & 0.3 & 0.35 & 0.18 & 0.38 & 0.4\\
& \_sym & 0.33 & 0.31 & 1  & 0.19 & 0.5 & 0.4 & 0.27 & 0.51 & 0.52 & {\bf 0.89} & 0.57 & 0.24 & 0.27\\
& \_sep &  0.25 & 0.4 & 0.19 & 1 & 0.26 & 0.26 & 0.52 & 0.25 & 0.25 & 0.19 & 0.18 & 0.4& 0.41\\ \hline
Balanced & \_none & 0.51 & 0.7 & 0.5 & 0.26 & 1 & 0.48 & 0.54 & {\bf 0.72} & 0.27 & 0.53 & 0.6 & 0.14 & 0.18\\
Cuts & \_sym  & 0.51 & 0.39 & 0.4 & 0.26 & 0.48 & 1 & 0.31 & 0.41 & 0.49 & 0.4 & 0.45 & 0.11 & 0.43 \\\hline
\multirow{2}{*}{SPONGE} &  \_none & 0.52 & {\bf 0.74} & 0.27 & 0.52 & 0.54 & 0.31 & 1 & {\bf 0.62} & 0.22 & 0.31 & 0.33 & 0.21& 0.53\\
& \_sym & \underline{0.6} & {\bf 0.7} & 0.51  & 0.25  & {\bf 0.72}  & 0.41 & {\bf 0.62}  & 1  & 0.18 & 0.55 & 0.36 & 0.23 & 0.3\\ \hline
Power & GM  & 0.32 & 0.3 & 0.52 & 0.25 & 0.27 & 0.49 & 0.22 & 0.18 & 1  & 0.41 & 0.31 & 0.16 & 0.33\\
Means & SPM & 0.37 & 0.35 & {\bf 0.89} & 0.19 & 0.53  & 0.4 & 0.31 & 0.55 & 0.41 & 1 & 0.54 & 0.28 & 0.24\\ \hline
FCSG & & 0.31 & 0.4 & 0.57 & 0.18 & 0.6 & 0.45 & 0.33 & 0.36 & 0.31 & 0.54 & 1 & 0.19 & 0.23\\ \hline
SSSnet & & 0.13 & 0.18 & 0.24 & 0.4 & 0.14 & 0.11 & 0.21 & 0.23 & 0.16 & 0.28 & 0.19 & 1 & 0.24\\ \hline
graphB\_km & & 0.41 & 0.38 & 0.27 & 0.41 & 0.18 & 0.43 & 0.53  & 0.3 & 0.33 & 0.24 & 0.23 & 0.24  & 1 \\ \hline
\end{tabular}\caption{Detailed analysis of Adjusted Rand Index (ARI) for all twelve methods for {\bf sampson} dataset. We have marked \underline{\bf best} ARI and \underline{second best} ARI for recovering ground truth (first row and first column). We emphasize {\bf mutual} ARI scores higher than ground truth recovery. }\label{tab:ARISampson}
\end{table}
  
\begin{figure}[!ht]
\centering
\includegraphics[width=2in]{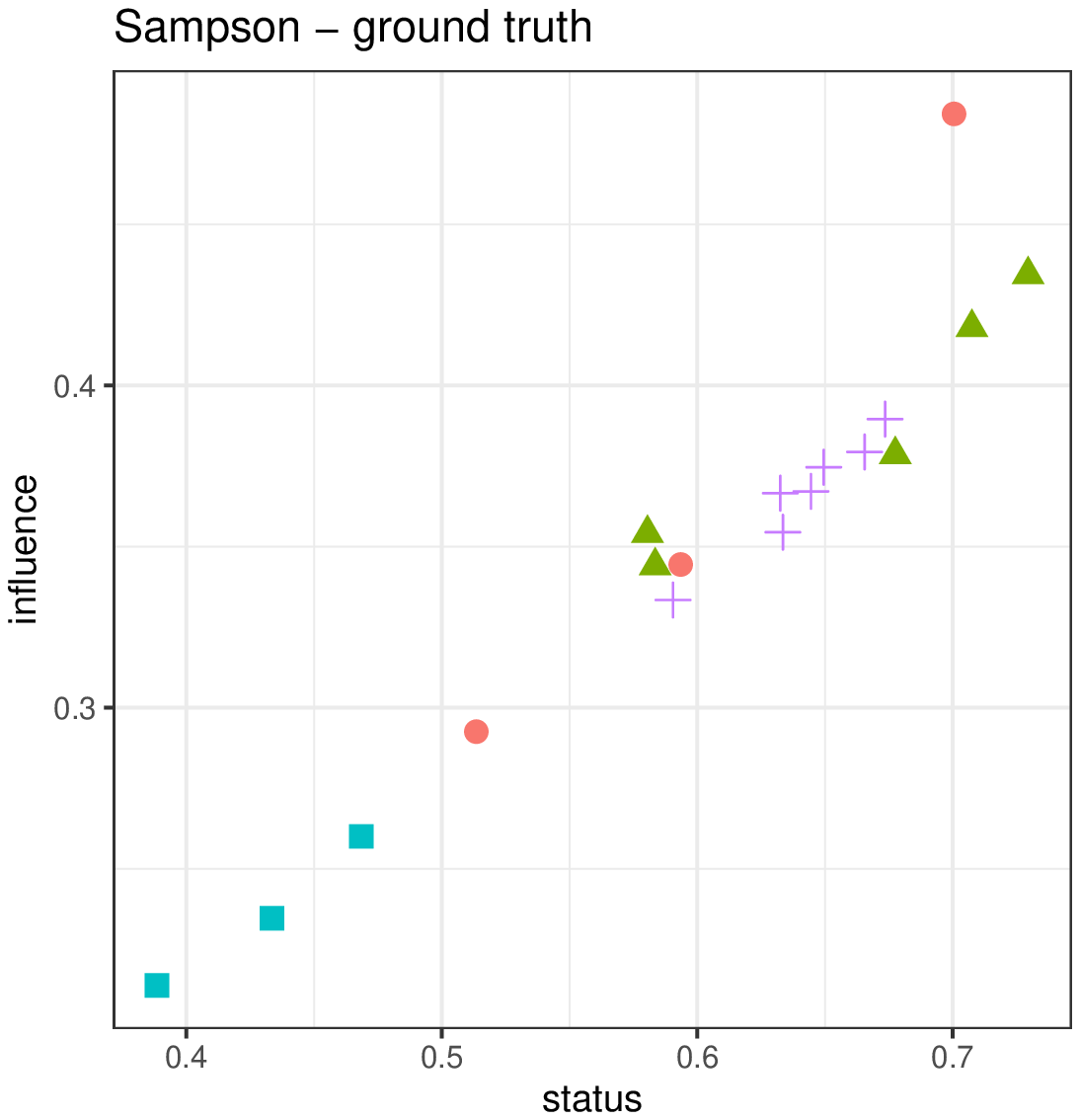}
\includegraphics[width=2in]{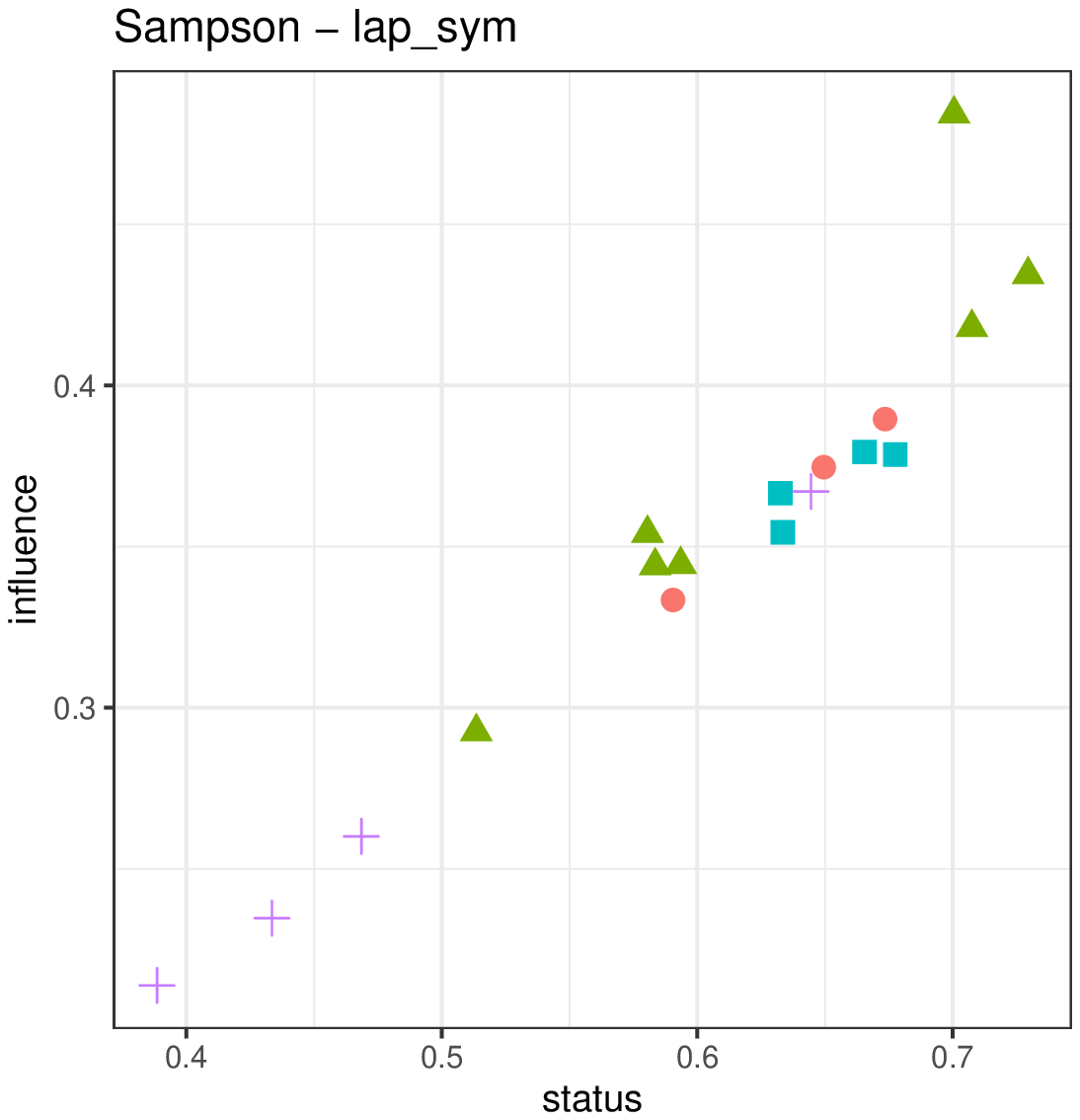}
\includegraphics[width=2in]{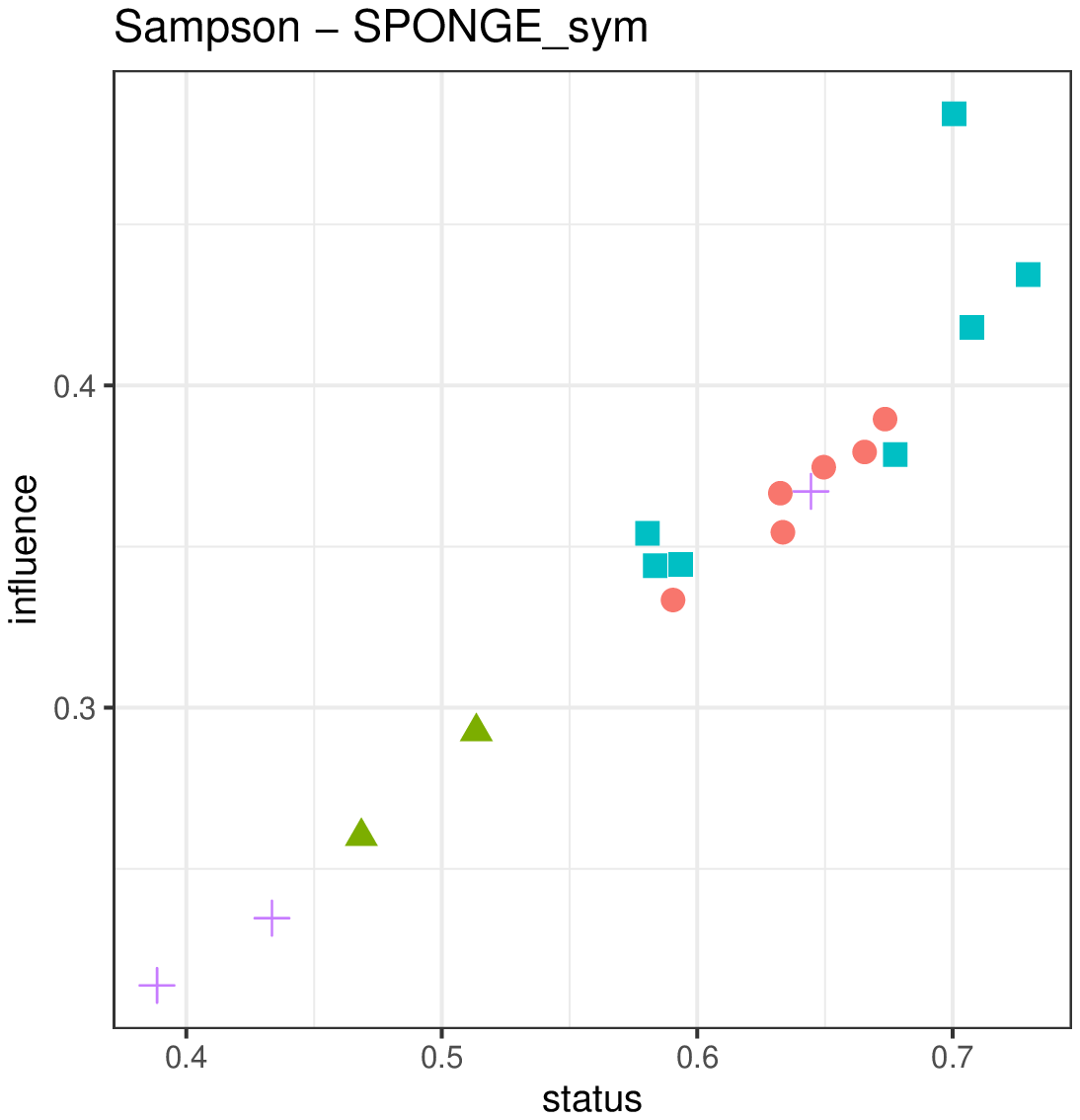}
\caption{Vertex correspondence of ground truth (left), Laplacian symmetric (lap\_sym) method \citep{kunegis_spectral_2010} (center), and SPONGE (\_sym) to graphB clustering in status/influence space \cite{2020Cloud} for {\bf sampson} dataset. Methods derive very different results for $k$=4.} \label{fig:SampsonMethods}
\end{figure}

The Sampson data set is also complex in combining multiple sentiments into a single edge weight.  In the original Sampson data set, surveys were administered in which the monks were asked to rank their top three and bottom three peer choices on four qualities.  To create the signed graph, we look at sentiments as measured by the surveys in each monk-monk pair.  If each quality was ranked positively, we assign a +1 edge.  If all qualities were ranked negatively, we assign a -1 edge.  In the case of mixed sentiments, we use a weighted average for the scores to determine the edge sign.  If the weighted average is positive, we assign a positive edge; if it is negative, we assign a negative edge; and if the average is 0, we consider the relation ambivalent and assign no edge between the monks.  The signed graph resulting from Sampson's monk survey, \textbf{sampson}, has 18 vertices, 61 positive, and 51 negative edges as illustrated in Figure~\ref{fig:Sampson}(right). The density of the graph is 0.732, and it has 60\% of triangles that are balanced. The Sampson Monks group dynamic is more complex than Highland Tribes, and the mapping of the ground truth to signed graph communities is not as clean cut, as illustrated in Fig.~\ref{fig:Sampson}(right).  The data has 4 ground community labels, clear community separation (98\% of negative edges are between communities) and \emph{poor} clusterability as 52\% of the positive edges among vertices are \emph{not} captured by ground truth community labels. 

The performance of the twelve approaches on the Sampson data greatly varies with the highest ARI score of 0.61 for ground truth, and great variation in the mutual ARI scores Table~\ref{tab:ARISampson}. The greatest agreement is between basic SPONGE and spectral clustering using signed graph Laplacians 0.75. This was expected as they are both spectral methods. Large variations in performance scores show that the constructed signed network \emph{and} ground truth community labeling does not capture the complexity in the relations sufficiently to decipher assigned labels (see Figure~\ref{fig:Sampson}(left) for the ground truth). Figure~\ref{fig:SampsonMethods} shows the labeling of ground truth and two methods in status-influence space. In Figure~\ref{fig:SampsonMethods}(right), vertex 17 is far from its community in status-influence space, and no approach can recover that. Figure~\ref{fig:SampsonMethods} also shows that signed spectral clustering using symmetric Laplacian and SPONGE symmetric struggle with coherently forming agreeable groups with minimal sentiment disruption that recover ground truth labels. FCSG underlying assumption that the positive-only subgraph of the network must be a single connected component recovers 2 out of 4 communities for sampson and results in low ARI. graphB methodology performs at the level of existing methods but provides additional data resolution and features for analysis. Spectral methods show the best results for {\bf sampson} dataset clustering. 

\subsection{Signed Graph Clustering for Communities Not Reflected in a Signed Sparse Graph: Correlates of War} 
\label{ssec:CoW}

\begin{figure}[!ht]
\centering
\includegraphics[width=2in]{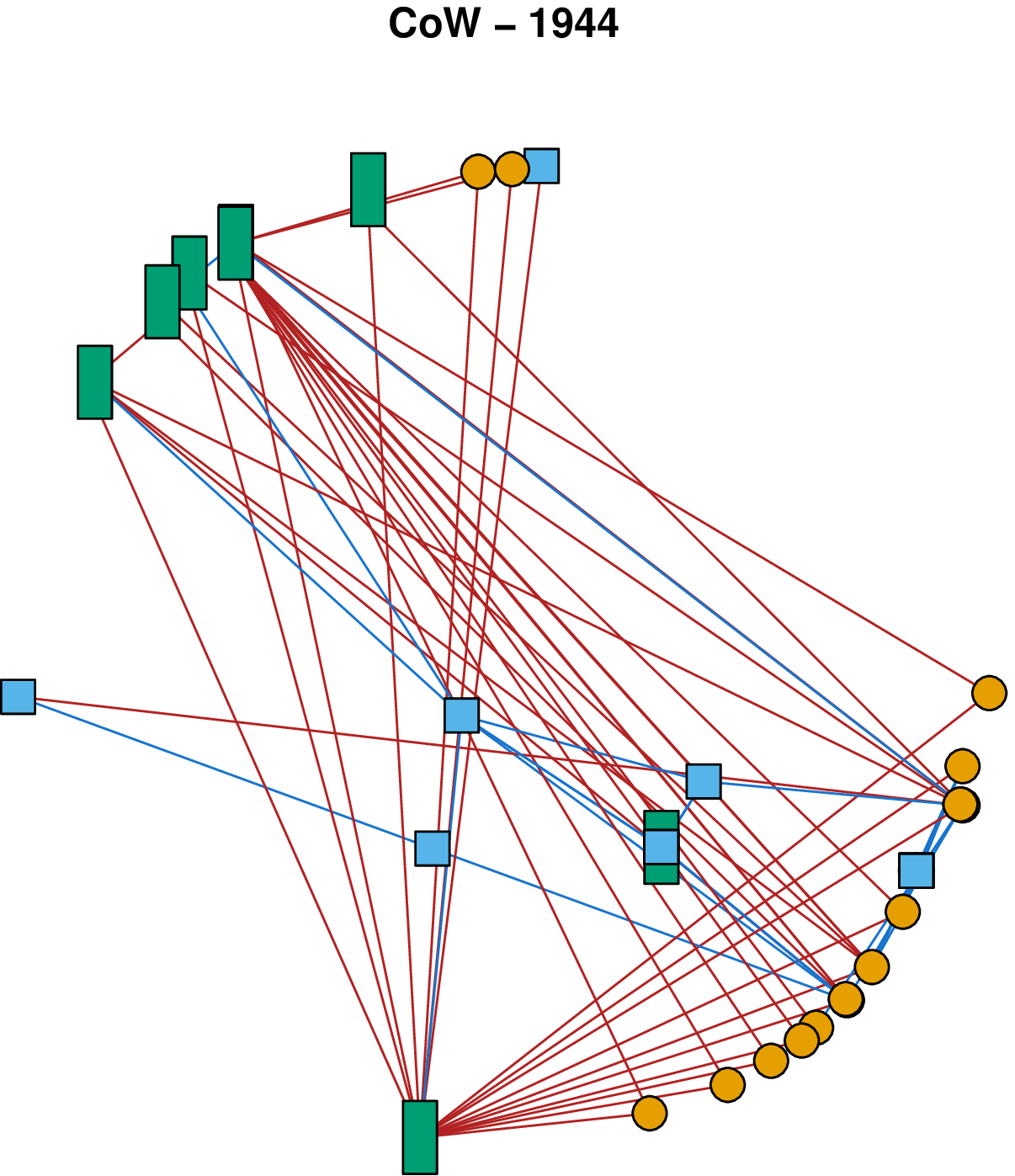}
\includegraphics[width=2in]{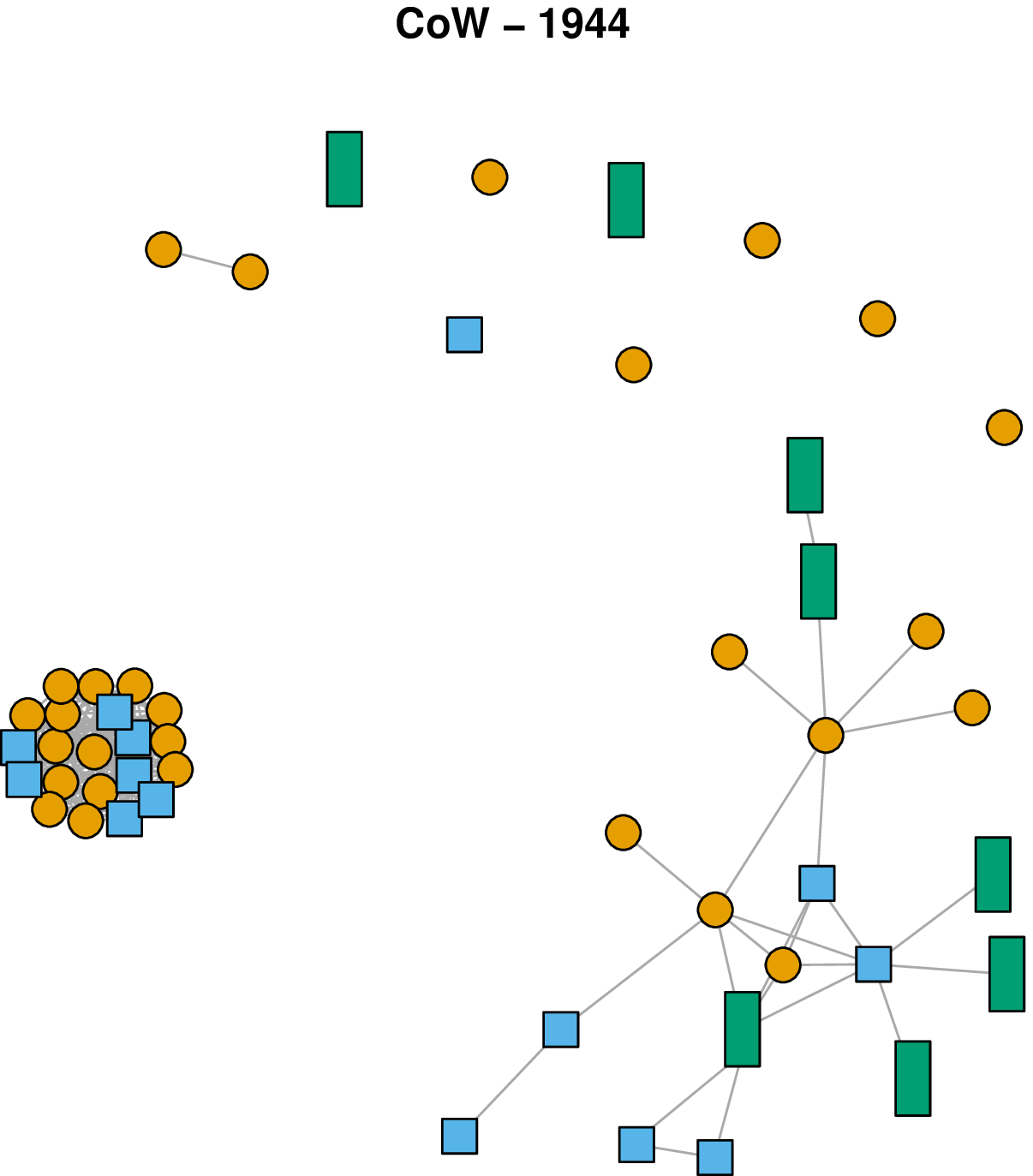}
\includegraphics[width=2in]{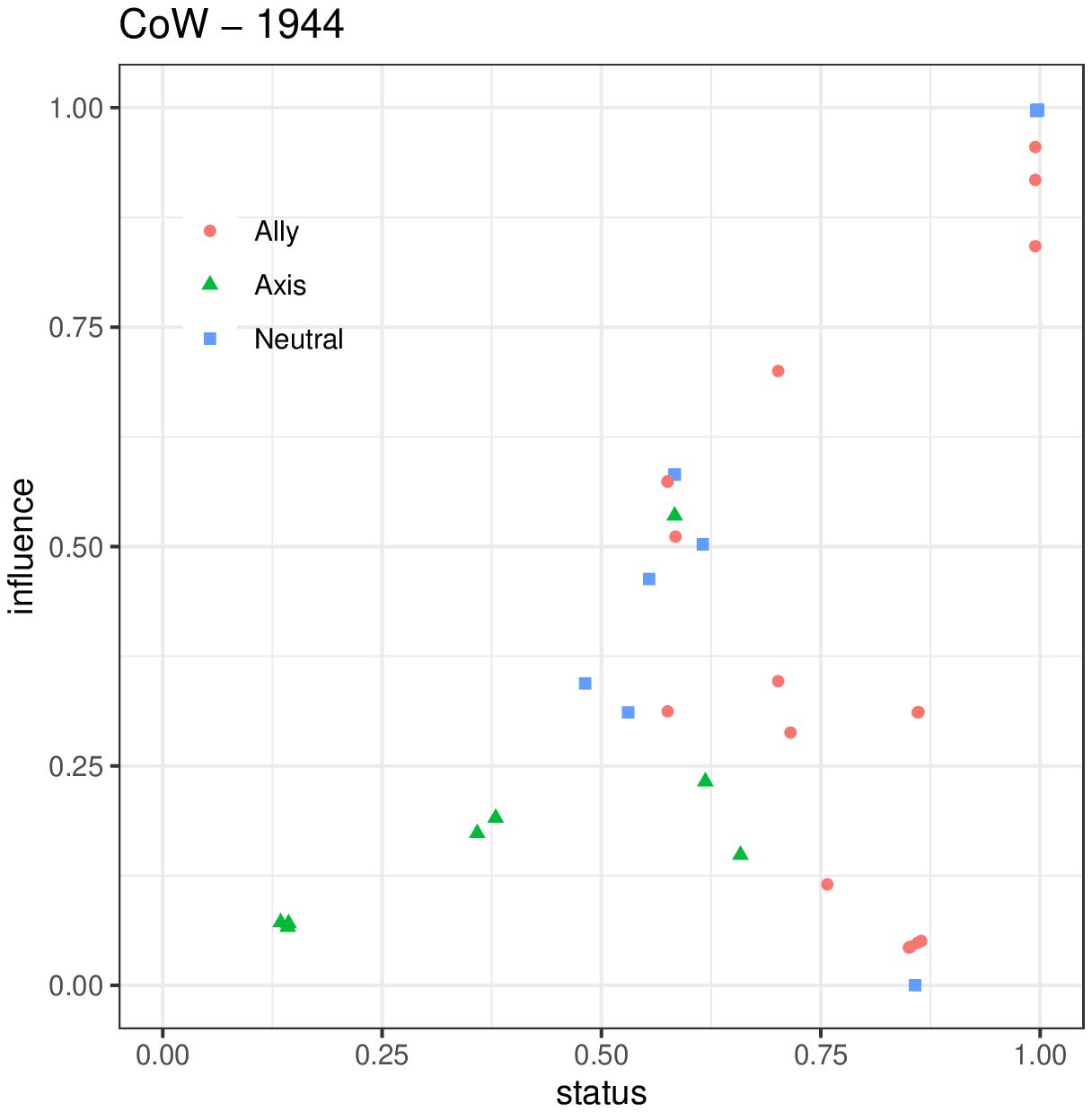}
\caption{\textbf{cow} signed graph with positive edges in blue and negative edges in red (left) and ground-truth communities labelled; the positive-edge-only subgraph of CoW with ground-truth communities labelled (center); and ground-truth labelling in graphB status-influence space (right)}\label{fig:CoW}
\end{figure}

The Correlates of War data set contains records of alliances, wars, and militarized interstate disputes between 50 countries from 1816 to 2014 \cite{COW}. In this study, we chose to focus on the year 1944, assuming that it was clear which side the participants were on during the Second World War. Resulting signed graph from the records for 1944, \textbf{cow}, has countries were represented by vertices $v=50$, and edges represented relationships between countries $e=276$. If two countries were allies, we assigned a positive edge between them. If two countries were at war or experienced a militarized interstate dispute, we assigned a negative edge between them. If there was no record of two countries interacting that year, there was no edge in the signed graph. Since we picked 1944, we assigned one of the three labels to a country: Allied Powers, Axis Powers, or remaining neutral.

\begin{table}[!ht]
\setlength\tabcolsep{2pt}
\begin{tabular}{|ll||l|l|l|l|l|l|l|l|l|l|l|l|l|}
\hline
\textbf{cow} &  & ground & \multicolumn{3}{c|}{Laplacians} & \multicolumn{2}{c|}{Balanced Cuts} & \multicolumn{2}{c|}{SPONGE}& \multicolumn{2}{c|}{Power Means} & FCSG & SSS & graphB\\
 &  & truth & \_none & \_sym & \_sep & \_none & \_sym & \_none & \_sym & GM & SPM & & net & \_km \\ \hline \hline
ground & truth & 1 & 0.06 & \underline{0.12} & -0.02 & 0.02 & 0.12 & 0.06 & 0 & 0.06 & 0.06 & -0.01 & \underline{\bf 0.22} & 0.1\\ \hline
\hline
\multirow{3}{*}{Laplacian} & \_none & 0.06 & 1 &  0.72 & 0.36 & 0.84 & 0.04 & {\bf 0.91} &  0.8 & 1 & 0.65 & 0.84 & -0.01 & 0.65\\
& \_sym & \underline{0.12} & 0.72 & 1 & 0.34 & {\bf 0.84} & 0.17 & 0.68 & 0.74 & 0.72 & 0.71 & 0.77 & 0.13 & 0.71\\
& \_sep & -0.02 & 0.36 & 0.34 & 1 & 0.38 & 0.38 & 0.36 & {\bf 0.44} & 0.36 & 0.38 & 0.33 & 0.03 & 0.4\\ \hline
Balanced  & \_none & 0.02 & 0.84 &  0.84 & 0.38 & 1 & 0.03 & 0.77 & 0.87 & 0.84 &  0.69  & {\bf 0.92} & 0.02 & 0.66\\ 
Cuts  & \_sym & 0.12 & 0.04 & 0.17 & {\bf 0.38} & 0.03 & 1 & 0.06 & 0.1 & 0.04 & 0.14 & 0 & 0.16 & 0.25\\ \hline
\multirow{2}{*}{SPONGE} & \_none & 0.06 & {\bf 0.91} & 0.68 & 0.36 & 0.77 & 0.06 & 1 & 0.74 & {\bf 0.91} & 0.67 & 0.77 & 0.01 & 0.65\\
 & \_sym & 0 & 0.8 & 0.74 & 0.44 &  0.87 & 0.1 & 0.74 & 1 & 0.8 & 0.65 & {\bf 0.88} & 0.03 & 0.65\\ \hline 
Power & GM & 0.06 & 1 & 0.72 & 0.36 & 0.84 & 0.04 & {\bf 0.91} & 0.8 & 1 & 0.65 & 0.84 & -0.01 & 0.65\\
Means & SPM & 0.06 & 0.65 & {\bf 0.71} & 0.38 &  0.69 & 0.14 & 0.67 & 0.65 & 0.65 & 1 & 0.64 & 0.06 & 0.62\\ \hline
FCSG & & -0.01 &  0.84 & 0.77 & 0.33 & {\bf 0.92} & 0 & 0.77 &  0.88 &  0.84 & 0.64 & 1 & 0 & 0.62\\ \hline
SSSnet & & \underline{\bf 0.22} & -0.01 & 0.13 & 0.03 & 0.02 & 0.16 & 0.01 & 0.03 & -0.01 & 0.06 & 0 & 1 & 0.1\\ \hline
graphB\_km & &0.1 & 0.65 & {\bf 0.71} & 0.4 &  0.66 & 0.25 & 0.65 & 0.65 & 0.65 & 0.62 & 0.62 & 0.1 & 1\\ \hline
\end{tabular}\caption{Detailed analysis of Adjusted Rand Index (ARI) for all twelve methods for the  \textbf{cow} dataset. We have marked \underline{\bf best} ARI and \underline{second best} ARI for recovering ground truth (first row and first column). Note that most of {\bf mutual} ARI scores are better than ground truth recovery, and we emphasize top {\bf mutual} ARI scores higher than ground truth recovery for each method.}\label{tab:ARICoW}
\end{table}

The Correlates of War {\bf cow} signed graph has 50 vertices, 236 positive edges and 40 negative edges as illustrated in Figure~\ref{fig:CoW}. 25 vertices have less than 9 edges, while one vertex interacts with half of the vertices (25). The reason we selected the {\bf cow} dataset is that it exhibits similar traits as social network signed graphs at a smaller scale, with skewed vertex density distribution, over 80\% positive edges, and a low density (0.225). 98.7\% of all the triangles in the graph are balanced. Correlates of War provide 3 ground community labels for the graph; see Fig.~\ref{fig:Sampson}(right). The ground truth on the signed graph shows solid community separation (88\% of negative edges are between communities) and \emph{poor} clusterability as 48\% of the positive edges among vertices are not captured by ground truth community labels.

The {\bf cow} clustering products are illustrated in Table~\ref{tab:ARICoW}. ARI scores for ground truth are low for \emph{all} methods. We conclude that ground truth labeling does not capture communities in the signed graph.  This makes sense as the WWII countries listed are from multiple continents, and the dynamic between the countries is more complex than their WWII siding. Next, we focus on the method performance and how they compare in recovering data to one another.  The Laplacian\_none and Sponge\_none ARI score is $0.91$, the balanced cuts non-symmetric method and FCSG ARI score is $0.92$, and the other mutual ARI scores for this dataset are in Table~\ref{tab:ARICoW}. 

\begin{figure}[!ht]
    \centering
    \includegraphics[width=2in, trim=50 50 50 50, clip]{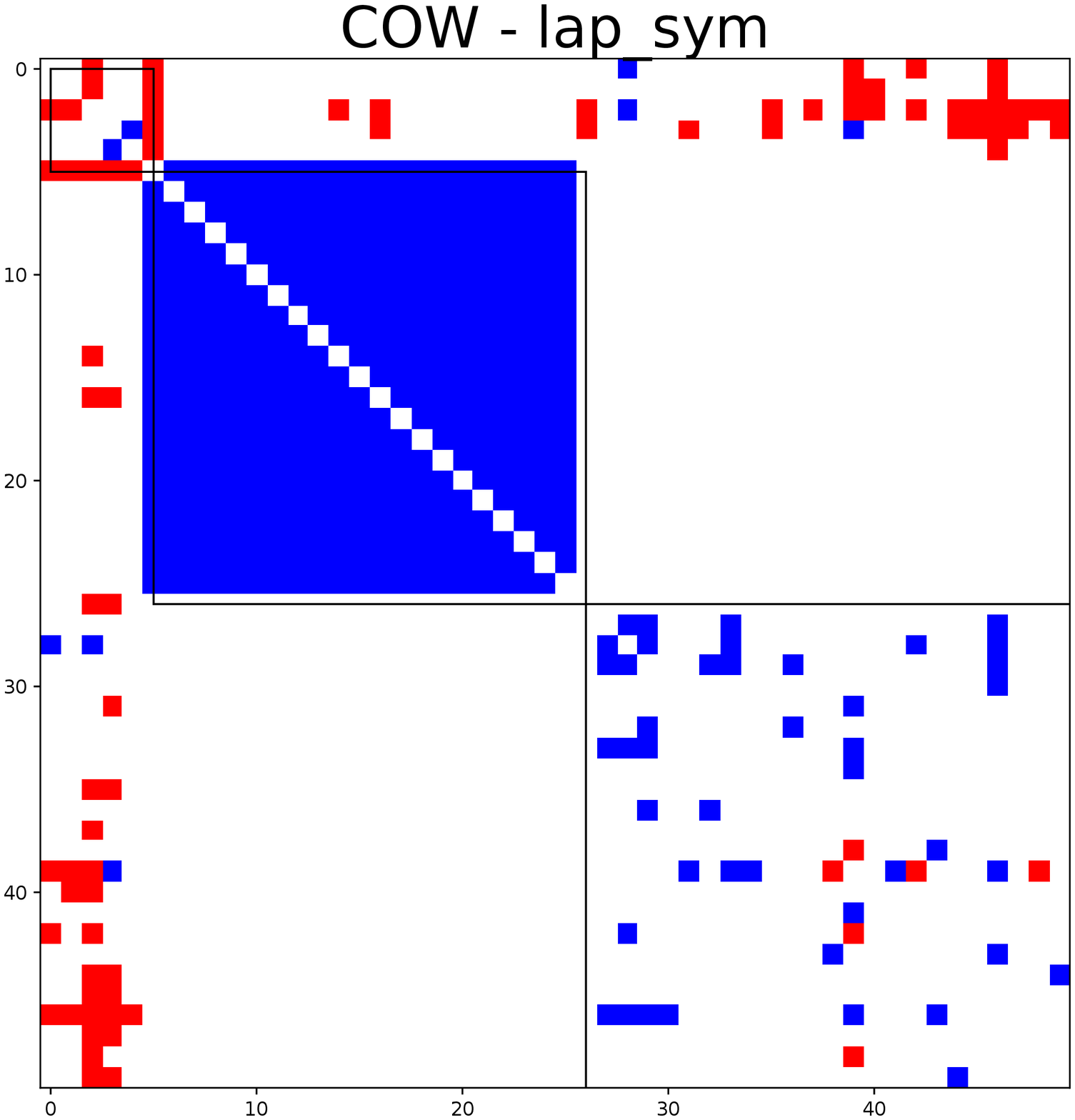}
    \includegraphics[width=2in, trim=50 50 50 50, clip]{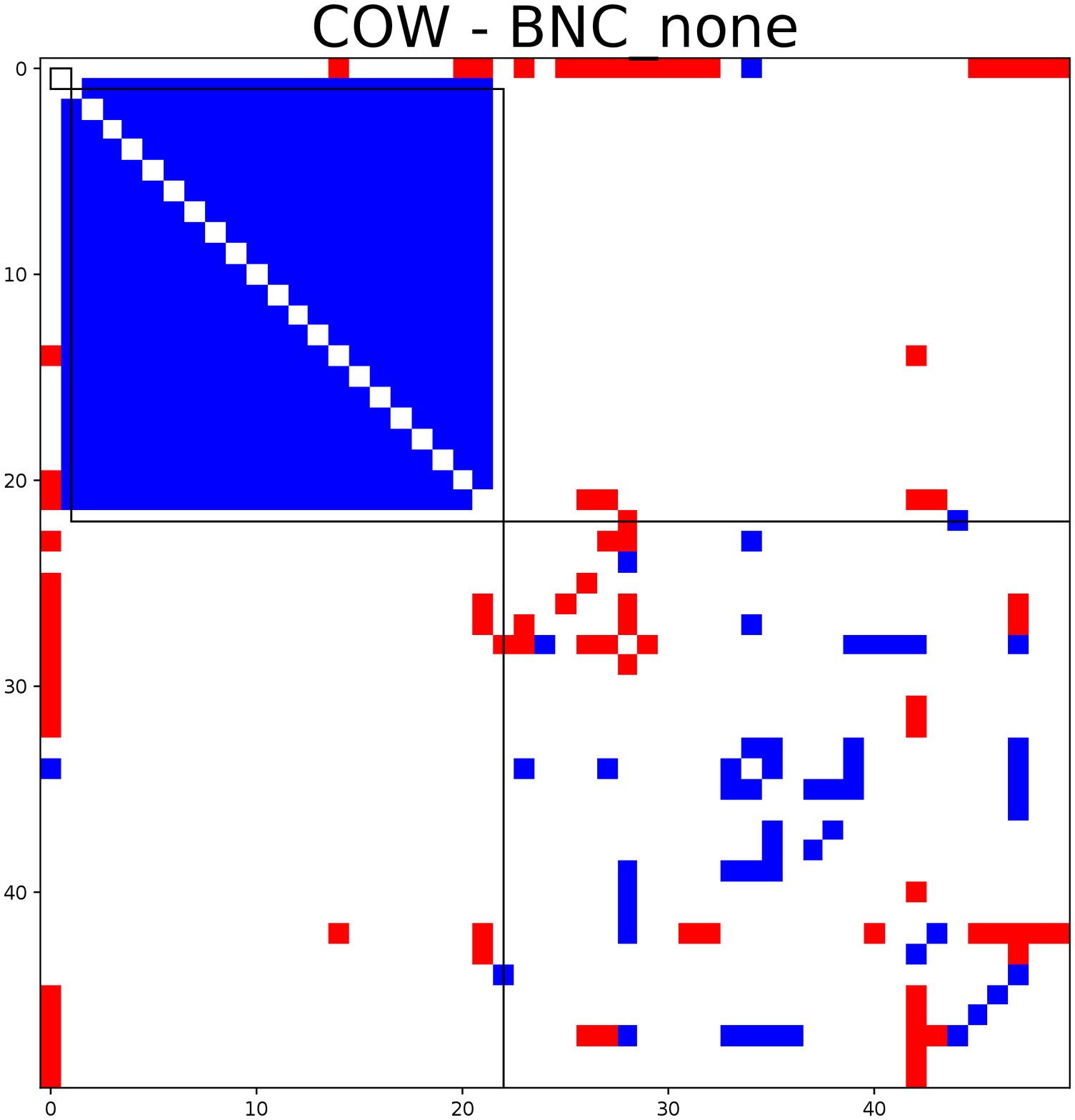}
    \includegraphics[width=2in, trim=50 50 50 50, clip]{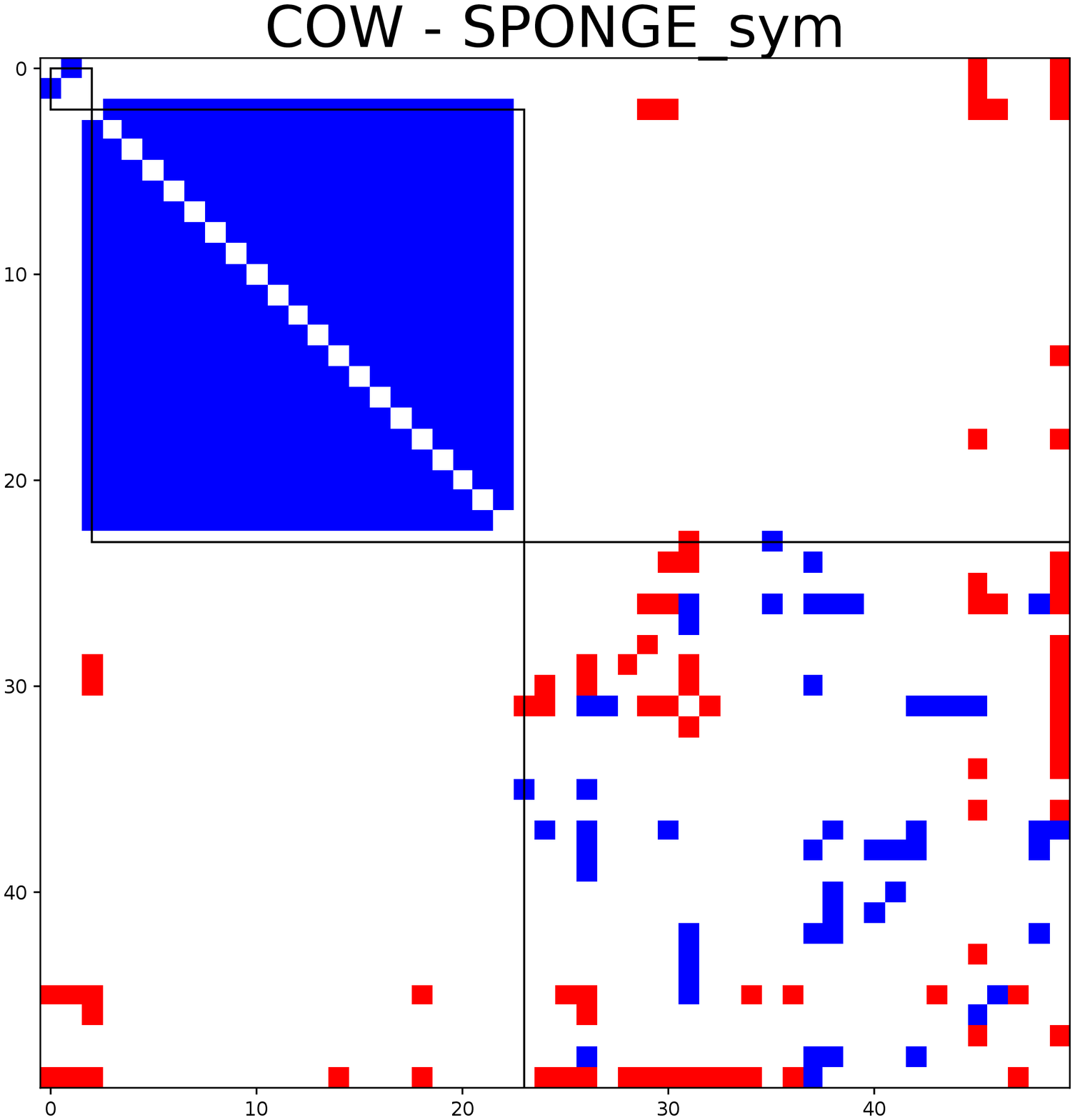}
    \caption{Adjacency matrices sorted by clustering labels for (left) symmetric Laplacian (center) balance normalized cut, and (right) symmetric SPONGE for {\bf cow} dataset.}
    \label{fig:cow_sorted}
\end{figure}

\begin{table}[!ht]
\setlength\tabcolsep{2pt}
\begin{tabular}{l||l|l|l|l|l|l|l|l|l|l|l|l|l}
\multirow{2}{*}{method} & ground & \multicolumn{3}{c|}{Laplacians} & \multicolumn{2}{c|}{Balanced Cuts} & \multicolumn{2}{c|}{SPONGE}& \multicolumn{2}{c|}{Power Means} & FCSG & SSSnet & graphB\_km\\
        & truth & \_none & \_sym & \_sep & \_none & \_sym & \_none & \_sym & GM & SPM &  &  &  \\ \hline \hline
pos\_in & 0.52  & 0.98 & \underline{\bf 0.99} & 0.94 & 1.0 & 0.99 & 0.97 & 1.0 & 0.98 & \underline{0.97} & 0.92  & 0.51 & 0.98 \\ \hline
neg\_out & 0.88  & 0.2 & \underline{\bf 0.88} & 0.73 & 0.55 & 0.78 & 0.2 & 0.3 & 0.2 & \underline{0.88} & 0.20  & 0.88 & 0.58 \\ 
\end{tabular}\caption{Percentage of positive edges in the detected communities (pow\_in) and negative edges outside the detected communities (neg\_out) per signed clustering method for {\bf cow} dataset. }\label{tab:cowCluster}
\end{table}

Next, we visualize the performance of three methods in terms of positive (blue) and negative (red) edges in adjacency matrix in Figure~\ref{fig:cow_sorted}. Adjacency matrices for {\bf cow} are sorted by clustering labels for (a) symmetric Laplacian: API 0.12 pos\_in 0.99 neg\_out 0.88 (b)  balance normalized cut: API 0.12, pos\_in 1 neg\_out 0.55 and (c) symmetric SPONGE: API 0.06, pos\_in 1 neg\_out 0.3. We can see that all three methods successfully identified the \emph{ Allied Powers} label, and it is consistently the second largest cluster out of three. Their pos\_in scores are almost perfect. We conclude that the methods work well in recovering ground truth labels, as the API score is not a great representation of the recovery measure. Where the methods differ is the ability to keep negative edges outside of the clusters. We expand our experiment analysis to all twelve methods in Table~\ref{tab:cowCluster}. All methods but SSSnet achieve near-perfect results in recovering positive edges within assigned communities better than ground truth. The performance on the negative edges separates methods into 2 groups: the ones that separate negative edges well for a fixed $k$ and ones that do not. FCSG  underlying assumption that \emph{the positive-only subgraph of the network must be a single connected component} recovers only 1 community for cow (out of 3) resulting in low neg\_out score.

\subsection{Signed Graph Clustering for Sparse Social Network Data: Sports Communities} 
\label{ssec:Sports}

\begin{figure}[!ht]
\centering
\includegraphics[width=3in]{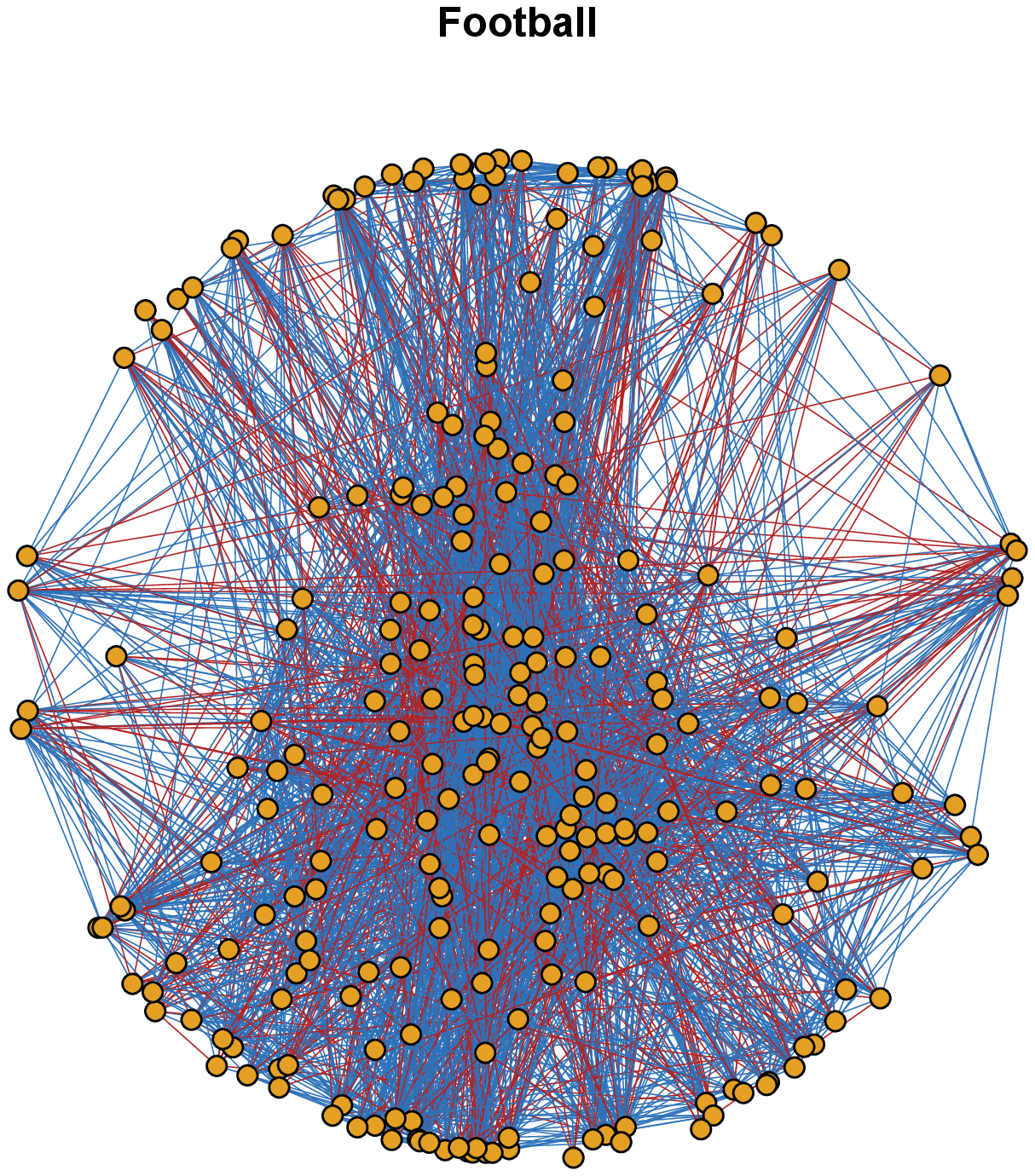}
\includegraphics[width=3in]{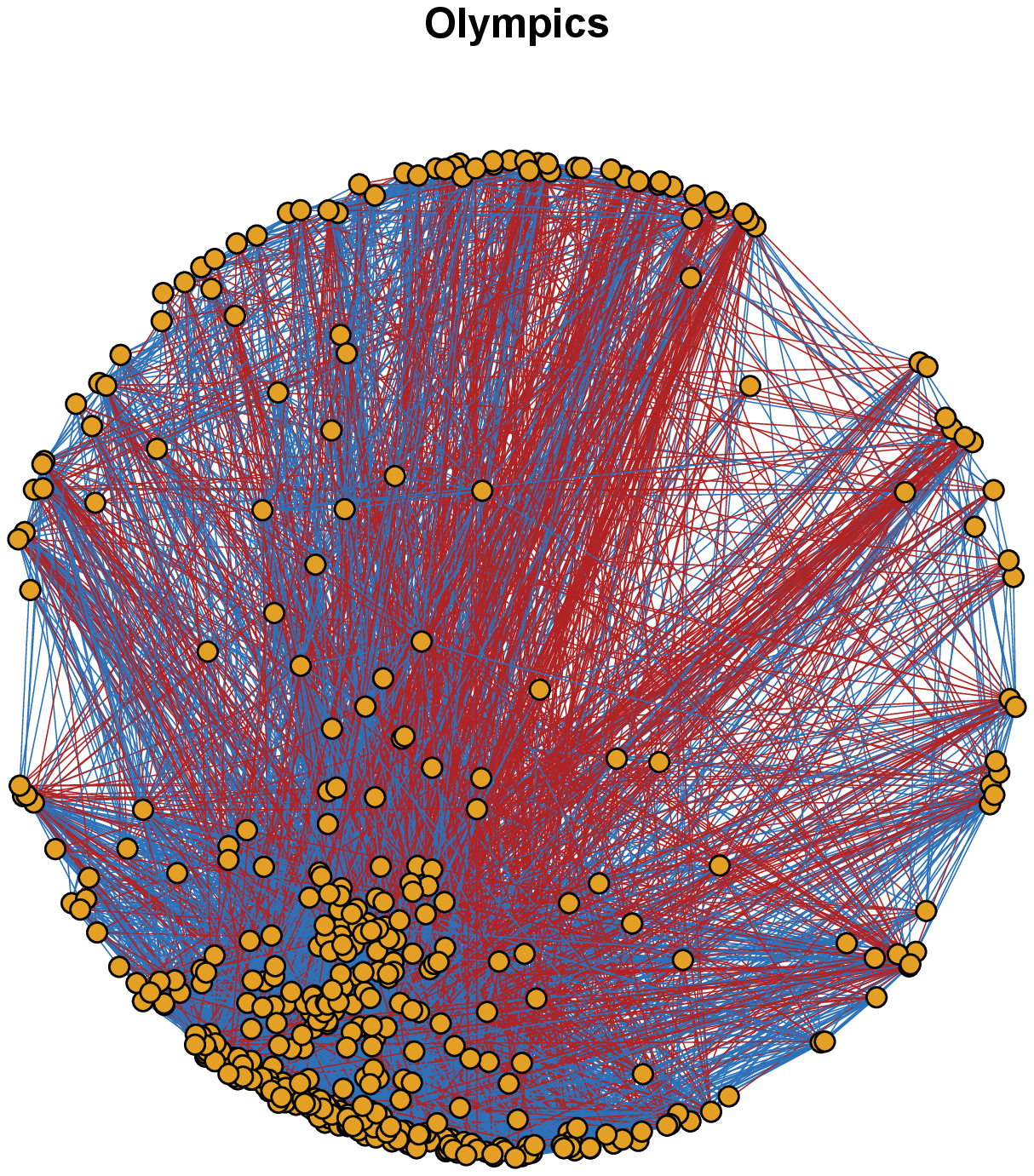}
\caption{Illustration of the network structure with blue positive edges and red negative edges. The original network have no negative edges \cite{2013greene}, and the 20\% of negative edges was augmented based on the community label separation for {\bf football} (left) and {\bf olympics} (right) data sets \cite{2013greene}.
}\label{fig:sportsGraphs}
\end{figure}

For this experiment, we have used the network graph datasets constructed in \cite{2013greene}. They have used cosine similarity to produce edges, and we are using the unweighted directed follower graph the authors produced for {\bf football} and {\bf olympics} dataset forms positive only graph basis construction. Since the sports graphs do not have negative edges, we augment them. In {\bf football} data, the graph captures club fan communities, so it is appropriate to use randomized block sampling to insert negative edges among communities. In {\bf olympics} data, the graph captures athletes competing in a sport.  Since they were not likely to compete in other sports, we have used randomized block sampling to insert negative edges among labeled groups.  We enforce the separation among the communities in the following fashion: first, we randomly select communities to insert a negative edge and then randomly select nodes within those communities for that negative edge to be added between selected nodes. All generated random edges were checked against previously generated negative edges and existing positive edges to avoid duplication. The number of negative edges to add was determined as a percent relative to the number of positive edges. In this case, the number of negative edges was set to be 20 percent of the number of positive edges. Negative edges are only added between communities. 

\subsubsection{Football}

\begin{table}[!ht]
\setlength\tabcolsep{2pt}
\begin{tabular}{|ll||l|l|l|l|l|l|l|l|l|l|l|l|l|}
\hline
\textbf{football} &  & ground & \multicolumn{3}{c|}{Laplacians} & \multicolumn{2}{c|}{Balanced Cuts} & \multicolumn{2}{c|}{SPONGE}& \multicolumn{2}{c|}{Power Means} & FCSG & SSS & graphB\\
 &  & truth & \_none & \_sym & \_sep & \_none & \_sym & \_none & \_sym & GM & SPM & & net & \_km \\ \hline \hline
ground& \_truth & 1 & 0.03 & \underline{\bf 0.76} & 0.09 & 0.68 & 0.49 & \underline{0.74} & 0.71 & 0.08  & 0.27 & 0 & 0.7 & 0.01\\ \hline \hline
\multirow{3}{*}{Laplacian} & \_none     & 0.03          & 1         & 0.03     & 0.01          & 0.02      & 0.09     & 0.04         & 0.04        & 0.05  & 0.07  & 0.18  & 0.04   & 0.01  \\
& \_sym      & \underline{\bf 0.76}          & 0.03      & 1        & 0.09          & 0.68      & 0.48     & {\bf 0.84}         & 0.7         & 0.08  & 0.28  & 0     & 0.66   & 0.01       \\
& \_sep & 0.09          & 0.01      & 0.09     & 1             & 0.09      & 0.09     & 0.08         & 0.12        & 0.01  & 0.04  & 0     & 0.1    & 0          \\ \hline
Balanced & \_none     & 0.68          & 0.02      & 0.68     & 0.09          & 1         & 0.46     & 0.65         & 0.66        & 0.07  & 0.28  & 0     & 0.63   & 0.03       \\
Cuts & \_sym      & 0.49          & 0.09      & 0.48     & 0.09          & 0.46      & 1        & 0.48         & 0.51        & 0.07  & 0.23  & 0.03  & 0.46   & 0.01       \\ \hline
\multirow{2}{*}{SPONGE} & \_none  & \underline{0.74}          & 0.04      & {\bf 0.84}     & 0.08          & 0.65      & 0.48     & 1            & 0.67        & 0.07  & 0.27  & 0.01  & 0.66   & 0.02 \\
& \_sym   & 0.71          & 0.04      & 0.7      & 0.12          & 0.66      & 0.51     & 0.67         & 1           & 0.08  & 0.27  & 0.01  & 0.64   & 0.01       \\ \hline
Power & GM            & 0.08          & 0.05      & 0.08     & 0.01          & 0.07      & 0.07     & 0.07         & 0.08        & 1     & 0.15  & -0.01 & 0.08   & 0          \\
Means & SPM           & 0.27          & 0.07      & 0.28     & 0.04          & 0.28      & 0.23     & 0.27         & 0.27        & 0.15  & 1     & -0.01 & 0.25   & 0          \\
FCSG & & 0             & 0.18      & 0        & 0             & 0         & 0.03     & 0.01         & 0.01        & -0.01 & -0.01 & 1     & 0.01   & 0          \\
SSSnet & & 0.7           & 0.04      & 0.66     & 0.1           & 0.63      & 0.46     & 0.66         & 0.64        & 0.08  & 0.25  & 0.01  & 1      & 0.01       \\
graphB\_km & & 0.01          & 0.01      & 0.01     & 0             & 0.03      & 0.01     & 0.02         & 0.01        & 0     & 0     & 0     & 0.01   & 1 \\ \hline
\end{tabular}\caption{Detailed analysis of Adjusted Rand Index (ARI) for all twelve methods for the Football dataset with emphasis on \underline{\bf best} ARI and \underline{second best} ARI for recovering ground truth (first row and first column). One {\bf mutual} ARI method score is better than ground truth recovery, and we observe a great variations in ARI scores per methods.}\label{tab:ARIfootball}
\end{table}

The Football data set represents football players and clubs from the English Premier League. The data contains 248 Twitter users grouped into one of 20 clubs in the league. Positive edges in the graph are constructed based on the computed co-occurrence of two players from a larger Twitter community of 7,814 followers \cite{2013greene}. Figure~\ref{fig:sportsGraphs} (left) illustrates the network with  for the network structure. The graph has 2644 positive edges, and we have added 530 negative edges among 20 communities. Vertex degree distribution is converging to power law, as one vertex is connected to almost half of the dataset, while half of vertices are connected to less than 10\% of the vertex set. This is a very sparse graph (0.104) with 87.8\% of the triangles balanced and all the characteristics of a real social networks. 

The dataset has 20 labels, and since we augmented negative edges only between the communities, we get the perfect neg\_out score.  Only 41\% of positive edges are captured within clusters by ground truth. The high number of positive edges among players in different communities indicates that they all know each other. The labeled community here is more defined by augmented negative edges than by true positive ones.  Mutual ARI between methods greatly varies in Table~\ref{tab:ARIfootball}. Laplacian\_sym, Balance Cuts, SPONGE and SSSnet seem to recover some ground truth, while other methods fail.  FGSC yielded isolated vertices at the end of runtime and they were placed in an 'outcast' cluster together to ensure communities sizes of at least two, similar to the sampson dataset.  As graphB builds positive clusters, applying fixed $k$ in $k$-means only distorts its outcome, and hierarchical clustering is better suited. We conclude that experiment needs Twitter edge sentiment data to construct a more truthful signed network graph.

\begin{table}[!ht]
\setlength\tabcolsep{2pt}
\begin{tabular}{|ll||l|l|l|l|l|l|l|l|l|l|l|l|l|}
\hline
\textbf{olympics} &  & ground & \multicolumn{3}{c|}{Laplacians} & \multicolumn{2}{c|}{Balanced Cuts} & \multicolumn{2}{c|}{SPONGE}& \multicolumn{2}{c|}{Power Means} & FCSG & SSS & graphB\\
 &  & truth & \_none & \_sym & \_sep & \_none & \_sym & \_none & \_sym & GM & SPM & & net & \_km \\ \hline \hline
ground& truth & 1             & 0.02      & 0.72    & 0.21          & 0.3       & 0.11     & 0.64         & \underline{\bf 0.85} & 0.38 & 0.46 & 0.03  & \underline{0.8}  & 0.05  \\
\multirow{2}{*}{Laplacian}  &  \_none     & 0.02          & 1         & 0.02     & 0.02          & -0.05     & 0.22     & 0.03         & 0.03        & 0.06 & 0.06 & 0.14  & 0.03   & 0          \\
& \_sym      & 0.72         & 0.02      & 1        & 0.18          & 0.25      & 0.08     & 0.68         & 0.7         & 0.31 & 0.38 & 0.02  & 0.66   & 0.05       \\
& \_sep & 0.21          & 0.02      & 0.18     & 1             & 0.08      & 0.06     & 0.17         & 0.2         & 0.15 & 0.16 & 0.01  & 0.2    & 0.02       \\ \hline
Balanced & \_none     & 0.3           & -0.05     & 0.25     & 0.08          & 1         & 0.02     & 0.21         & 0.28        & 0.19 & 0.23 & -0.06 & 0.28   & 0.04       \\
Cuts & \_sym      & 0.11          & 0.22      & 0.08     & 0.06          & 0.02      & 1        & 0.09         & 0.11        & 0.14 & 0.13 & 0.3   & 0.12   & 0.01       \\ \hline
\multirow{2}{*}{SPONGE} & \_none  & 0.64          & 0.03      & 0.68     & 0.17          & 0.21      & 0.09     & 1            & 0.64        & 0.29 & 0.35 & 0.03  & 0.63   & 0.04       \\
& \_sym   & \underline{\bf 0.85}          & 0.03      & 0.7      & 0.2           & 0.28      & 0.11     & 0.64         & 1           & 0.35 & 0.43 & 0.03  & 0.75   & 0.05       \\ \hline
Power & GM            & 0.38          & 0.06      & 0.31     & 0.15          & 0.19      & 0.14     & 0.29         & 0.35        & 1    & 0.5  & 0.03  & 0.37   & 0.03       \\
Means & SPM           & 0.46          & 0.06      & 0.38     & 0.16          & 0.23      & 0.13     & 0.35         & 0.43        & 0.5  & 1    & 0.01  & 0.43   & 0.02       \\
FCSG    &      & 0.03          & 0.14      & 0.02     & 0.01          & -0.06     & 0.3      & 0.03         & 0.03        & 0.03 & 0.01 & 1     & 0.03   & 0.01       \\
SSSnet & & \underline{0.8}           & 0.03      & 0.66     & 0.2           & 0.28      & 0.12     & 0.63         & 0.75        & 0.37 & 0.43 & 0.03  & 1      & 0.05       \\
graphB\_km &   & 0.05          & 0         & 0.05     & 0.02          & 0.04      & 0.01     & 0.04         & 0.05        & 0.03 & 0.02 & 0.01  & 0.05   & 1  \\ \hline
\end{tabular}\caption{Detailed analysis of Adjusted Rand Index (ARI) for all twelve methods for {\bf olympics} dataset with emphasis on \underline{\bf best} ARI and \underline{second best} ARI for recovering ground truth (first row and first column). Olympics data show great variations in ARI scores per method.}\label{tab:ARIOlympics}
\end{table}

\subsubsection{Olympics}  The Olympics data set features athletes and organisations that were part of the London 2012 Summer Olympics. The data contains 464 Twitter users grouped into 28 different sports based on an analysis of the profile content of 4942 users, whom they follow, and their 725662 tweets \cite{2013greene}. Each vertex (athlete) belongs to only one community (sport); see Figure~\ref{fig:sportsGraphs} (right) for the network structure of 7784 positive edges (in blue) and augmented 1561 negative edges in red \cite{2013greene}. Vertex degree distribution converges to power law as half of the vertices are connected with less than 7\% of the vertex set, and one super user is connected to 45\% of all other athletes. 92\% of the triangles are balanced, and this graph is sparse with a density of $<0.09$. There are 28 communities, and ground truth captures 45\% of the positive labels within the groups. When we augment negative edges only between the communities, we get the perfect neg\_out score. The high number of positive edges among athletes in different communities indicates they are likely connected by country or adjacent sport. This dataset is more suitable for overlapping labeling, and the labeled community here is more defined by augmented negative edges than by true positive ones. The method ARI scores are very similar to the football dataset. 

\begin{table}[!ht]
\setlength\tabcolsep{3pt}
\begin{tabular}{l||l|l|l||l|l|l||l|l||l|l|l}
\textbf{labeled} & {\bf vertices} & \multicolumn{2}{c||}{{\bf edges}} &  \multicolumn{3}{c||}{\bf vertex degrees} & \multicolumn{2}{c||}{\bf attributes} &  \multicolumn{3}{c}{\bf communities} \\
\textbf{dataset} & $v$ & $e$ & \% positive & average & median & max & density $d$ & $bal_3$ & $l$ & pos\_in & neg\_out \\
 \hline  \hline
highland \cite{1954Read}& 16 & 58 & 50 & 7.25 & 7.5 & 10 & 0.483 & 0.868 & 3 & 0.93 & 1\\  \hline
sampson \cite{sampson} & 18 & 112 & 54.4 & 12.44 & 12.50 & 16 & 0.732 & 0.6 & 4 & 0.48 & 0.98 \\  \hline
cow \cite{COW} & 50 & 276 & 85.51 & 11.04 & 9 & 25  & 0.225 & 0.987 & 3 & 0.52 & 0.88 \\ \hline
football \cite{2013greene}  & 248 & 3174 & 83.3 & 25.6 & 22.5 & 121 & 0.104 & 0.878 & 20 & 0.41 & 1 \\ \hline
olympics  \cite{2013greene} & 464 & 9345 & 83.3 & 40.28 & 33.00 & 207 & 0.087 & 0.920 & 28 & 0.45 & 1 \\
\end{tabular}
\caption{Data set attributes: $v$ is a number of vertices; $e$ is a number of edges in a graph; \emph{ \% positive} is the number of positive edges divided by $e$; vertex degree statistics is computed in terms of average, mean, and max node degree; graph density $d$ is calculated as in \ref{eq:d}, and $bal_3$ is the percent of triangles in the graph that are balanced; $l$ is a number of communities,  pos\_in is the percentage of positive edges in the ground truth communities, and neg\_out is the percentage of negative edges between ground truth communities.} \label{tab:exp1}
\end{table}


\begin{figure}[!ht]
\centering
\includegraphics[width=2.7in]{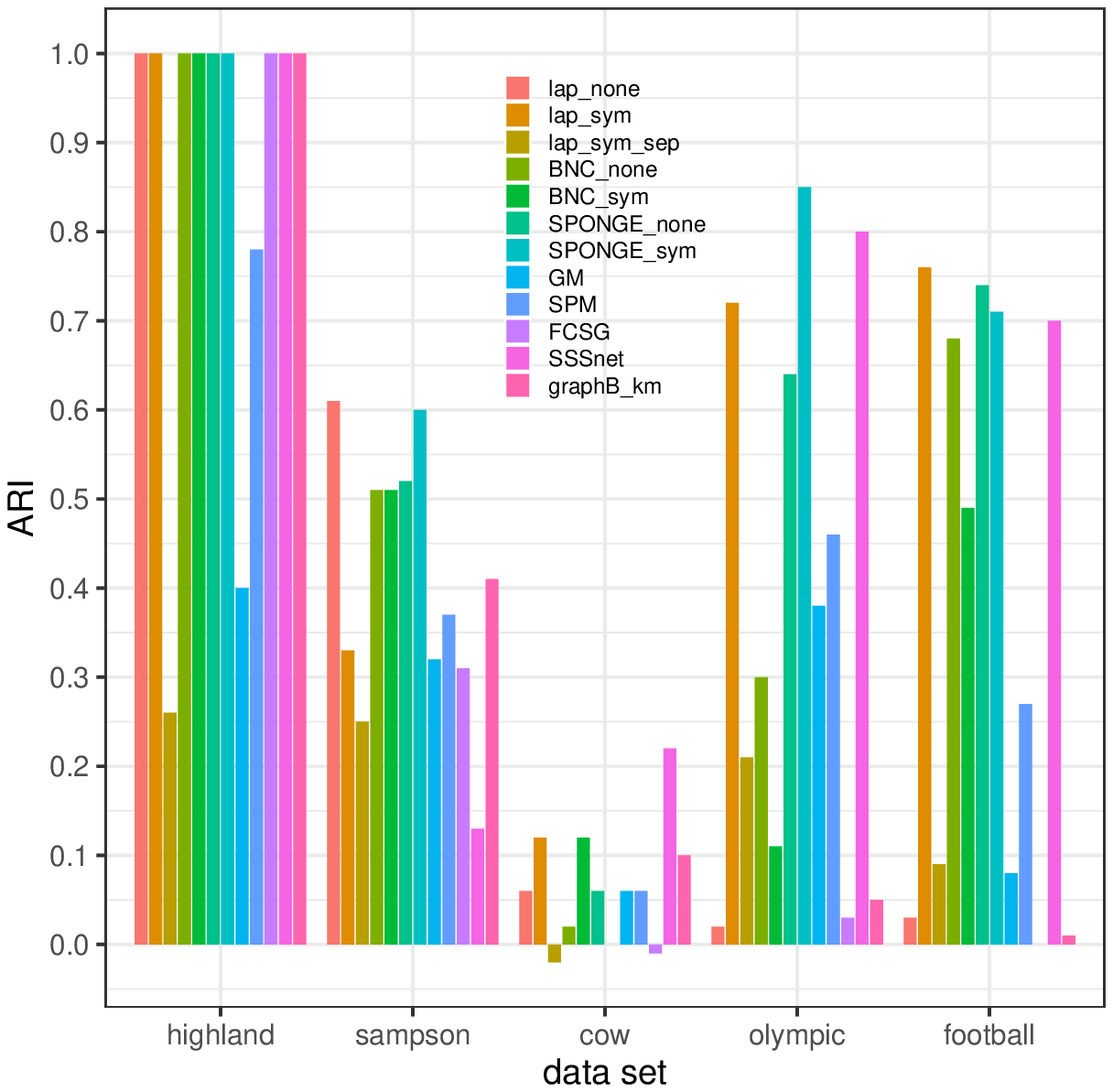}
\includegraphics[width=3.35in]{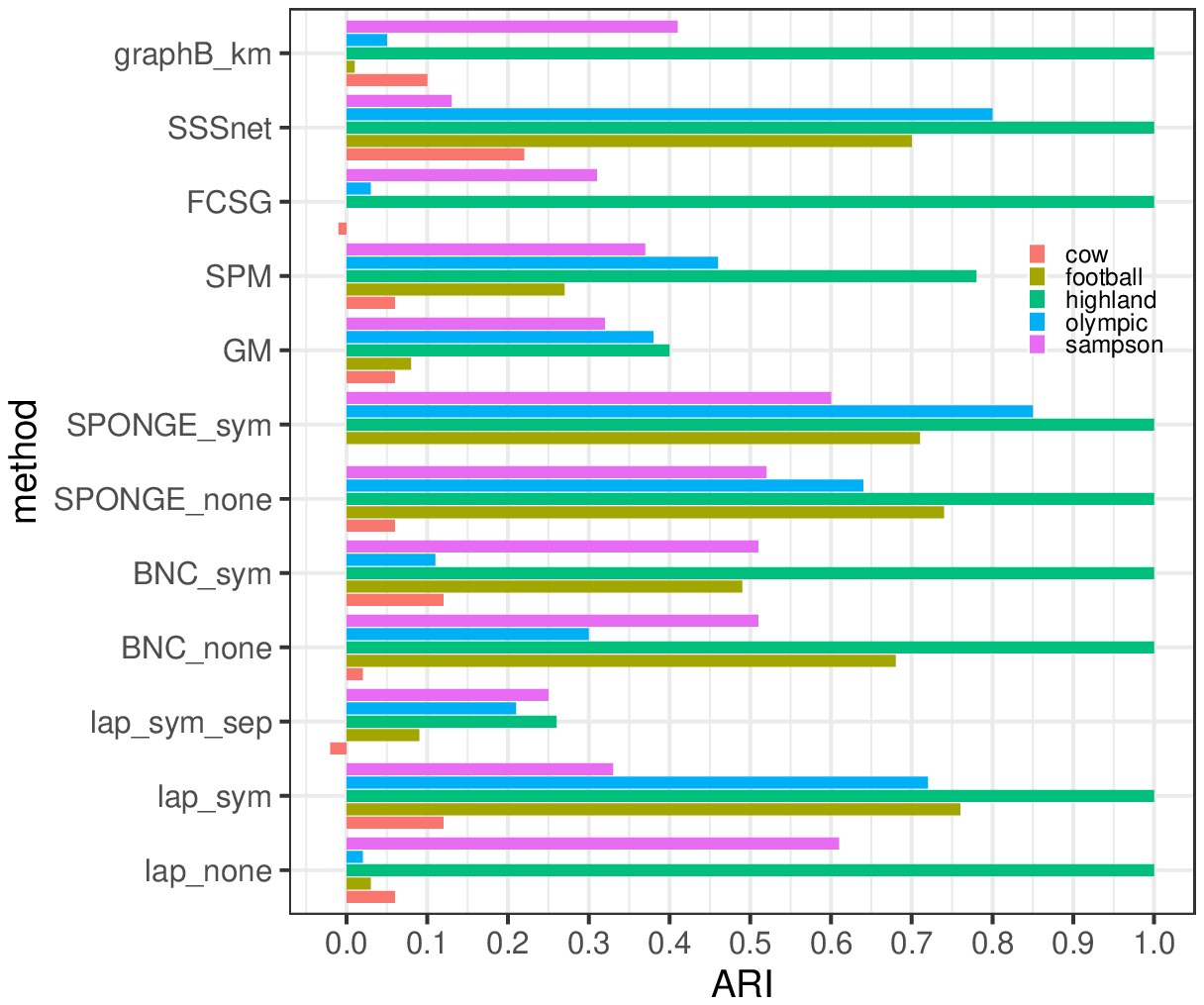}
\caption{The Adjusted Rand Index (ARI) \cite{ari_article} based comparison of the accuracy of the methods: by data set (left), and by method (right). A  value near 0 represents a random pairing and 1 represents perfect recovery of ground truth labels.}\label{fig:labeledGraph}
\end{figure}

\begin{table}[!ht]
\setlength\tabcolsep{6pt}
\begin{tabular}{l||l|l|l|l|l|l|l|l|l|l|l|l}
\textbf{method/}  & \multicolumn{3}{c|}{Laplacians} & \multicolumn{2}{c|}{Balanced Cuts} & \multicolumn{2}{c|}{SPONGE}& \multicolumn{2}{c|}{Power Means} & FCSG & SSS & graphB \\
\textbf{dataset} & \_none & \_sym & \_sep & \_none & \_sym & \_none & \_sym & GM & SPM &  & net & \_km \\ \hline \hline
highland & \textbf{1}  & \textbf{1} & 0.26 & \textbf{1}  & \textbf{1}  & \textbf{1}  & \textbf{1} & 0.4 & 0.78  & \textbf{1}  & \textbf{1}  & \textbf{1}\\ \hline
sampson & \underline{\bf 0.61} & 0.33 & 0.25  & 0.51 & 0.51 & 0.52 & \underline{0.6} & 0.3 & 0.37 & 0.41 & 0.31 & 0.13\\\hline
cow & 0.06 & \underline{0.12} & -0.02 & 0.02 & 0.12 & 0.06 & 0 & 0.06 & 0.06 & -0.01 & \underline{\bf 0.22} & 0.1\\
 \hline
football & 0.03 & \underline{\bf 0.76} & 0.09 & 0.68 & 0.5 & \underline{0.74} & 0.71 & 0.08 & 0.27 & 0 & 0.7 & 0.01\\ \hline
olympics & 0.02 & 0.72 & 0.21 & 0.3 & 0.11 & 0.64 & \underline{\bf 0.85} & 0.38 & 0.46 & 0.03 & \underline{0.8} & 0.07\\
\end{tabular}
\caption{Adjusted Rand Index (ARI) illustrated in Figure~\ref{fig:labeledGraph} for twelve methods over five data sets considered with emphasis on \underline{\bf best} ARI and \underline{second best} ARI for recovering ground truth.
}\label{tab:exp1ARI}
\end{table}

\subsection{Experiment analysis and Conclusion}
\label{ssec:all}

Network attributes of the five graphs we have applied twelve signed clustering methods to are outlined side-by-side in Table~\ref{tab:exp1}. Figure~\ref{fig:labeledGraph} captures ARI score comparisons per dataset (left) and per method (right) on five datasets. The bolded scores in Table~\ref{tab:exp1ARI} show the top performer per data set for the ground truth recovery. 

We have analyzed the clustering success with respect to the size, sparseness, balance, and percent of the positive edges of these data sets, and the results are summarized in Table~\ref{tab:exp1ARI}. This experiment does not show clear trends in the effectiveness of a clustering algorithm with respect to the characteristics of the dataset as listed in Table~\ref{tab:exp1}; dataset size, ratio of positive edges, and density do not influence the top performing algorithm. 
Clusterable datasets (Table ~\ref{tab:exp1} higher pos\_in and neg\_out scores for ground truth) tend to give better results across most methods (see Table~\ref{tab:exp1ARI}), especially for symmetric Laplacian, symmetric SPONGE, and Balanced Cuts. The SPM (matrix power means) \cite{mercado2019} approach was an update to the originally proposed geometric means (GM) \cite{mercado2016}, and SPM has shown in our experiments to do a better job across the board than the GM, as illustrated in Table~\ref{tab:exp1ARI}.

The Highland dataset is generally the easiest small dataset to recover ground truth, and all methods handled this well except for Laplacian\_sep and Power Means. Sampson serves as a contrasting small dataset with poor ground truth recovery with singed Laplacian and symmetric SPONGE producing the best (but only serviceable) results, and SSSnet falling furthest behind. For Correlates of War, only SSSnet was able to recover some ground truth. The Football and Olympics dataset evaluation shows symmetric Laplacian and symmetric SPONGE as the best performers.

We have made several interesting observations in this section. First, the success of SSSnet seems related to the percentage of balanced triangles, but not the density of edges. This is interesting as the entire balance for the signed graph is reclaimable via triangles in complete graphs. Second, balanced normalized cuts and graphB (which uses a generalization of balanced cuts via the frustration cloud) provide competitive results when they are more suited for hierarchical applications.

\section{Scaling Considerations For Signed Graph Community Discovery in Real Networks}
\label{sec:exp2}
\begin{table}[!ht]
\setlength\tabcolsep{2pt}
\begin{tabular}{l||l||l}
\textbf{Criteria} &  \textbf{Method Name} & \textbf{Table and Figure label}\\ \hline \hline
\multirow{3}{*}{Effectiveness} & Symmetric Laplacian \cite{Kunegis2009} & lap\_sym \\
& Balanced Normalized Cuts \cite{chiang_scalable_2012} & BNC\_none\\ 
& SPONGE \cite{cucuringu_sponge_2019}  & SPONGE\_sym \\ \hline \hline
\multirow{3}{*}{Scalability} & Fast Clustering for Signed Graphs  \cite{hua_fast_2020} & FCSG\\ 
& graph Balancing \cite{2020Cloud} & graphB\_km\\ 
\end{tabular}
\caption{Five methods used in Section~\ref{sec:exp2} experiments.}\label{tab:select}
\end{table}

In this section, we select five out of the twelve methods discussed and evaluate their performance on real signed social networks \cite{snapnets}. We also evaluate how these methods fare against large sparse networks. Symmetric Laplacian, Symmetric SPONGE, and balanced normalized cuts were selected as they demonstrated the most robust performance in Section~\ref{sec:exp1} experiments. We have selected Fast Clustering for Signed Graphs (FCSG) \cite{hua_fast_2020} and graph Balancing (graphB) \cite{2020Cloud} methods as they both claim to have the potential to deal with the large sparse graphs that are social networks. Table~\ref{tab:select} summarizes five methods evaluated in this section. We wanted to include SSSnet also, but we were not able to successfully run provided code on large graphs due to underlying OpenBlas library error for large dataset. First, we access the effectiveness and the efficiency of the spectral methods in the proposed framework in Section~\ref{ssec:modularity}. Second, we analyze implementation and scalability of FCSG and graphB methods, and discuss avenues for the improvement in Section~\ref{ssec:scale}. 

\begin{table}[!ht]
\setlength\tabcolsep{6pt}
\begin{tabular}{l||l||l|l||l|l|l||l|l}
\textbf{labeled} & {\bf vertices} & \multicolumn{2}{c||}{{\bf edges}} &  \multicolumn{3}{c||}{\bf vertex degrees} & \multicolumn{2}{c}{\bf attributes}  \\
\textbf{dataset} & $v$ & $e$ & \% positive & average & median & max & density $d$ & $bal_3$ \\
 \hline  \hline
 cow & 50 & 276 & 85.51 & 11.04 & 9 & 25  & 0.225 & 0.987 \\ \hline
wiki & 7,468 & 105,160 & 73.33 & 28.16 & 5 & 1,007 & 0.004 & 0.798 \\  \hline
slashdot & 82,140 & 500,481  & 77.03 & 12.19 & 2 & 2,548 & $< 0.001$ & 0.856\\  \hline
epinions & 131,828 & 711,210 & 83.23 & 11.82 & 2 & 3,558 & $< 0.001$ & 0.890 \\ 
\end{tabular}
\caption{Data set attributes: $v$ is a number of vertices; $e$ is a number of edges in a graph; \emph{ \% positive} is the number of positive edges divided by $e$; vertex degree statistics is computed in terms of average, mean, and max node degree; graph density $d$ is calculated as in \ref{eq:d}, and $bal_3$ is the percent of triangles in the graph that are balanced; $l$ is a number of communities,  pos\_in is the percentage of positive edges in the ground truth communities, and neg\_out is the percentage of negative edges between ground truth communities.}\label{tab:exp2}
\end{table}

We are using four signed networks: Correlates of War (cow) \cite{COW}, Wikipedia Elections (wiki) \cite{snapnets}, Slashdot \cite{snapnets}, and Epinions \cite{snapnets}. Their attributes are listed in Table~\ref{tab:exp2}. The max vertex degree node reflects the existence of influencers in the network. From Table~\ref{tab:exp2}, we see that the median vertex degree for Wikipedia Elections is 5, and for Slashdot and Epinions is 2. Median degree in sparse networks is usually much lower than the average degree, unlike Table~\ref{tab:exp1} data.

\paragraph{Wikipedia Elections} The Wikipedia Elections data set was created from data curated by the Stanford Network Analysis Project (SNAP) Wikipedia vote network \cite{snapnets}.  It consists of 7066 Wikipedia users running for and voting in adminship elections prior to January 3, 2008.  There were a total of 103,663 votes cast in 2794 elections. Users are represented by nodes, and votes are represented by edges, with a positive edge between $i$ and $j$ indicating that either $i$ voted for $j$ or vice versa.  Negative edges indicate a negative vote between users.  Duplicate and self-referencing edges were removed, and users who were both voters and candidates were represented by separate voter nodes and candidate nodes, with the voter node adjacent only to the edges representing votes they cast and the candidate node adjacent only to the edges representing votes they received.  Before clustering, the greatest connected component (GCC) of the graph was taken to eliminate the trivial clustering problem of isolated vertices and/or disconnected components, which will always be placed in their own cluster. Pre-processing resulted in the signed graph w 7468 nodes, $77,117$ positive and $28,043$ negative edges. The dataset has low density (0.04) and 79.8\% of the triangles in the network are balanced.

\paragraph{Slashdot} Slashdot is a technology website that allows users to tag each other as “friends" or “foes."  The data set was curated by SNAP and contains 82,140 vertices and 549,202 edges \cite{snapnets}.  Duplicate and self-referencing edges were removed, and all vertices are connected in this dataset.  After pre-processing, the graph has the same number of vertices, 82,140, and now has $385,515$ positive, and $114,966$ negative edges. Half of the nodes have two or less edges, and the graph density is 0.1 \%. The graph qualifies as a large and sparse graph with a power-law degree distribution \cite{2021dallamico}. 

\paragraph{Epinions} Epinions.com is a website that hosts consumer reviews and employs a "trust/distrust" metric among users. Users may rate each other as trusted or not, and the resulting "Web of Trust" is used to weigh product reviews based on the reputations of the authors. The SNAP signed graph has $131,828$ vertices and $841,372$ edges \cite{snapnets}. Duplicate and self-referencing edges were removed from the data set.  After pre-processing, the Epinions dataset had $131,828$ vertices, $592,236$ positive and $118,974$ negative edges, and low density (under 0.1\%). The graph qualifies as a large and sparse graph with a power-law degree distribution \cite{2021dallamico}. 

\begin{table}[!ht]
\setlength\tabcolsep{1pt}
\begin{tabular}{l||c||c|c|c|c|c|c|c|c|c|c||c|c|c}
&  & \multicolumn{2}{c|}{Laplacian} & \multicolumn{2}{c|}{Balanced Cuts} & \multicolumn{2}{c|}{SPONGE } & \multicolumn{2}{c|}{FCSG} & \multicolumn{2}{c||}{graphB} & \multicolumn{3}{c}{graphB}\\
dataset & k & pos\_in & neg\_out & pos\_in & neg\_out &  pos\_in & neg\_out & pos\_in & neg\_out & pos\_in & neg\_out & k  & pos\_in & neg\_out\\ \hline \hline
cow  & 3 & \underline{\bf 0.99} & \underline{\bf 0.89} & 1.0 & 0.55 & 1.0 & 0.3  & 0.84& 0.25 & 0.98 & 0.58 & 3& \underline{0.98} & \underline{0.58}\\ \hline
wiki  & 30 & \underline{\bf 0.63} & \underline{\bf 0.67} & 0.9 & 0.17 & \underline{0.59} & \underline{0.71} & 0.49 & 0.52 & 0.05 & 0.96 & 4 & 0.29 & 0.73\\ \hline
slashdot  & 100 & 1.0 & 0.0 & \underline{0.96} & \underline{0.19} & 1.0 & 0.0 & N/A & N/A & 0.02 & 0.98 & 10 & \underline{\bf 0.22} & \underline{\bf 0.78} \\ \hline
Epinions  & 100 & 1.0 & 0.0 & \underline{\bf 0.96} & \underline{\bf 0.19} & 1.0 & 0.0 & N/A & N/A & 0.03 & 0.97 & 10 & \underline{0.13} & \underline{0.88} \\ 
\end{tabular}
\caption{Ratio of positive edges within clusters and negative edges between clusters for Laplacian symmetric \cite{Kunegis2009}), Balanced Normalized Cuts\cite{chiang_scalable_2012}), and symmetric SPONGE \cite{cucuringu_sponge_2019} on the CoW \cite{COW}, Wiki, Slashdot, and Epinions \cite{snapnets}. We present results for recommended $k$ \cite{mercado2019} and lower $k$ for graphB as fixed $k$ is not applicable for the method with emphasis on \underline{\bf best} and \underline{second best} scores.
}\label{tab:edge}
\end{table}

\subsection{How do best performing methods fare when scale and sparsity of the graphs increase?} 
\label{ssec:modularity}

\begin{figure}[!ht]
\centering
\includegraphics[width=2.8in]{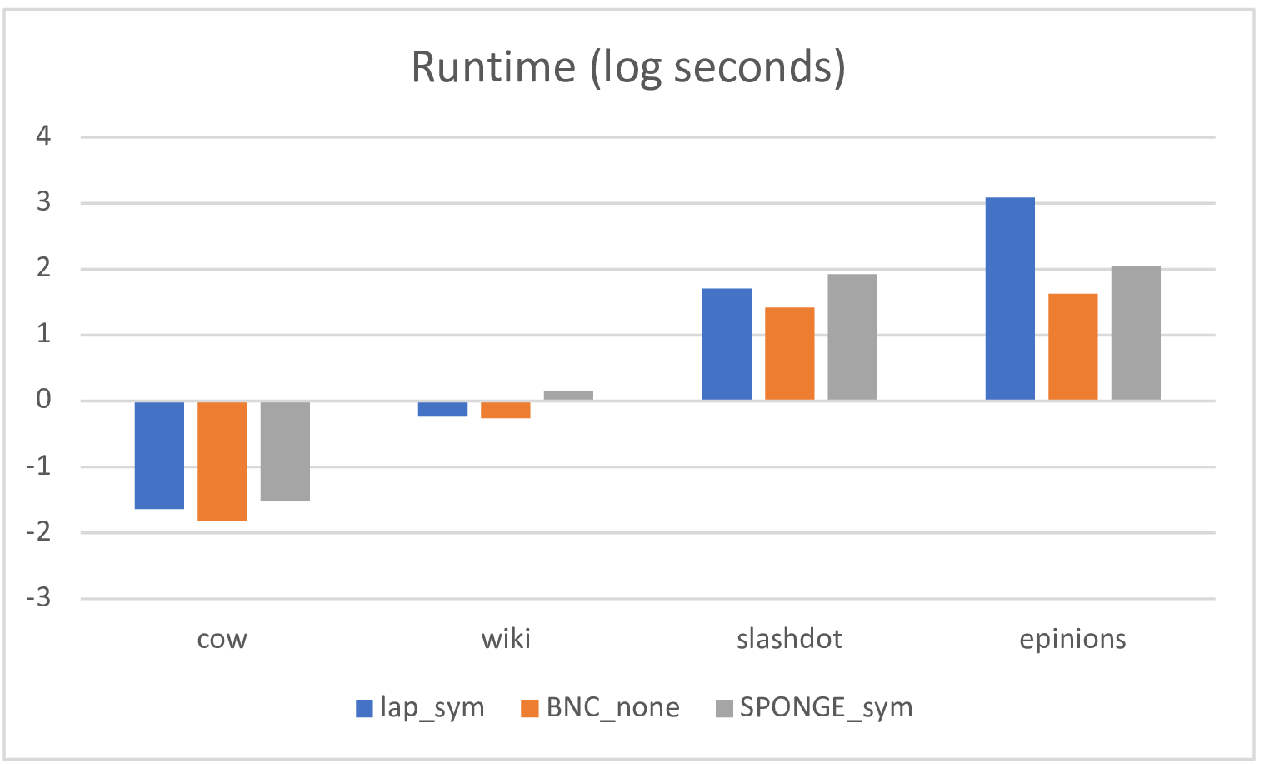}
\includegraphics[width=3.2in]{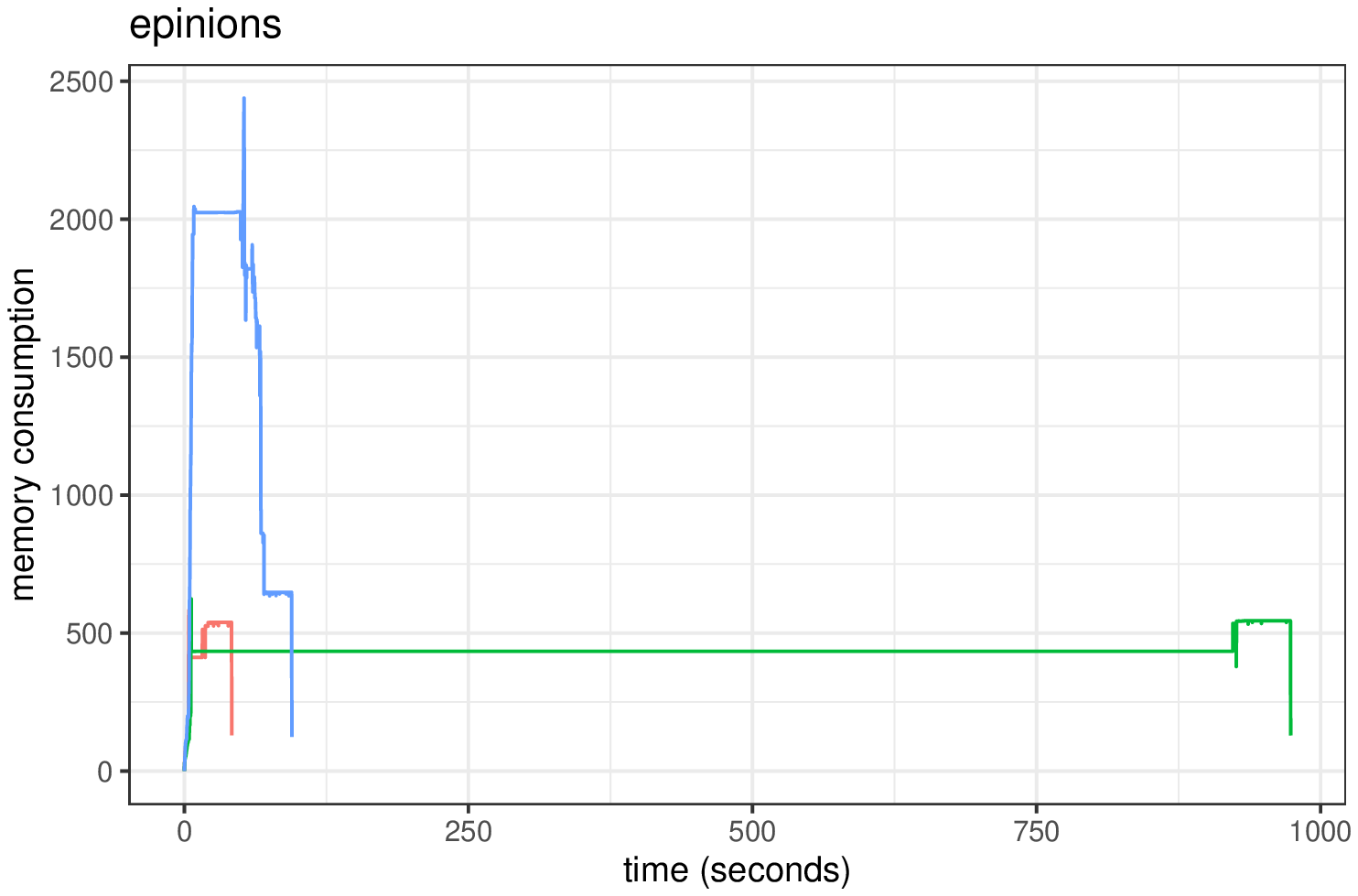}
\caption{Log run-times (left) for Laplacian symmetric (lap\_sym (\cite{Kunegis2009}),  Balanced Normalized Cuts (BNC\_none \cite{chiang_scalable_2012}), and SPONGE symmetric (SPONGE\_sym \cite{cucuringu_sponge_2019}) on the CoW (Correlates of War (\cite{COW}), Wiki, Slashdot, and Epinions (\cite{snapnets}) datasets; Memory consumption for all 4 methods on Epinion dataset (right). \label{fig:timing}}
\end{figure}

First, we evaluate the effectiveness of three most effective methods from Section~\ref{sec:exp1}, namely symmetric Laplacian, symmetric SPONGE and compare it to normalized Balanced Cuts. Community labels are not available so we are using pos\_in and neg\_out measures to access method's performance. We choose $k=30$ based on prior work \cite{mercado2019}. The performance of the methods is measured by computing the percent of positive edges placed within a cluster and the percent of negative edges placed between clusters for the newly discovered communities. The ratio of positive edges within clusters and negative edges between clusters that each algorithm produces is outlined in Table ~\ref{tab:edge}, and their runtime is outlined in Figure~\ref{fig:timing}. 

In a ground-truth clustering of a perfectly modular graph, both metrics would be 1 (100\%). Symmetric Laplacian and SPONGE show degradation in performance in Table ~\ref{tab:edge} on the Wikipedia Elections data as the sparsity of the graph increases and the vertex degree distribution begins converging to power law.  The Slashdot and Epinions clustering results in Table \ref{tab:edge} producing similar results. The low success metrics for negative edges between Epinions and Slashdot, combined with the low community size, suggest a breakdown in effectiveness across the three methods. While they classify almost all positive  edges within clusters, close to zero edges remain outside of the clusters. So for large datasets, the community cut is the random negative edge.  All the methods show relative fast run times in Figure~\ref{fig:timing}, with normalized balanced cuts being the most efficient method runtime and memory wise. 

Cucuringu et al. \cite{cucuringu_sponge_2019} note the difficulty in recovering community structure in sparse networks with spectral methods, even when normalization is introduced.  Dalla'mico \cite{2021dallamico} observes the following for large sparse network spectral analysis: (1) the eigenvalues of a sparse network tend to spread, which can obscure the largest and smallest eigenvalues and makes the informative eigenvalues difficult to isolate; and (2) high heterogeneity in the degree distribution modifies the $i^{th}$ entry of the informational eigenvectors in proportion with the degree of node $i$, known as "eigenvector pollution" by the authors \cite{2021dallamico}. Symmetric Laplacian $lap\_sym$ and symmetric SPONGE $SPONGE\_sym$ method implementations rely on the eigenvalue approximation. Eigenvalue approximation is notoriously \emph{unstable} for large matrices \cite{Knyazev2001TowardTO}, and the results we have obtained suggest the calculations in these three methods were so prone to error on the large data sets that meaningful community labels were not found. From this experiment, we conclude that the effectiveness of the spectral methods significantly degrades for large sparse networks, likely due to a combined effect of "eigenvector pollution" and cumulative error in approximate eigenvector computations for large sparse datasets. The Balanced Cuts method $BNC\_none$, does not fully break, but its capability to separate clusters by negative edges degrades for networks with a few thousand nodes. While it preserves the high pos\_in score for larger datasets, neg\_out score severely degrades for networks with few thousand nodes. 

\begin{figure}[!ht]
    \centering
    \includegraphics[width=2in, trim=50 50 50 50, clip]{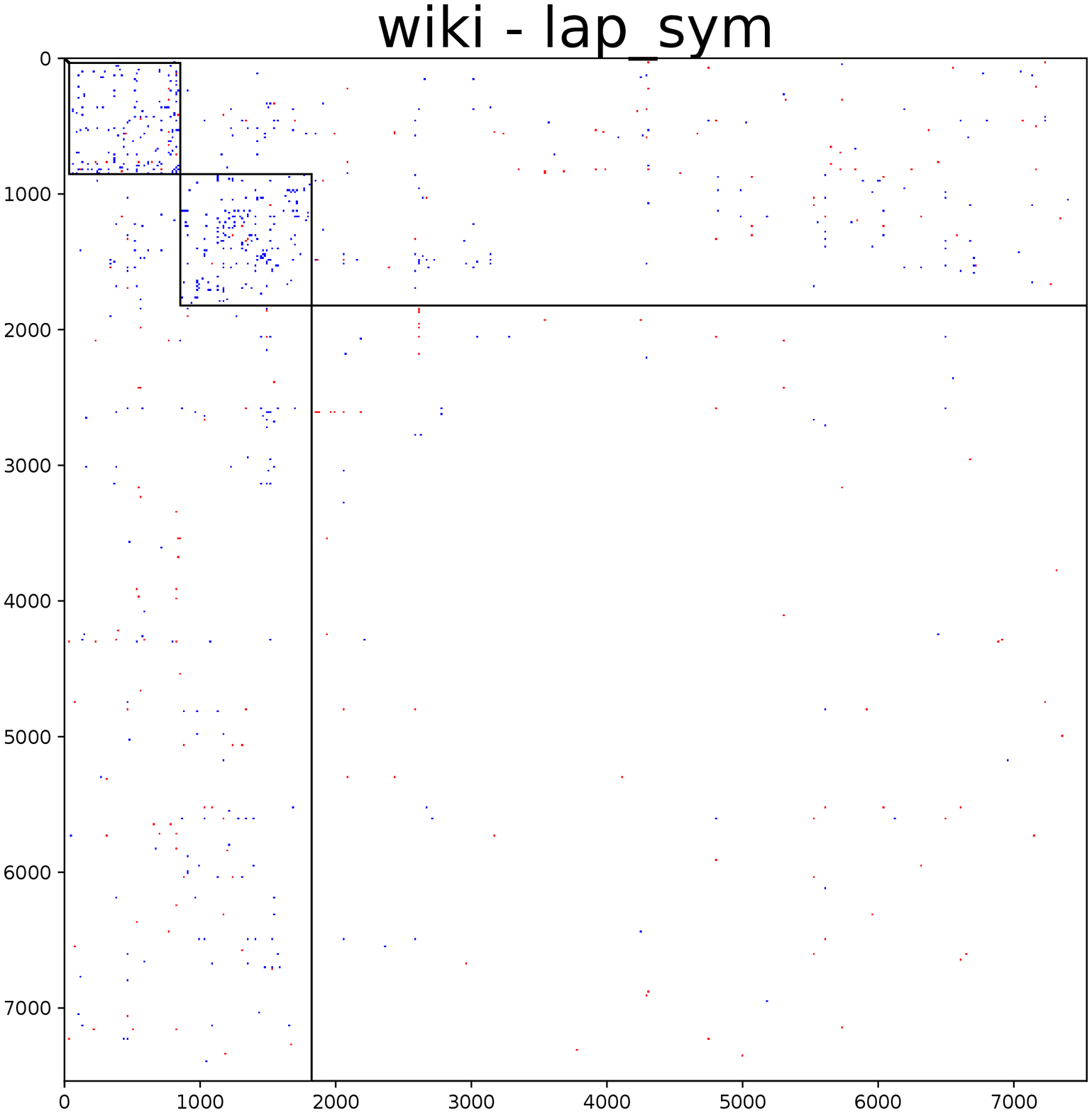}
    \includegraphics[width=2in, trim=50 50 50 50, clip]{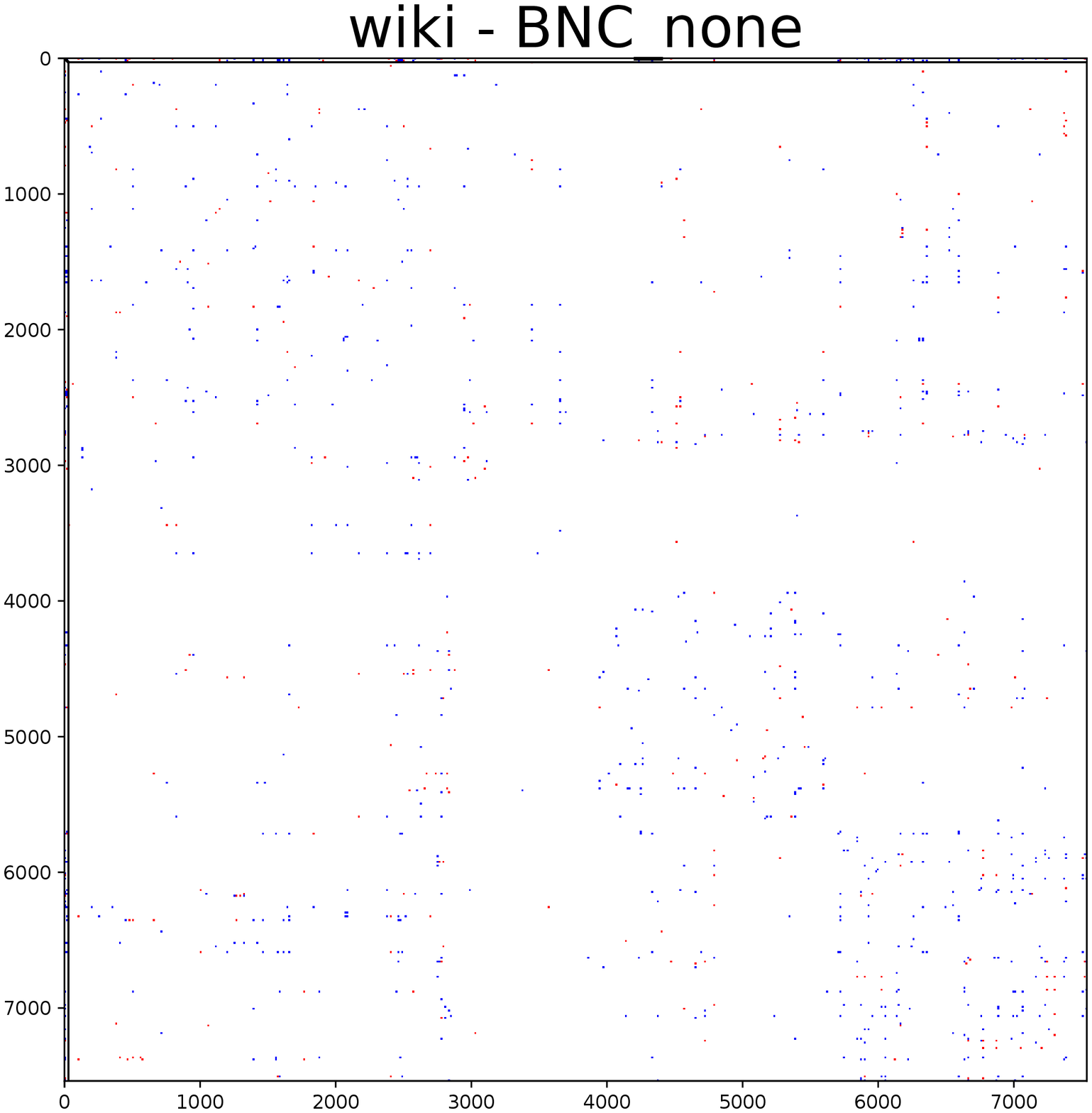}
    \includegraphics[width=2in, trim=50 50 50 50, clip]{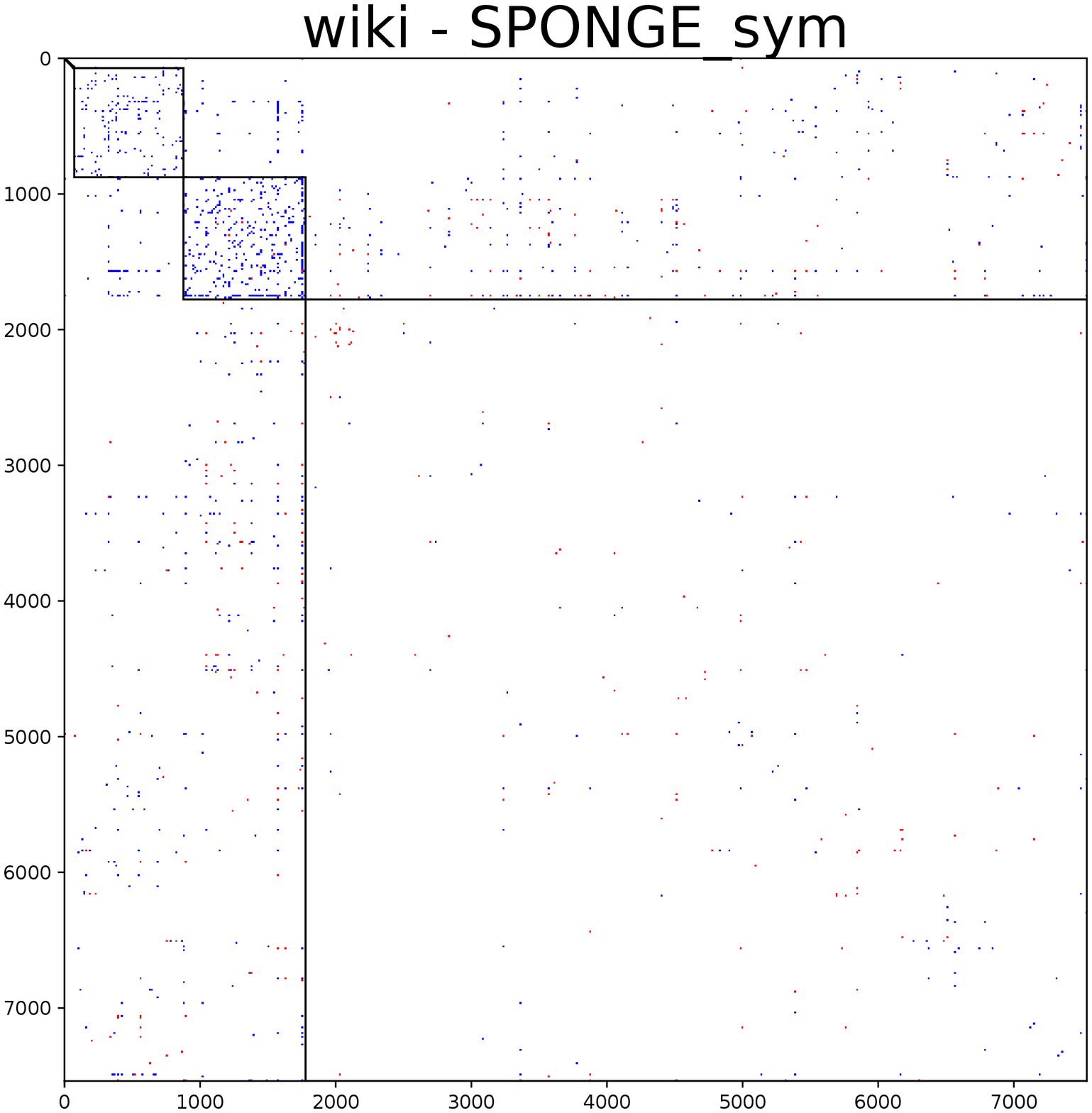}
    \caption{Blockmodels for Wiki \cite{snapnets} community discovery methods for (left) Laplacian symmetric \cite{Kunegis2009}, (center) Balanced Normalized Cuts\cite{chiang_scalable_2012}, and (right) symmetric SPONGE symmetric \cite{cucuringu_sponge_2019}. \label{fig:wikiBlock}
    }
\end{figure}

\begin{figure}[!ht]
    \centering
    \includegraphics[width=2in]{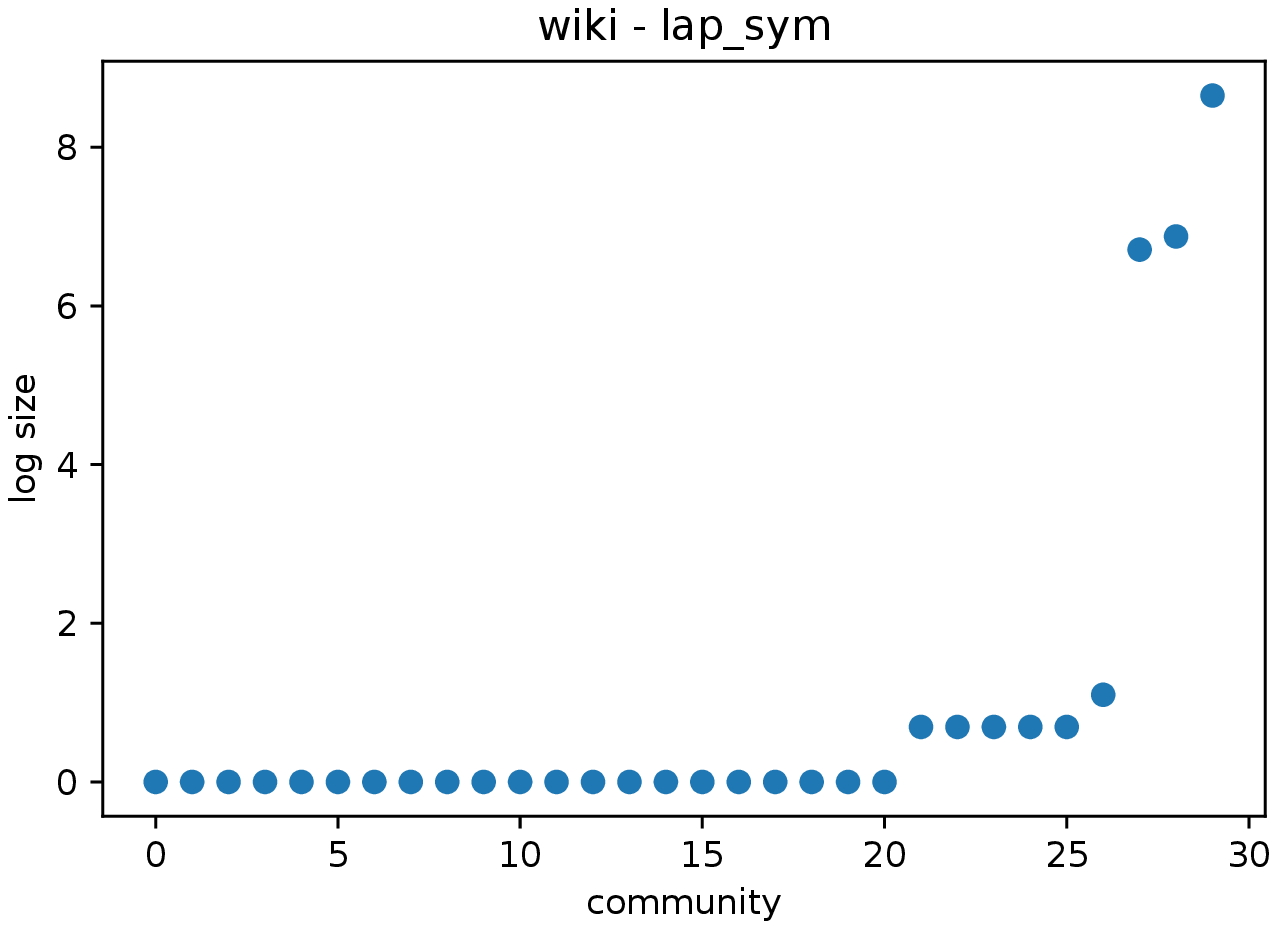}
    \includegraphics[width=2in]{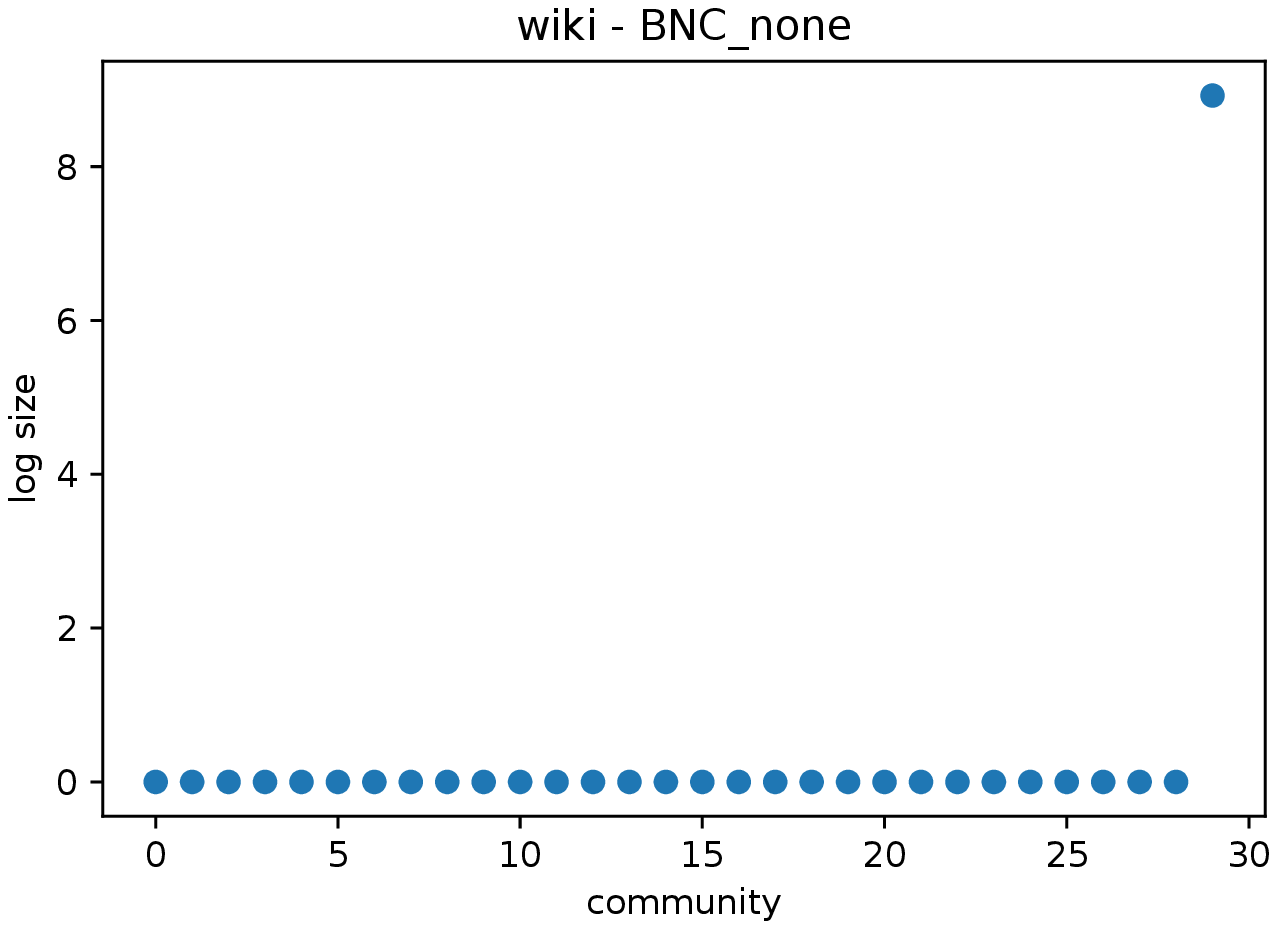}
    \includegraphics[width=2in]{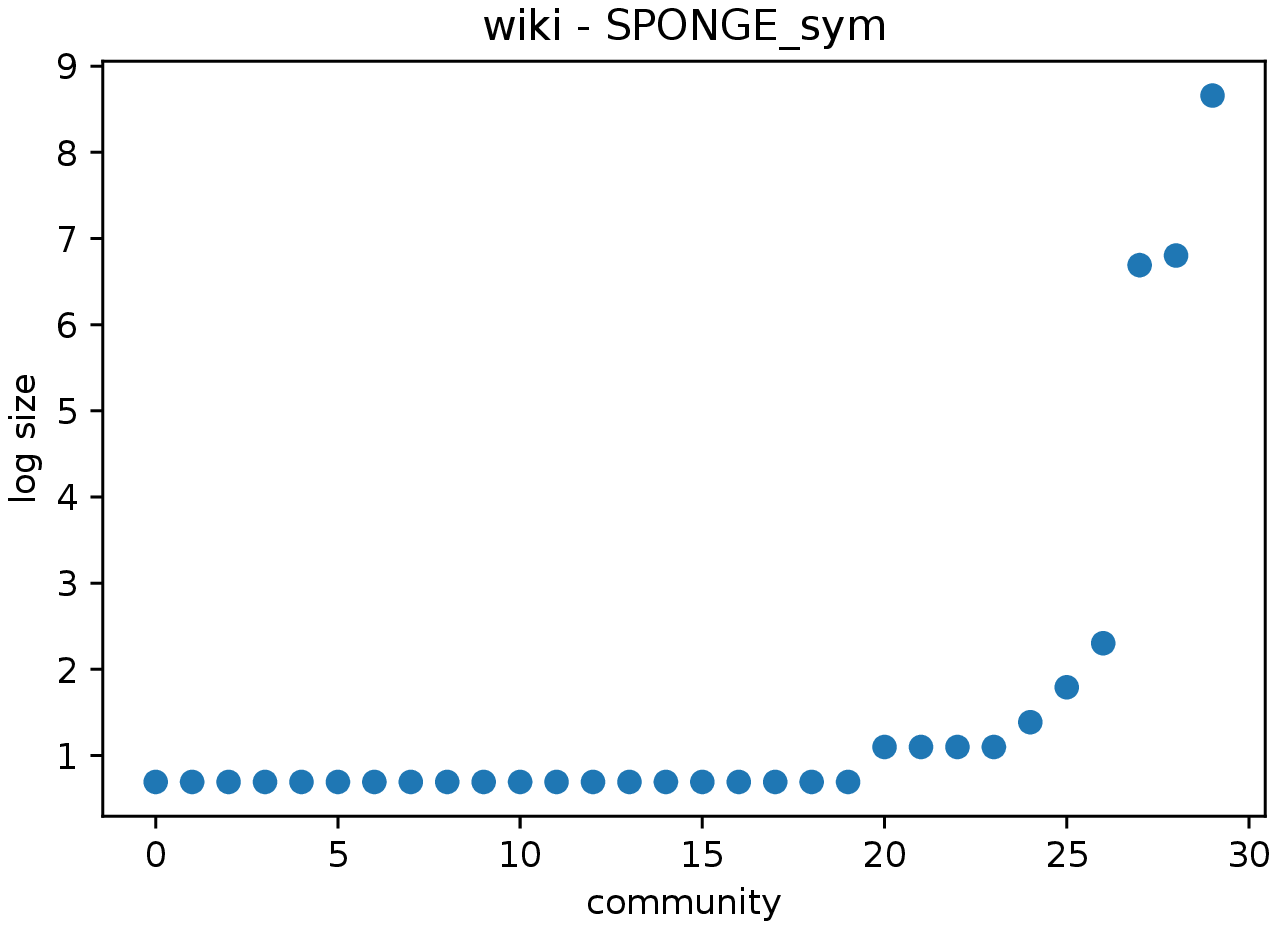}
    \vspace*{-1em}
    \caption{Community size on log scale in ascending order for Wiki community discovery for (left) Laplacian symmetric \cite{Kunegis2009}, (center) Balanced Normalized Cuts\cite{chiang_scalable_2012}, and (right) symmetric SPONGE symmetric \cite{cucuringu_sponge_2019}. \label{fig:wiki_cluster_size}}
\end{figure}

Next, we analyze the degradation of community assignments these three methods produced for Wiki dataset. The Wikipedia Elections data set clustering using the three methods is hard to visualize using blockmodeling due to the sparseness of the network, as illustrated in Figure~\ref{fig:wikiBlock}. The log scale of discovered community size for all three methods is illustrated in Figure~\ref{fig:wiki_cluster_size}. Figure~\ref{fig:wiki_cluster_size} reveals that all three methods struggled with small cluster sizes for large sparse graphs.  Figure~\ref{fig:wikiBlock}(center) contains 29 tiny communities and 1 large community, so the community lines are not visible as for  Figure~\ref{fig:wikiBlock}(left) and Figure~\ref{fig:wikiBlock}(right). The introduction of the signed normalized cut, as an alternative to the signed ratio cut, was intended to prevent clustering algorithms from producing trivial or near-trivial optimization solutions with either single nodes or pairs of nodes isolated into a cluster, but we can see that this remedy fails on highly sparse networks such as the Wikipedia Elections data set. While spectral clustering is a powerful technique for the detection of graph communities, the eigenvalues of signed graphs present a substantial obstruction in the development of a parallel spectral theory that is meaningful for the real social network data \cite{2021dallamico}. Normalized balanced cuts approach did not scale to sparse networks and trivial communities are prioritised in the clustering.  We conclude that none of the approaches that works well on the small dense graphs scales to real large sparse networks.  

\subsection{Evaluation of scalable methods for community detection in large sparse graphs} 
\label{ssec:scale}

\paragraph{FCSG} Fast Clustering for Signed Network (FCSG) implementation and its limitations are described in Section~\ref{ssec:randomwalk}. The authors did not provide an implementation, and the paper did not discuss efficiency strategies \cite{hua_fast_2020}. Our in-house implementation follows the paper guidelines when possible, as outlined in  Alg.~\ref{alg:RWG} and Alg.~\ref{alg:FCSG}. The approach assumes that the positive-only subgraph of the network must be a single connected component. This places a large constraint on running the algorithm on a real data set, and reduced FCSG ARI for all the data but highland (ARI=1.0) in Section~\ref{sec:exp1}.  FCSG results are outlined in Table~\ref{tab:exp2}. For the {\bf cow} dataset, the number of vertices in largest positive connected component (see Alg.~\ref{alg:RWG}) is 21, and 29 vertices are all assigned the same community label. This explains the high pos\_in score (0.84) and low neg\_out score (0.25) as the algorithm does not consider most of the vertices. 

Sparse graphs do not conform to small-world hypothesis, a theory that most users are linked by no more than 5 degrees of separation in a social network. The diameter of largest positive connected component for the {\bf wiki} dataset is 8, for the {\bf slashdot} dataset is 11 and for the {\bf Epinion} is 14 \cite{snapnets}. The authors recommend that ${L}$ be set to 5 and warn that the algorithm begins to degrade in quality for ${L} > 5$ and is theoretically unsound for ${L}$ greater than or equal to 10. The recommended value of ${L}$ cannot be used for Slashdot and Epinion for step 4 of Alg.~\ref{alg:RWG} as the parameter used in the random walk gap matrix calculation ${L}$ must be greater than or equal to the diameter of the all-positive subgraph of the input graph. For the {\bf wiki} dataset, FCSG considers 6530 out of total of 7468 wiki graph vertices for community clustering. The diameter of this positive-subcommunity is 8. Because the diameter value we are using is higher than 5, we see that algorithm begins to degrade for the $wiki$ data as pos\_in score is 0.49 and neg\_out score is 0.52. 

FCSG algorithm runs under a minute for small graph processing on a local workstation. The implementation did not scale to the thousands of vertices in the Wikipedia Election data set \cite{snapnets}, and experiments were run for \emph{over 2 weeks}  on the Texas State University LEAP system \citep{LEAP}. The Dell PowerEdge C6320 cluster node consists of two (14-core) 2.4 GHz E5-2680v4 processors with 128 GB of memory each, and two 1.5TB memory vertices with four (18-core) 2.4 GHz E7-8867v4 Intel Xeon processors \citep{LEAP}. The implementation ran out of memory when tried on {\bf slashdot} and {\bf Epinion}. The algorithm is irregular, so we used a serial implementation for proof-of-concept as parallelization was non-trivial. We had to implement a custom merge function for this algorithm that did not scale, resulting in N/A entries in Table~\ref{tab:exp2} for slashdot and Epinion data. 

\paragraph{graphB} Graph balancing approach uses Balanced Cuts to bipartition the network into agreeable (all positive clusters) \cite{2020Cloud}; a posteriori application of $k$-means will necessarily reduce positive edges. Table~\ref{tab:exp2} shows that graphB performance on {\bf cow} is comparable to Balanced Cuts, pos\_in is 0.98, neg\_out is 0.58. For the wiki dataset, for chosen $k=100$ graphB is forcefully separating established clusters resulting in pos\_in is 0.05, neg\_out is 0.96. As implementation of hierarchical clustering is beyond the scope of this paper, we chose $k=4$ and graphB results improve to pos\_in 0.29, neg\_out 0.73. We observed the similar trend for slashdot and Epinion dataset for $k=100$ and $k=10$. graphB clustering performance improves when more optimal $k$ is chosen, see Table~\ref{tab:exp2} for more details. 

Graph Balancing community discovery takes under a minute on the highland, wiki, and sampson datasets on a local workstation. The implementation (underlying NetworkX calls) did not scale to the thousands of vertices in the Wikipedia Election data set \cite{snapnets}. Since we typically process 1000 trees, one of the bottlenecks was reading and writing the h5 files \cite{graphB}. We mitigated this bottleneck using Apache Spark to parallelize the file operations and achieved a speedup of 12.5 over serial processing. Wiki election experiments were completed in 96 minutes on the Texas State University LEAP system \citep{LEAP} using this released timing reproducibility benchmark of the code \cite{graphB}. Processing the slashdot and Epinions datasets on LEAP took approximately 100 hours on LEAP for these datasets and exposed excessive memory consumption of the implementation. Since the number of fundamental cycles tends to grow with the size of the graph, this quickly became excessive. The authors have proposed leaner more scalable solutions to be used for networks with millions of users and billions of edges \cite{2021Alabandi}.

FCSG and graphB scores are complementary on the wiki data, and further studies are needed to determine the right synergy of the methods, and how to overcome small world assumption for FCSG that does not hold for real graphs.  \emph{graphB} offers dual metrics for vertices and edges \cite{2020Cloud}, and that is guiding us in the direction of hierarchical clustering for community discovery for large sparse graphs. 

\section{Conclusion and Future Work}
\label{sec:conclusion}

\begin{table}[!ht]
\setlength\tabcolsep{2pt}
\begin{tabular}{p{0.14\textwidth}||p{0.3\textwidth}|p{0.4\textwidth}}
\textbf{Scope} &  \textbf{Issue} & \textbf{Mitigation Approach}\\ \hline \hline
Evaluation & No clear preferred method across some datasets &  Community-agreed large and sparse dataset benchmarks \\
Reproducibility & Cannot reproduce the claimed results & Artifacts for code availability and result reproducibility to accompany research products.\\
Scale & Methods don't scale to large and/or sparse networks &  Methods advancement needed, good starting points \cite{hua_fast_2020,2022sssnet,2022Cluster}.\\ 
\end{tabular}
\caption{This survey helped us identify three issues with state-of-art signed graph clustering, and propose mitigation plan.}\label{tab:conclusion}
\end{table}

Scalable, effective, and reproducible signed clustering algorithms for community detection on the networks have been a prominent focus in social network analysis research in the past decade.  In this paper, we have selected eight real world graph datasets and characterized signed graphs in terms of percent of positive edges, vertex degrees, percentage of balanced triangles, vertex degrees, and density. These real-world networks and their summative and comparative statistics provide new benchmarks to build synthetic datasets to examine the strengths and weaknesses of each method. We have compared \emph{twelve} signed graph clustering methods for community discovery. We have evaluated them in terms of effectiveness: ARI scores when ground truth is available, and percentages of positive edges in (pos\_in) and negative edges among (neg\_out) the discovered clusters when ground truth is not available. 

First, we have compared the efficiency of the algorithms on five real social media graphs with ground truth community labels, and found the top three performers over a wide range of data with ground-truth labels in Section~\ref{sec:exp1}. Second, we showed in Section~\ref{ssec:modularity} that the state-of-the-art methods of approximate eigenvector calculation do not scale as they retrieve trivial communities. The cumulative errors in the algorithm prevent the meaningful interpretation of output results, as shown in Section~\ref{ssec:scale}.  We used author-provided code when available, but many of the techniques discussed in this paper did not have the original code used to perform the published experiments. This was especially notable for the large real signed networks Epinions and Slashdot, which performed poorly across the board in our experiments but have more promising results published in the literature. The computationally expensive family of techniques that did not rely on local minima or maxima showed potential, but the provided implementation details were too sparse to reproduce the results, as shown in Section~\ref{ssec:scale}. Another central flaw is that algorithms make assumptions that limit the applicability of some techniques to real-world data, such as the small world assumption in Section~\ref{ssec:scale}.  

We can group the conclusions from the study in three items, as outlined in Table~\ref{tab:conclusion}.
This real-world network study provided us many new avenues to explore. We hope to expand and inform the generation of synthetic signed networks to highlight the differences between each of these methods, implementing a scalable process similar such as \cite{2020Jung}. The next step is to \emph {quantify the network attributes that will predict if spectral methods will fail for community detection} We are looking into the improvement of non-spectral, hierarchically-robust, methods to synergize with the spectral methods we have studied.

\end{document}